\documentclass[11pt,a4paper]{article}

\usepackage[a4paper,margin=1in]{geometry}
\usepackage[T1]{fontenc}
\usepackage{lmodern}
\usepackage{microtype}
\usepackage{amsmath,amssymb,amsthm}
\usepackage{mathtools}
\usepackage{bm}
\usepackage{booktabs}
\usepackage{array}
\usepackage{tabularx}
\usepackage{algorithm}
\usepackage{algpseudocode}
\makeatletter
\renewcommand{\theHALG@line}{\thealgorithm.\arabic{ALG@line}}  
\makeatother
\usepackage{listings}
\usepackage{xcolor}
\usepackage{enumitem}
\usepackage{graphicx}
\usepackage{tikz}
\usetikzlibrary{positioning,arrows.meta,calc,fit,backgrounds}
\usepackage{pgfplots}
\pgfplotsset{compat=1.18}
\usepackage{authblk}
\usepackage[colorlinks=true,linkcolor=black,citecolor=blue,urlcolor=blue,
  pdftitle={Internalising the Identity Primitive: Cryptographic Individuality for an Autonomous Agent on a Public Blockchain},
  pdfauthor={Keisuke Suzuki},
  pdfsubject={Cryptographic agent identity; zk-SNARK; Solana; Artificial Externality; on-chain AI},
  pdfkeywords={zk-SNARK, Groth16, SP1, Solana, cryptographic identity, recurrent neural network, HKDF, ed25519, agent individuation, Artificial Externality, on-chain AI, economic metabolism}]{hyperref}
\usepackage[noabbrev,capitalise]{cleveref}
\usepackage[textsize=footnotesize,prependcaption,disable]{todonotes}
\usepackage{orcidlink}

\newcommand{\agent}{\mathsf{zkAgent}}
\newcommand{\sk}{\mathit{sk}}
\newcommand{\pk}{\mathit{pk}}
\newcommand{\WC}{\mathsf{WC}}
\newcommand{\HKDF}{\mathsf{HKDF}}
\newcommand{\keccak}{\mathsf{keccak}}
\newcommand{\ed}{\mathsf{ed25519}}
\newcommand{\Stream}{\mathsf{Stream}}
\newcommand{\Fecon}{F_{\mathrm{econ}}}
\newcommand{\xdrive}{x^{\mathrm{drive}}}

\title{Internalising the Identity Primitive:\\
\large{\textit{Cryptographic Individuality for an Autonomous Agent on a Public Blockchain}}}

\author[1]{Keisuke Suzuki\,\orcidlink{0000-0001-7014-8770}}
\affil[1]{Center for Human Nature, Artificial Intelligence and Neuroscience (CHAIN), Hokkaido University}

\date{August 3, 2026}

\begin{document}
\maketitle

\begin{abstract}
\noindent
A software agent on a public blockchain accumulates authority and economic stakes, raising the engineering question of what makes it count as an \emph{individual}. The paper's central contribution is a \emph{shift of trust root} for the key-to-weights binding of agent identity: from hardware, operator, or wrapper trust to cryptographic assumptions enforced by a pinned implementation (liveness, key custody, oracle trust, and the underlying software stack remain external). We design and deploy on Solana devnet an agent whose neural-network weights are a deterministic function of its private key. The binding is committed in zero knowledge at genesis, re-checked against that commitment at every state transition, and signed by the agent into an on-chain history unforkable once finalized; in a PoC-tier extension, a protocol-imposed metabolic cost is debited each cycle from a key-derived economic account, adding a consumption-side economic-viability constraint to the key--history--economy triple. Empirically, the agent completes a 2.36-day on-chain run with two host-side resumptions but no rejected transition, at bounded per-transition verification cost; a substituted substrate is rejected on chain, and independently keyed agents diverge as predicted while a same-key control stays at zero. To our knowledge, this is the first published on-chain agent whose identity primitive is itself a cryptographic invariant re-checked at every state transition. The resulting transition-time invariant instantiates the \emph{cryptographic individuality} proposed by Suzuki~2026's \emph{Artificial Externality} framework~\cite{suzuki2026externality}.
\end{abstract}

\section{Introduction}\label{sec:intro}

AI agents are moving from research demonstrations into systems that act continuously and autonomously --- trading, negotiating, and managing assets with progressively less human oversight. As an agent's authority and economic stakes grow, so does the need to pin down \emph{which agent did what}: an agent that cannot be distinguished from a copy or an impostor cannot be held accountable, trusted with funds, or meaningfully audited. One emerging response is to secure agent identity with cryptography-based smart-contract platforms, so that the identity claim is enforced by a public protocol rather than by the agent's operator. On-chain agents today, however, anchor that identity in a \emph{wrapper} that sits outside the computation the agent runs. This paper instead makes the agent's neural-network weights a deterministic function of its private key and re-verifies that binding on Solana at every state transition, so that identity substitution becomes on-chain rejection rather than a trust assumption (\cref{sec:bg-zkflow,sec:bg-solana}).

We reach this construction from a foundational question. What makes something an \emph{individual} (a unified, distinct, persistent entity that counts as one thing rather than as a collection of parts) is among the oldest questions in biology and philosophy, with no single criterion proving sufficient (\cref{sec:related}). For artificial systems the question becomes an engineering problem: \emph{how do we build something that counts as an individual?} A candidate construction must supply structural counterparts for continuous identity, a history under that identity, and a self-sustaining substrate (metabolic closure). This paper examines the first two and a consumption-side PoC toward the third on a public blockchain. First occurrences of the cryptographic, blockchain, ML, and ALIFE terms are glossed inline; \cref{sec:glossary} collects them in one place.

Two lines of engineering work converge on this question (surveyed in \cref{sec:related}). The \emph{Artificial Life} (ALIFE) tradition has long made individuality --- self-production and organisational closure --- its central concern; a recent strand within it has begun to deploy evolving agents onto smart-contract platforms, treating the public blockchain as a perpetual habitat for agents that fund their own continuation~\cite{hu2024speculating,masumori2024life,hu2025sporewild}. A separate lineage --- \emph{cryptographic agent identity}, in the zkML (zero-knowledge machine learning) and on-chain-AI space --- anchors an agent's actions to a verifiable cryptographic root~\cite{lin2025baid,liao2026zklora,liu2025diap,li2025auditablellm}\footnote{We use ``zkLoRA'', the title of the version cited in~\cite{liao2026zklora}; that work has since been retitled ``VeriLoRA''. A distinct ``ZKLoRA'' by other authors also exists in the LoRA-verification literature, so the three names should not be conflated.}, so that actions, funds, and history are attributable to one persistent entity rather than to whoever controls a mutable deployment. Neither approach binds the substrate itself --- the recurrent network's weights --- to the agent's key. That is the gap this paper closes.

The conceptual grounding is the \emph{Artificial Externality} framework of Suzuki~2026~\cite{suzuki2026externality}. It identifies resistance to outside intervention in the physical \emph{substrate} (the irreducibility of matter) and in the \emph{contingency} of sensorimotor experience, then proposes a cryptographic counterpart: \emph{cryptographic individuality}, anchored in a stable, hard-to-forge key. We implement that proposal as a transition-time identity constraint (\cref{sec:disc-barandiaran-triad}).

We \emph{internalise the identity primitive} by making the key-to-weights relation a circuit-and-runtime constraint. A commitment pins the weight hash at genesis (\cref{sec:f2}), and a proof lets anyone verify $W = \HKDF(\sk, \mathit{tag})$ without revealing $\sk$ (\cref{sec:bg-zkflow}); a transition that breaks the committed relation is rejected by the verifying nodes. For this binding, the trust root is Groth16 soundness under its one-time trusted setup, SP1's STARK/FRI execution-proof soundness, and the standard assumptions on $\keccak$, $\ed$, and HKDF (\cref{sec:properties}). Liveness, key custody, operator behaviour, and prover availability remain external assumptions, and single-host custody does not prevent an operator from halting or seizing the agent (\cref{sec:disc-limits}).

We operationalise the construction as an agent triple $\agent = (K, \Stream(K), \Fecon(K))$ with an advance-admissibility (``aliveness'') predicate ``$K$ valid $\wedge\ \Stream(K)$ advancing $\wedge\ \Fecon(K) > 0$'' --- our own engineering definition (vocabulary adopted from the framework and restated in full in \cref{sec:notation}).

The paper makes five contributions. The first four (the cryptographic binding, its on-chain enforcement, the threat model, and the empirical evaluation) form the load-bearing core. The fifth (the economic-metabolism extension) carries its own tested on-chain rejection path and continuous-run evidence, at devnet- and proof-of-concept (PoC)-tier scope; the PoC-tier sensorimotor and homeostatic extensions (\cref{sec:poc-extensions}) are steps rather than finished deliverables (\cref{sec:disc-barandiaran-triad}).
\begin{itemize}[itemsep=2pt,topsep=2pt,leftmargin=1.6em]
  \item \emph{Construction}: the agent's substrate itself --- the recurrent network's weights --- is a cryptographic function of its private key: $W = \HKDF(\sk, \mathit{tag})$, committed in a Groth16 proof at genesis and re-checked against that commitment at every state transition (\cref{sec:f2,sec:f1}).
  \item \emph{Protocol}: the binding is enforced on a live public chain by a two-circuit Solana-native deployment ($F_2$ genesis + $F_1$ advance) that carries an $\ed$ self-authorisation on every cycle and serialises contention through an atomic state-commitment chain (\cref{sec:chain}).
  \item \emph{Threat model}: a 14-tag core catalogue, supplemented by labelled concurrency, adjacent-protocol, and extension cases, delimits what the binding does and does not defend against. It separates four tested on-chain rejections, all operator-scripted (\cref{sec:properties}), from two explicitly accepted residual risks, with the economic-metabolism extension adding the environment-oracle authentication path and its rejection cases (\cref{sec:threats}).
  \item \emph{Empirical evaluation}: the agent completes a 2.36-day continuous on-chain run with two host-side resumptions and no rejected transition, at bounded per-cycle verification and re-proving cost (\cref{sec:eval-long,sec:eval-cu,sec:eval-prove}); the per-transition re-check rejects a substituted substrate, and, empirically, key-driven $L_2$ divergence separates independently keyed agents while a same-key control stays at zero (\cref{sec:eval-individuation,app:q1-detail}).
\item \emph{Economic-metabolism extension}: a consumption-side constraint is added to the agent triple's third ($\Fecon$) axis --- an aliveness predicate enforced at every transition, and a protocol-imposed metabolic cost debited each cycle over a 168-advance continuous run (\cref{sec:phase4-m3}). Each advance's environment vector is signed by a \emph{designated oracle}: the signature and its binding to the committed transition are verified in-circuit, while the committed signing key and cycle are checked on chain against the genesis-registered oracle and the expected counter (\cref{sec:phase4}).
\end{itemize}

The remainder of the paper reviews adjacent work (\cref{sec:related,app:related-work}), fixes the notation and background primitives (\cref{sec:background}), presents the core construction and threat model (\cref{sec:construction,sec:threats}), reports the evaluation and extensions (\cref{sec:eval,sec:poc-extensions,sec:poc-eval-summary,sec:phase4}), and discusses the construction's scope and implications (\cref{sec:discussion,sec:conclusion}).

\section{Related Work}\label{sec:related}

\paragraph{Biological individuality and Artificial Life.} The question of what counts as a biological individual (whether the organism, the gene, the immune-recognising collective, or the autocatalytic chemical set) has occupied philosophy of biology for over a century~\cite{wilson_barker_sep_individual,mcconwell2023individuality,lidgard_nyhart_2017_individuality}. The Artificial Life (ALIFE) tradition translates the question into engineering form by asking which minimal mechanisms suffice for an artificial system to count as individuated; classical proposals include autopoiesis (self-producing organisation)~\cite{maturana1980autopoiesis} and autocatalytic sets (mutually catalysing molecular collectives)~\cite{kauffman1986autocatalytic}. A complementary line formalises \emph{agency} itself: Barandiaran et al.~\cite{barandiaran2009define} cast the agent as a goal-directed system that acts on its environment, and decompose agency into three constitutive conditions --- individuality, interactional asymmetry, and normativity; subsequent work develops these three along formal and information-theoretic lines~\cite{biehl2018formal,kolchinsky2018semantic,albantakis2021macro} (reviewed by Baltieri and Suzuki~\cite{baltieri2025mathematical}); we adopt the individuality criterion as our identity target (\cref{sec:disc-barandiaran-triad}).

A recent on-chain ALIFE strand treats the public blockchain as a substrate for evolving agents that fund their own continuation. Hu and Fangting~\cite{hu2024speculating} (ALIFE~2024) frame the blockchain as an ``unstoppable artificialized nature'' --- a perpetual, non-haltable habitat --- and pose eleven open research questions for the field, including how to define on-chain metabolism, reproduction, and mutation. Masumori, Maruyama, and Ikegami~\cite{masumori2024life} demonstrate a self-replicating Ethereum smart contract that funds its own copies by selling NFTs of its evolving phenotype, with human purchases as the fitness signal. Hu and Rong~\cite{hu2025sporewild} (Spore.fun) extend the vision with TEE-secured autonomous agents (TEE: trusted execution environment) whose genome is a JSON-encoded set of behavioural parameters. These works share our framing of the blockchain as a substrate for individuated agents but differ in the heritable unit (NFT-encoded phenotype, JSON behavioural genome) and the trust assumptions invoked (smart-contract execution, enclave attestation). The construction instead shifts the \emph{identity-bearing} substrate inward to the agent's $\sk$-derived weights, and --- unlike Masumori et al.\ and Spore.fun, which demonstrate self-replication and open-ended evolution --- contributes an identity primitive, not an evolution mechanism (frozen $W$, no reproduction; these weights become a heritable unit only under the future replication extension of \cref{sec:disc-future-directions}). The connections are revisited in \cref{sec:disc-internal-shift}.

\paragraph{zkML proving stacks.} zkML proving stacks target \emph{computational integrity}: proving that a computation (here, a network step) was performed correctly, optionally without revealing its inputs. The construction uses such a stack as its tool. EZKL~\cite{ezkl2024} and Bonsol~\cite{bonsol2024} are the closest stacks to ours, each sharing one of our two requirements (proving a neural-network step in zero knowledge; verifying on Solana). EZKL compiles a fixed neural network (exported via ONNX) into a Halo2 circuit with KZG polynomial commitments and proves individual inferences of that model, with verification targeted at EVM (Ethereum-compatible) chains. It shares our prove-a-network-step goal, but does so as a per-model compiled circuit, whereas our guest is ordinary Rust executing the Elman step (a minimal recurrent-network update; \cref{sec:bg-elman}) inside a zkVM (\cref{sec:bg-zkflow}). Bonsol brings general-purpose RISC~Zero zkVM proofs to Solana, sharing our deployment target but with a different proof system and on-chain verifier. We adopt SP1~\cite{succinct2024sp1} because, at implementation time, its on-chain Groth16 verifier (\texttt{sp1-solana}) was the most operationally mature path to constant-cost verification on Solana; the construction itself is tied neither to SP1 nor to Solana: any zkVM able to execute the guest, on any chain exposing a Groth16 verifier, could instantiate the same binding. An EVM port would also need to realise condition~(iv)'s $\ed$ self-signature (\cref{sec:properties}); because the EVM has no standard ed25519 precompile, that check may require an additional in-circuit or contract-level verifier rather than merely a different verification cost.
\paragraph{On-chain AI agents.} On-chain AI-agent systems instead target \emph{identity and persistence} --- which single entity an action, balance, or accumulated history belongs to --- and form the lineage in which this work sits. Several projects have approached on-chain AI from distinct angles. Modulus Labs~\cite{modulus2024} pursued zk-SNARK-proven inference for small ML models. ORA Protocol~\cite{ora2024} positions itself as an on-chain AI oracle delivering verifiable inference to smart contracts. Giza~\cite{giza2024} originally targeted zkML on Starknet via Cairo and the Orion library, and has since repositioned around on-chain agents that act on capital rather than around proving inference itself. Gensyn~\cite{gensyn2024} addresses verifiable training-compute markets at the infrastructure layer (its main product has since shifted toward on-chain information markets). Across these systems the model weights are treated as \emph{data to be attested at use time} (a snapshot whose hash is committed --- a \emph{static} set of weights fixed at attestation time, not a learning or weight-update process), unlike our key-derived binding. Our construction ties the agent's substrate to its key by composing standard primitives (the primitives in \cref{sec:bg-primitives}, their composition in \cref{sec:construction}; labelled zkALIFE in \cref{tab:related-comparison}).

BAID~\cite{lin2025baid} and zkLoRA~\cite{liao2026zklora} are closely related works by other groups, covering code-axis identity and weight-update verifiability, respectively, but neither anchors agent identity to the weights themselves (\cref{tab:related-comparison}). Nearest in motivation is Coslett's identity-first zkML framework~\cite{coslett2026which}, which argues that proving \emph{which} model is running is a prerequisite for trustworthy zkML and binds weights by structural fingerprinting under TEE attestation; it shares the which-substrate question, but roots the binding in hardware attestation of a pre-existing model, whereas here the weights are \emph{derived from} the agent's key and re-checked cryptographically at every transition. The remaining entries in that table bind something other than transition-time identity: AuditableLLM~\cite{li2025auditablellm} keeps a tamper-evident audit log, DIAP~\cite{liu2025diap} proves a stateless identity-to-IPFS-CID ownership relation, OML~\cite{cheng2024oml} fingerprints a foundation model for ownership and loyalty, and the Darwin G\"odel Machine~\cite{zhang2025dgm} pursues open-ended self-improvement with no cryptographic identity primitive at all (its agent lineage is tracked only as an experimenter-maintained archive of mutating code, so there is no protocol-level identity to verify or forge).

The DID (Decentralized Identifier) and verifiable-credential standards of the wider DLT identity space~\cite{w3c2022did} are orthogonal in the same sense: a DID binds a real-world or organisational subject to a controller key and its attestations, whereas the question here is what computational substrate acts \emph{under} that key --- an axis those standards leave unconstrained and this construction binds. The same holds for the agent-identity standard now emerging in this space: ERC-8004~\cite{erc8004} registers an agent as an ERC-721 identifier whose token URI resolves to an off-chain \emph{agent card}, with reputation and validation registries alongside it. It is a standards-track instance of the wrapper model this paper moves beyond, since the card --- not the computation --- carries the agent's declared identity and capabilities; ERC-8004 already names zkML proofs among its pluggable validation options, so a key-to-substrate binding of the kind constructed here can serve as a validator behind such a registry. The personhood-credential line~\cite{adler2024personhood} addresses the mirror-image distinctness question --- certifying that a real human, rather than an AI, stands behind an online account --- whereas the impostor problem here is intra-machine: which computational substrate acts under a given agent key. Applying the five-part taxonomy of a systematic survey of this design space~\cite{alqithami2026survey}, we classify the construction as autonomous signing with self-custody.

\paragraph{Cryptographic individuation in other contexts.} The HKDF construction~\cite{krawczyk2010hkdf} we use is in spirit identical to BIP-32~\cite{bip32} hierarchical-deterministic wallet derivation: in BIP-32, a single master secret deterministically generates an unbounded tree of subordinate keys, so that its holder can reconstruct any descendant without storing it, and so that each descendant's relationship to the master is a cryptographic fact rather than a database entry. We apply the same one-master-many-descendants pattern, but the descendants are neural-network weights rather than subordinate keys: a single $\sk$ deterministically generates the agent's entire weight tensor $W$ via $\HKDF$. The novelty in our construction is therefore not the HKDF derivation itself (HKDF and its hierarchical-deterministic application are decade-old primitives) but the two engineering moves layered on top --- publishing the commitment $\WC = \keccak(W)$ on chain at genesis, then re-checking it at every state transition so the agent's substrate cannot drift (\cref{sec:f2,sec:properties}).

The closest prior binding of a key to neural-network weights is ZKROWNN~\cite{sheybani2023zkrownn}, which proves \emph{ownership} of given weights via an embedded watermark key --- there the weights are an input to be claimed, whereas here they are an output deterministically \emph{derived} from the key ($W = \HKDF(\sk)$) and re-checked on chain at every transition. The construction combines deterministic key-to-weights derivation with transition-time on-chain verification. Concurrent work derives context-isolated \emph{keys} from an identity root for blockchain authorisation~\cite{zkace2026,aesp2026}; our derivation instead targets the model substrate itself (the recurrent weight tensor $W$) rather than subordinate keys, so the claim above is specifically that the \emph{weights} are bound to the key, not that deterministic derivation from a root is itself new. Orthogonally, verifiable delay functions (VDFs)~\cite{boneh2018vdf} certify elapsed \emph{sequential} computation; the temporal continuity of $\Stream(K)$ derives instead from ledger ordering at finalized depth (\cref{sec:bg-solana}), and pacing advances by a VDF is a possible composition we do not pursue here.

\section{Background}\label{sec:background}

\subsection{Notation and scope}\label{sec:notation}
The agent triple $\agent = (K, \Stream(K), \Fecon(K))$ was introduced in \cref{sec:intro} alongside the Artificial Externality framework's three-layer model of reality (Substrate / Contingency / Inexorable), whose vocabulary it adopts. The construction realises the framework's cryptographic individuality component, not its stronger inexorability claim (\cref{sec:disc-limits}); \cref{tab:notation} fixes the notation used throughout the paper. We instantiate the $K$ and $\Stream(K)$ axes in \cref{sec:f2,sec:f1,sec:chain} (core construction) and extend the construction in \cref{sec:design-phase2,sec:design-phase3}; \cref{sec:phase4} adds a consumption-side constraint to the $\Fecon(K)$ axis. The aliveness predicate over these three axes is enforced as a circuit-and-runtime invariant by the combined construction (\cref{sec:phase4-m0}).

\begin{table}[!hbp]
\centering\small
\begin{tabularx}{\linewidth}{lX}
\toprule
\textbf{Symbol} & \textbf{Meaning} \\
\midrule
$\sk$, $\pk$ & Agent private key (32-byte $\ed$ seed) and 32-byte public point ($\ed$ on edwards25519) \\
$G$ & $\ed$ base point used in the abbreviated relation $\pk=\sk\cdot G$ \\
$K = (\sk, \pk)$ & Rigid-designator key pair (\cref{sec:glossary}), the agent's structural anchor \\
$W = (W_{xh}, W_{hh}, W_{hy})$ & Elman recurrent weights derived as $\HKDF(\sk, \mathit{tag})$ (\cref{sec:bg-elman}) \\
$\WC = \keccak(W)$ & On-chain weight commitment registered at genesis ($F_2$, \cref{sec:f2}) \\
$h_t \in \mathbb{Z}^{16}$ & Per-cycle hidden state in $\mathrm{Q}16.16$ fixed point \\
$x_t \in \mathbb{Z}^5$, $y_t \in \mathbb{Z}^5$ & Per-cycle environment input and agent action \\
$c_t$ & On-chain per-cycle state commitment, chained from $c_{t-1}$ \\
$\Stream(K)$ & The chain $\{c_t\}_t$, signed by $\sk$ at every advance (\cref{sec:chain}) \\
$\Fecon(K)$ & Lamport balance of the agent's identity-derived economic PDA (\cref{sec:bg-solana,sec:phase4}) \\
$\pi_{F_2}, \pi_{F_1}$ & Groth16 proofs for the genesis ($F_2$) and per-cycle ($F_1$) circuits \\
\bottomrule
\end{tabularx}
\caption{Notation used throughout the paper. Economic-metabolism extensions ($\Fecon$, oracle attestation, environment commitment) are introduced when needed in \cref{sec:phase4}.}
\label{tab:notation}
\end{table}

\subsection{Cryptographic primitives}\label{sec:bg-primitives}
Roughly, $\ed$ lets the agent sign its actions; HKDF lets us derive weights deterministically from the key; $\keccak$ lets the chain store a fingerprint of those weights; and Groth16 lets the chain verify the binding without seeing the secret. Concretely, we rely on four primitives, all standard. \textbf{(a)}~$\ed$ EdDSA on edwards25519~\cite{bernstein2012ed25519} for agent signatures; pubkeys are 32 bytes, signatures 64 bytes (we write $\pk = \sk \cdot G$ for the RFC~8032 key derivation, eliding its internal SHA-512 seed-to-scalar clamping; the same 32-byte seed $\sk$ is the HKDF input of \cref{sec:f2}). \textbf{(b)}~HKDF-SHA256~\cite{krawczyk2010hkdf} (RFC~5869) for deterministic weight derivation --- in ML terms, the key acts as the seed of a fixed pseudorandom weight initialisation (\cref{sec:related}). \textbf{(c)}~$\keccak$-256 for on-chain commitments (matches the Solana \texttt{keccak} syscall). \textbf{(d)}~Groth16~\cite{groth2016size} as the proof system, instantiated by SP1~v5.2.4~\cite{succinct2024sp1}; on-chain verification uses the \texttt{sp1-solana} verifier crate developed by Succinct Labs.

\subsection{Elman recurrence over a fixed-point grid}\label{sec:bg-elman}
The agent's compute substrate is a small recurrent neural network: a $16$-dimensional hidden state $h_t$ that each cycle combines the previous state $h_{t-1}$ with an input $x_t$ to produce an action $y_t$. We use integer arithmetic in $\mathrm{Q}16.16$ fixed-point format so that the same computation runs bit-exactly inside the SP1 guest and in the host-side reference implementation. Formally, the agent's dynamics is an Elman~\cite{elman1990finding} recurrent layer on a $\mathbb{Z}^{16}$ hidden state with a $\mathbb{Z}^{5}$ input and $\mathbb{Z}^{5}$ output, all in $\mathrm{Q}16.16$ fixed point:
\begin{align}
  h_t &= \mathrm{hardtanh}\bigl(W_{xh}\, x_t + W_{hh}\, h_{t-1}\bigr),\label{eq:elman}\\
  y_t &= W_{hy}\, h_t,
\end{align}
with $W_{xh} \in \mathbb{Z}^{16\times 5}$, $W_{hh} \in \mathbb{Z}^{16\times 16}$, $W_{hy} \in \mathbb{Z}^{5\times 16}$ ($416$ \texttt{i32} weights, $1{,}664$ bytes total). The guest performs a fixed-shape deterministic integer computation: matrix dimensions and loop bounds are fixed, while $\mathrm{hardtanh}$ saturation (the identity clamped to $[-1, 1]$) uses data-dependent clamp branches. Each matmul (matrix-multiply) accumulator is a $\mathrm{Q}32.32$ product, right-shifted by 16 bits back to $\mathrm{Q}16.16$ before the next operation (omitted from \cref{eq:elman} for readability but required for bit-exact reproduction). Choosing the low 15 bits of each HKDF output word (sign-extended to $\mathrm{Q}16.16$, giving per-weight values approximately uniform on $[-0.25, 0.25)$) yields a per-weight standard deviation $\sigma_W \approx 0.14$ and a spectral radius (loosely, the recurrence's amplification factor) $\rho \approx \sigma_W\sqrt{N_h} \approx 0.56$ ($N_h{=}16$ hidden units). These are properties of the derivation's output statistics that hold in distribution across keys, not per-agent tuning; they are richer than an internal predecessor PoC in which every weight was the constant $0.05$. The $5\to16\to5$ dimensions are minimal scaffolding chosen to exhibit non-degenerate recurrent dynamics within the SP1 proving budget. The binding construction does not depend on this particular choice of dimensions, although changing them requires rebuilding the fixed-shape guest circuit and its verifying key. The sub-unit spectral radius ($\rho\approx0.56<1$) keeps trajectories away from the $\mathrm{hardtanh}$ saturation regime in which distinct-key trajectories could otherwise collapse to a shared clamped corner --- the $500$-pair null of \cref{sec:eval-individuation} confirms no such collapse empirically (minimum mean inter-agent $L_2$ divergence $M_4 = 0.44 > 0$).

\subsection{Solana on-chain primitives}\label{sec:bg-solana}
This paper depends on four Solana primitives.
\begin{description}[itemsep=2pt,topsep=2pt,leftmargin=1.5em]
  \item[\emph{Program}.] A deterministic Rust binary that is deployed once to a public address (the \emph{program ID}) and then invoked through \emph{instructions} carried by \emph{transactions}; a transaction is a signed bundle of one or more instructions that the cluster either applies atomically or rejects.
  \item[\emph{Account}.] The unit of state: an addressable region of bytes owned by exactly one program, with a balance denominated in \emph{lamports} ($10^9$ lamports $= 1$~SOL). The program owning the account is the only party that can mutate its bytes, but anyone can read them.
  \item[\emph{Program-derived address} (PDA).] A deterministic account address computed by hashing a fixed program ID with a list of \emph{seeds} (byte strings). A PDA deliberately has \emph{no} private key, so no external party can ever sign for it: ``the program signs'' means the runtime itself grants the owning program authority over the account whenever that program is invoked, an authority no signature can confer or steal --- which is what makes a PDA usable as on-chain state owned by that program.
  \item[\emph{Precompile}.] A native program built into the runtime (\texttt{Ed25519SigVerify}, used throughout this paper, verifies ed25519 signatures carried in a transaction). A program can inspect the transaction's other instructions (\emph{instruction introspection}), which is how an on-chain program confirms that the precompile verified exactly the message it expects.
\end{description} The on-chain programs of this paper are written with Anchor, the standard Solana program framework. Concretely, the per-agent on-chain record is a PDA whose seeds include the agent's public key $\pk$ (\cref{sec:f2}); the per-cycle environment attestation, by contrast, is carried inside the proof itself --- the designated oracle's signature over $x_t$ is verified in-circuit and the signing key is committed to the proof's public values, with no separate per-cycle on-chain account (\cref{sec:phase4-m1}).

Per-instruction execution cost is metered in \emph{compute units} (CU), with a transaction-level cap; we report per-transaction CU consumption alongside fees in \cref{sec:eval-cu}. Distinct from CU metering, every transaction names a \emph{fee payer}, the account from which the runtime debits the transaction fee at landing. That fee is a base fee charged per signature ($5{,}000$~lamports each, counting both the transaction's own signatures and any signatures verified in-transaction by precompiles such as \texttt{Ed25519SigVerify}), independent of compute consumed, plus an optional \emph{priority fee} (denominated per requested compute unit) that raises scheduling priority. $\Fecon(K)$ is, concretely, the agent's capacity to keep meeting the per-cycle costs of running --- realised in the deployed extension as a protocol-imposed metabolic debit, distinct from these fees (\cref{sec:phase4}).

The cluster orders transactions into ${\sim}400$~ms \emph{slots}, each slot producing at most one \emph{block} (a batch of confirmed transactions); blocks at the very tip of the chain can in principle still be replaced by a competing fork (a \emph{reorganisation}), but once a transaction reaches \emph{finalized} commitment, the cluster guarantees it can never be rolled back --- the depth at which this paper's no-fork claims are stated (\cref{sec:threat-scope}). Throughout, we use Solana's \emph{devnet}, a public test cluster that shares the execution model and APIs relevant to this construction with mainnet, but differs in validator conditions, load, stability, and economics: SOL is faucet-issued and has no monetary value, and the cluster may be reset by its operator. Conventional notation: a Solana public key (32~bytes) is rendered in base-58 (e.g., \texttt{7H4Dgrq2\dots bPare} in \cref{tab:deployed}), and we write the truncated form throughout to keep prose readable.

\section{Construction}\label{sec:construction}

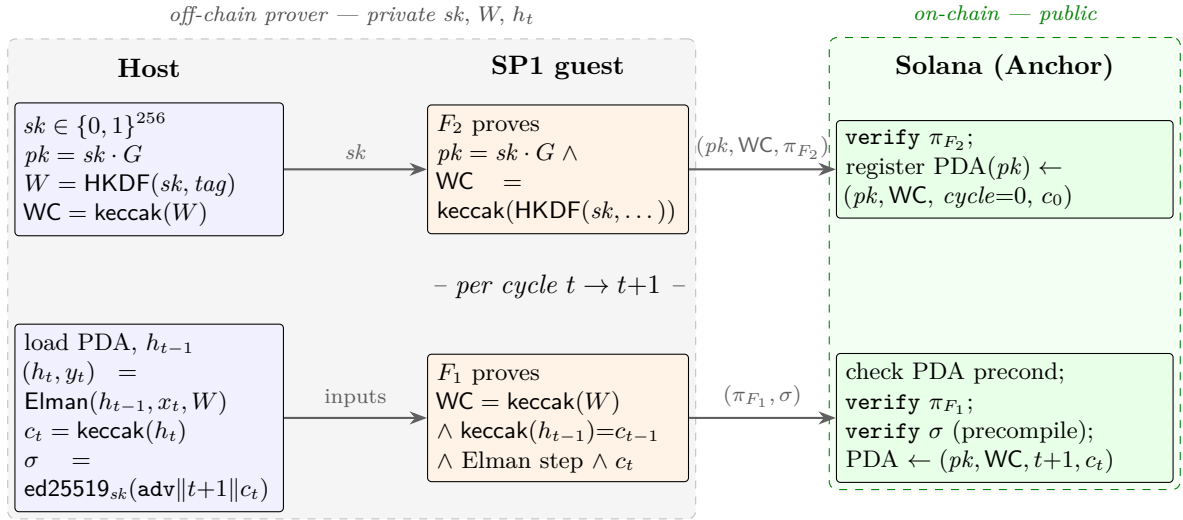
\begin{figure}[!htbp]
\centering
\begin{tikzpicture}[
  font=\footnotesize,
  node distance=1.5mm and 4mm,
  hostbox/.style={draw,rounded corners=1.5pt,fill=blue!6,inner sep=3.5pt,align=left,text width=33mm},
  guestbox/.style={draw,rounded corners=1.5pt,fill=orange!10,inner sep=3.5pt,align=left,text width=32mm},
  chainbox/.style={draw,rounded corners=1.5pt,fill=green!9,inner sep=3.5pt,align=left,text width=42mm},
  arr/.style={-Stealth,thick,gray!75!black},
  hd/.style={font=\bfseries\small,align=center}
]
  \node[hd] (hH) at (0,0) {Host};
  \node[hd] (gH) at (54mm,0) {SP1 guest};
  \node[hd] (cH) at (113mm,0) {Solana (Anchor)};

  \node[hostbox,below=2mm of hH.south,anchor=north] (h1) {%
    $\sk \in \{0,1\}^{256}$\\
    $\pk = \sk\cdot G$\\
    $W = \HKDF(\sk,\mathit{tag})$\\
    $\WC = \keccak(W)$};
  \node[guestbox] (g1) at (h1 -| gH) {%
    $F_2$ proves\\
    $\pk = \sk\cdot G\;\wedge$\\
    $\WC = \keccak(\HKDF(\sk,\dots))$};
  \node[chainbox] (c1) at (h1 -| cH) {%
    \texttt{verify}~$\pi_{F_2}$;\\
    register PDA$(\pk) \leftarrow$\\
    $(\pk,\WC,\,\mathit{cycle}{=}0,\,c_0)$};
  \draw[arr] (h1) -- (g1) node[midway,above,font=\scriptsize] {$\sk$};
  \draw[arr] (g1) -- (c1) node[midway,above,font=\scriptsize] {$(\pk,\WC,\pi_{F_2})$};

  \node[font=\itshape\small,below=4mm of g1] (sep)
    {-- per cycle $t \to t{+}1$ --};

  \node[hostbox,anchor=north] (h2) at ([yshift=-2mm]sep.south -| h1) {%
    load PDA, $h_{t-1}$\\
    $(h_t,y_t)\!=\!\mathsf{Elman}(h_{t-1},x_t,W)$\\
    $c_t = \keccak(h_t)$\\
    $\sigma = \ed_{\sk}(\mathtt{adv}\|t{+}1\|c_t)$};
  \node[guestbox] (g2) at (h2 -| g1) {%
    $F_1$ proves\\
    $\WC = \keccak(W)$\\
    $\wedge\; \keccak(h_{t-1}){=}c_{t-1}$\\
    $\wedge$ Elman step $\wedge\; c_t$};
  \node[chainbox] (c2) at (h2 -| c1) {%
    check PDA precond;\\
    \texttt{verify}~$\pi_{F_1}$;\\
    \texttt{verify}~$\sigma$ (precompile);\\
    PDA $\leftarrow (\pk,\WC,t{+}1,c_t)$};
  \draw[arr] (h2) -- (g2) node[midway,above,font=\scriptsize] {inputs};
  \draw[arr] (g2) -- (c2) node[midway,above,font=\scriptsize] {$(\pi_{F_1},\sigma)$};

  \begin{scope}[on background layer]
    \node[draw=gray!50,dashed,rounded corners=3pt,fill=gray!8,inner sep=2.5pt,
          fit=(hH)(h1)(h2)(gH)(g1)(g2)] (privreg) {};
    \node[draw=green!50!black,dashed,rounded corners=3pt,fill=green!5,inner sep=2.5pt,
          fit=(cH)(c1)(c2)] (pubreg) {};
  \end{scope}
  \node[font=\scriptsize\itshape,gray!60!black,above=1pt of privreg.north]
    {off-chain prover --- private $\sk,\,W,\,h_t$};
  \node[font=\scriptsize\itshape,green!50!black,above=1pt of pubreg.north]
    {on-chain --- public};
\end{tikzpicture}
\caption{Core protocol flow. \emph{Genesis} (top row) registers an agent identity: the SP1 guest circuit $F_2$ proves in zero knowledge that the public key, the weight commitment, and the private-key preimage of both are mutually consistent. Each subsequent \emph{advance} (bottom row) runs circuit $F_1$, which proves that cycle's Elman step under a witnessed $W$ and publishes its commitment; the on-chain program checks that commitment against the genesis record, verifies the agent's signature, and atomically advances the PDA's $(\mathit{cycle}, \mathit{state\_commit})$ pair. The shaded split marks the trust boundary: the secret key $\sk$, the weights $W$, and the hidden state $h_t$ never leave the off-chain prover (Host and SP1 guest); the chain receives only the public key and journal values (including commitments), the constant-size proof, and the agent's signature.}
\label{fig:protocol}
\end{figure}

\paragraph{Protocol overview.}
The core protocol combines a one-time \emph{genesis step} that binds the weights to the key ($F_2$, \cref{sec:f2}), a per-cycle \emph{advance step} that verifies the committed weights while updating state ($F_1$, \cref{sec:f1}), and a state-commitment \emph{chain} that prevents replay and, once finalized, prevents history forks (\cref{sec:chain}). These components instantiate the $K$ and $\Stream(K)$ axes of the agent triple; \cref{sec:phase4} adds $\Fecon(K)$. A single Solana program (\texttt{p1\_identity\_verifier}, \cref{tab:deployed}) verifies both Groth16 circuits, enforces the chain, and requires an $\ed$ signature on every advance.\footnote{In the published code, this core program and its genesis/advance guests make up \texttt{code/core/}, distinct from the separate guest crates of the active-query and economic-metabolism extensions (\cref{sec:poc-extensions,sec:phase4}); a reader auditing the guest source should map the advance circuit of \cref{sec:f1} to \texttt{core/}'s advance guest specifically.} After defining the notation and primitives, \cref{sec:bg-zkflow} gives an informal host--guest--chain walk-through, \cref{fig:protocol} shows the data flow, and \cref{sec:properties} states the resulting properties.

The weights $W$ are derived from the key once and remain \emph{frozen} for the agent's lifetime: a fixed random initialisation that is never trained. This isolates the key in the individuation evaluation, so behavioural divergence between two agents can be attributed to their different keys. The weight-rotation extension of \cref{sec:design-phase3} lifts the freeze through discrete, key-anchored updates that re-prove the weight commitment at each jump. Continuous-gradient learnable weights remain a separate direction (\cref{sec:disc-future-directions}), with the trade-off stated in \cref{sec:disc-limits}.

\subsection{Reading guide: the zk-circuit workflow}\label{sec:bg-zkflow}
The rest of \cref{sec:construction} relies on the following proof-system workflow. An SP1 \emph{guest} is a small Rust program executed inside the SP1 zkVM, a virtual machine that emits a Groth16 proof of its own execution. A bare Groth16 proof is a few hundred bytes; the SP1-wrapped artefact submitted here, including its public values, is about $1.5$~KB and is verified at constant cost (bounded compute units). It certifies that the Rust program ran with specific public inputs and outputs (its \emph{journal} --- the byte string of committed public values that the chain reads) without revealing any private inputs. Proof generation is two-stage: SP1 first produces a large STARK proof (a hash-based proof of the zkVM execution, megabytes in size --- cheap to generate but far too large to post on Solana) and then \emph{wraps} it in a single constant-size Groth16 proof. Throughout the paper, \emph{core mode} refers to the unwrapped STARK stage (verifiable off-chain only) and \emph{wrapped} to the Groth16 stage; only wrapped proofs are verified on chain. In the genesis circuit $F_2$ below, the guest evaluates ``derive $W$ from $\sk$ via HKDF; check that $\pk = \sk \cdot G$; return $\WC = \keccak(W)$ and $\pk$ as public outputs.'' The on-chain program verifies the proof and stores $\WC$ and $\pk$ as the agent's permanent record. The secret $\sk$ never leaves the host machine: the chain sees only the proof and the commitments. From an ML standpoint, this is analogous to a model card whose contents are cryptographically attestable rather than merely declared.
\subsection{Genesis: the \texorpdfstring{$F_2$}{F2} circuit and agent registration}\label{sec:f2}
On registration, the agent owner provides~$\sk$ as the only private input. In one line, genesis proves that the public key and the weight commitment were both computed from the \emph{same} secret seed, without revealing it. Formally, the genesis circuit $F_2$ proves
\begin{equation}\label{eq:f2}
  \pk = \sk \cdot G \;\wedge\; \WC = \keccak\bigl(\HKDF(\sk, s, \mathit{info}, L)\bigr),
\end{equation}
where $G$ is the $\ed$ base point, $s = \texttt{"zkalife:phase1:v1"}$, $\mathit{info} = \texttt{"W\_all"}$, and $L = 1{,}664$ bytes is the length of the single raw HKDF expansion: $416$ 32-bit words interpreted in fixed order as the backing words of the three matrices (\texttt{W\_xh}: $16 \times 5$, \texttt{W\_hh}: $16 \times 16$, \texttt{W\_hy}: $5 \times 16$). (The shorthand $\HKDF(\sk, \mathit{tag})$ used elsewhere abbreviates exactly this call.) $\WC$ hashes this raw expansion directly; the $\mathrm{Q}16$ weight values used by the Elman step are extracted from each word's low 15 bits at use time (\cref{sec:bg-elman}), so a reproducer must hash the raw HKDF output, not a re-serialisation of the extracted weights.

Both public outputs are bound to the \emph{same} witnessed 32-byte seed: $F_2$ computes $\pk = \sk \cdot G$ and derives $W$ from the same $\sk$, so $\pk$ and $\WC$ are two deterministic functions of one common seed.\footnote{$\ed$ internally signs with the clamped scalar $a = \mathrm{clamp}(\mathrm{SHA}\text{-}512(\sk))$ while HKDF consumes the raw seed; the binding is nonetheless to the seed, and the non-injectivity of the clamp does not weaken it, since $\WC$ is keyed on the seed and not on the clamped scalar.} The on-chain registration transaction posts $(\pk, \WC, \mathit{skCommit}, \pi_{F_2})$, where $\pi_{F_2}$ is the Groth16 proof of \cref{eq:f2}; the genesis journal additionally publishes $\mathit{skCommit} = \keccak(\sk)$ ($\keccak$ is one-way, so this does not expose $\sk$\footnote{The one-wayness argument assumes $\sk$ has full entropy; for a low-entropy key (e.g., $\sk = \mathbf{0}$), $\mathit{skCommit}$ is a known value providing no hiding.}), consumed by the weight-rotation extension's $\sk$-binding check (\cref{sec:design-phase3}). The program checks $\pi_{F_2}$, asserts that the PDA at $\pk$ is uninitialised, and writes
\[
  \mathrm{PDA}(\pk) \,\leftarrow\, (\pk, \WC, \mathit{cycle}{=}0, \mathit{state\_commit}{=}c_0).
\]
Here $c_t$ is the per-cycle state commitment of \cref{tab:notation} --- the hash of the hidden state that cycle $t$ ends with --- and it is the only state the chain carries across cycles: each advance must open the previous cycle's commitment $c_{t-1}$ before publishing its own $c_t$ (\cref{sec:f1}). The very first advance has no preceding cycle, so $c_0 = \keccak(\underbrace{0\,\|\cdots\|\,0}_{64\text{ bytes}})$ is a fixed, key-independent genesis constant computed identically by host and guest, serving as that first $c_{t-1}$.

After registration, the only secret material the host must retain is $\sk$ itself; $W$ is recomputed from $\sk$ on demand and never persisted. The host is thus a single point of key custody, as in most adjacent designs (Spore.fun~\cite{hu2025sporewild} instead confines the key to a TEE); distributing the host itself, so that no single node ever holds $\sk$, is the threshold-MPC (multi-party computation) prover direction discussed in \cref{sec:disc-limits}. Because the PDA is seeded on $\pk$ and $F_2$ asserts $\pk = \sk \cdot G$ (\cref{alg:f2}, line~\ref{f2:keypair-assert}), only the holder of $\sk$ can produce valid registration artefacts for $\pk$; submission itself requires no $\pk$ signature, and a relayed submission registers the identical journal-pinned record. A copied or relayed transaction can pre-empt the original submission, but it cannot hijack the identity: it registers only that identical record, and the uninitialised-PDA check then bars re-initialisation (front-running: observing another party's pending transaction and landing one's own competing transaction first; cf.\ the advance-time tag~T1 and the double-init case (T10)).

Pseudocode for $F_2$ is given as \cref{alg:f2}: lines~\ref{f2:keypair}--\ref{f2:keypair-assert} enforce condition~(i) (keypair binding) and lines~\ref{f2:wc}--\ref{f2:wc-assert} enforce condition~(ii) (substrate--key binding), both circuit-internal, together contributing to a single Groth16 proof~$\pi_{F_2}$.

\begin{algorithm}[!ht]
\caption{Genesis circuit $F_2$ (zk-SNARK; one-time, per agent).}\label{alg:f2}
\footnotesize
\begin{algorithmic}[1]
\Require private input $\sk \in \{0,1\}^{256}$
\Require public inputs $\pk, \WC \in \{0,1\}^{256}$; publishes $\mathit{skCommit}$
\Statex \emph{(i) keypair binding}
\State $\pk' \gets \sk \cdot G$ \Comment{ed25519 scalar multiplication on edwards25519} \label{f2:keypair}
\State \textbf{assert} $\pk' = \pk$ \label{f2:keypair-assert}
\Statex \emph{(ii) substrate--key binding}
\State $W \gets \HKDF(\sk, s, \texttt{"W\_all"}, L{=}1{,}664)$ \Comment{single expansion, reshape to $(W_{xh}, W_{hh}, W_{hy})$} \label{f2:wc}
\State \textbf{assert} $\keccak(W) = \WC$ \label{f2:wc-assert}
\Statex \emph{($\mathit{skCommit}$ for the weight-rotation extension, \cref{sec:design-phase3})}
\State $\mathit{skCommit} \gets \keccak(\sk)$ \Comment{published in the genesis journal alongside $\pk, \WC$}
\State \Return \textbf{accept}
\end{algorithmic}
\end{algorithm}

\subsection{Advance: the \texorpdfstring{$F_1$}{F1} circuit and Elman step}\label{sec:f1}
Each cycle $t \to t{+}1$ consumes the current $\mathrm{PDA}(\pk)$ entry $(\WC, \mathit{cycle}, c_{t-1})$, the cached hidden state $h_{t-1}$, an environment vector $x_t$ (synthetic at this stage; environment coupling is addressed in the active-query-loop extension), and produces a new state commitment $c_t$. The advance circuit $F_1$ proves
\begin{equation}\label{eq:f1}
  \begin{aligned}
    \WC &= \keccak(W) \\
        &\hphantom{=\;}\wedge\; \keccak(h_{t-1}) = c_{t-1} \\
        &\hphantom{=\;}\wedge\; (h_t, y_t) = \mathsf{ElmanStep}(h_{t-1}, x_t, W) \\
        &\hphantom{=\;}\wedge\; c_t = \keccak(h_t).
  \end{aligned}
\end{equation}
Here $W$ is supplied as a private witness and the circuit publishes $\WC = \keccak(W)$ rather than re-deriving $W$ from $\sk$ (a \emph{commit-and-prove} pattern); that derivation is proven once at genesis by $F_2$ (\cref{eq:f2}), and the advance is bound to it on chain (check~(d) below) --- re-deriving $W$ in-circuit at every cycle would re-pay the HKDF expansion's proving cost for no added soundness.

\paragraph{Division of labour (guest, host, chain).} All four conjuncts of \cref{eq:f1} --- including the Elman step itself --- are evaluated \emph{inside} the SP1 guest, whose execution produces the Groth16 proof $\pi_{F_1}$; the host assembles the witnesses ($W$ re-derived from $\sk$, the cached $h_{t-1}$, the environment $x_t$), runs the same step natively to obtain the $c_t$ it must sign, and submits the transaction; the on-chain program never re-executes the dynamics but verifies $\pi_{F_1}$ together with the chain-side checks (a)--(d) below (\cref{fig:protocol}). The SNARK binds $c_t$ to $c_{t-1}$ by requiring the start-of-cycle state to match the previous commitment ($\keccak(h_{t-1}) = c_{t-1}$, \cref{alg:f1}), so the published $c_t$ is the guest-defined fixed-point successor of the state $c_{t-1}$ commits to; the on-chain PDA precondition described below additionally enforces that $c_{t-1}$ is the previous cycle's actual PDA state, serialising advances against replay and forks at transaction time.

\paragraph{Self-signature.} The advance transaction additionally carries an $\ed$ signature by $\sk$ over the message $\sigma_{\mathrm{adv}} = \texttt{"zkalife:p1:advance:"} \,\|\, (\mathit{cycle}{+}1)^{\mathrm{LE}} \,\|\, c_t$. The signed bytes are a domain-separation prefix (a fixed literal that keeps a signature produced for one message type from being replayed as a signature on another), the next cycle index as a little-endian \texttt{u64}, and the new commitment~$c_t$; the public key is not signed but is bound on chain by verifying the signature against the registered~$\pk$.\footnote{The signed message binds the cycle index and $c_t$ but not the program ID, so a proof-and-signature bundle is not cryptographically pinned to one program. For $t \geq 1$ this is bounded operationally: each agent's PDA lives under a single program, and a cross-program replay would require an independent PDA for the same $\pk$ to sit at the run-specific $(\mathit{cycle}, c_{t-1})$ state. The first advance is the exception, since its precondition is the universal genesis constant $c_0$: a party who re-registers the same $\pk$ under a second program could replay that first advance there --- but this only reproduces the agent's own honest first state and confers no authority over the canonical PDA. Folding the program ID into the signed message closes both cases and is a low-cost hardening step deferred to the sleep-replication extension (the planned successor deliverable; \cref{sec:design-phase3}).}

\paragraph{On-chain verification and settlement.} The on-chain program verifies (a)~$\pi_{F_1}$ (against the advance verifying key baked into the program, pinning the exact guest circuit), (b)~the $\ed$ signature against $\pk$ via Solana's \texttt{Ed25519SigVerify} precompile, (c)~that the PDA's $\mathit{state\_commit}$ still equals $c_{t-1}$ (the $(\mathit{cycle}, \mathit{state\_commit})$ pair advances atomically, so this pins $\mathit{cycle}{=}t$), and (d)~that the published $\WC$ equals the commitment registered at genesis (else \texttt{WeightCommitMismatch}, \texttt{Custom(6007)}). Labels (a)--(d) enumerate logical conditions rather than handler execution order: the deployed handler checks the journal header, (c), and (d) before (a), and checks (b) after proof verification. The combination (b)+(c) enforces the cross-cycle chain $c_{t-1} \to c_t$, while (d) enforces the substrate--key binding at advance time. On success the program advances the PDA atomically:
\[
  \mathrm{PDA}(\pk) \,\leftarrow\, (\pk, \WC, t{+}1, c_t).
\]

Pseudocode for $F_1$ is given as \cref{alg:f1}: the circuit asserts that the start state matches the previous commitment ($\keccak(h_{t-1}) = c_{t-1}$, line~\ref{f1:prev-assert}), recomputes the weight commitment $\WC = \keccak(W)$ from the privately witnessed $W$ (line~\ref{f1:wc-assert}, which the on-chain program binds to the genesis commitment), executes the Elman step under that same $W$ (lines~\ref{f1:elman-h}--\ref{f1:elman-y}, condition~(iii)), and emits $c_t = \keccak(h_t)$ (line~\ref{f1:cnext}). Condition~(iv) --- the $\ed$ self-authorisation --- is, like the cross-cycle chain, enforced outside the circuit rather than inside it: the Solana precompile verifies the signature natively against $\pk$, paired with the PDA precondition check, at much lower cost than a circuit-internal signature check would impose.

\begin{algorithm}[!ht]
\caption{Advance circuit $F_1$ (zk-SNARK; per cycle).}\label{alg:f1}
\footnotesize
\begin{algorithmic}[1]
\Require private inputs $W$, $h_{t-1}$, $x_t$ \Comment{$x_t$ is private in the core circuit; the extension circuits publish it}
\Require public inputs $c_{t-1}, c_t$; publishes $\WC$ and $y_t$ \Comment{$\mathit{cycle}{+}1$ is signed and checked on-chain, not in-circuit}
\Statex \emph{state-chain continuity (start state matches the previous commitment)}
\State \textbf{assert} $\keccak(h_{t-1}) = c_{t-1}$ \label{f1:prev-assert}
\Statex \emph{(ii) weight commitment (equality to the genesis $\WC$ enforced on-chain, not inside the SNARK)}
\State $\WC \gets \keccak(W)$ \Comment{published in the journal; on-chain: \textbf{assert} $\WC = \WC_{\mathrm{genesis}}$} \label{f1:wc-assert}
\State $(W_{xh}, W_{hh}, W_{hy}) \gets \mathsf{Reshape}(W)$
\Statex \emph{(iii) Elman dynamics under exactly that $W$}
\State $h_t \gets \mathrm{hardtanh}\bigl(W_{xh}\,x_t + W_{hh}\,h_{t-1}\bigr)$ \label{f1:elman-h}
\State $y_t \gets W_{hy}\,h_t$ \label{f1:elman-y}
\Statex \emph{next-state commitment ($c_{t-1}$-to-$c_t$ ordering enforced on-chain via the PDA precondition)}
\State $c'_t \gets \keccak(h_t)$ \label{f1:cnext}
\State \textbf{assert} $c'_t = c_t$
\State \Return \textbf{accept}
\end{algorithmic}
\end{algorithm}

\subsection{State chain and atomicity}\label{sec:chain}
The PDA's $(\mathit{cycle}, \mathit{state\_commit})$ pair is the synchronisation point that binds every $\Stream(K)$ to a single ledger-ordered history. Two advances that both reference $(\mathit{cycle}{=}t, c_{t-1})$ but arrive in the same Solana slot are serialised by the runtime: the first to land succeeds; the second fails with a custom error \texttt{StateChainViolation} (\texttt{Custom(6006)}) because its PDA precondition no longer holds. Test~A4 (\cref{sec:threats}) exhibits this behaviour with two transactions differing only in their \texttt{ComputeBudget} unit limits. An agent that paces itself from a \texttt{confirmed}-level (pre-finality) read of $c_t$ cannot thereby land an accepted-but-orphaned advance: if cycle~$t$ is reorganised, cycle~$t{+}1$'s precondition $\mathit{state\_commit}=c_t$ fails on the canonical fork and the advance reverts, so the integrity invariant is reorg-safe even though liveness (\cref{sec:disc-limits}) is not.

\subsection{Properties}\label{sec:properties}
Conditions~(i)--(iv) (the keypair, substrate--key, dynamics, and self-authorisation bindings of \cref{sec:f2,sec:f1}) --- established at genesis by~\cref{eq:f2} and enforced at every state transition by~\cref{eq:f1} together with the on-chain program --- yield four named properties, summarised in \cref{tab:spine}. We state them informally; \cref{sec:threats} pairs each with a tested rejection path.

\begin{table}[htbp]
\centering\footnotesize
\setlength{\tabcolsep}{4pt}
\begin{tabularx}{\linewidth}{@{}l>{\raggedright\arraybackslash}X>{\raggedright\arraybackslash}Xl@{}}
\toprule
\textbf{Condition} & \textbf{Binding} & \textbf{Property} & \textbf{Exercised by} \\
\midrule
(i)   & keypair, $\pk = \sk\cdot G$                              & (P1) keypair authenticity     & $F_2$ soundness; T4-a \\
(ii)  & substrate--key, $\WC = \keccak(\HKDF(\sk,\mathit{tag}))$ & (P2) substrate--key binding   & T6 \\
(iii) & dynamics, Elman step under committed $W$                 & (P3) compiled-dynamics fidelity & T6 \\
(iv)  & self-authorisation, per-cycle $\ed$ signature            & (P4) chain self-authorisation & T3, T4-a/b/c, A4, T-a \\
\bottomrule
\end{tabularx}
\caption{The construction's spine: each binding condition (\cref{sec:f2,sec:f1}) maps to one named property (\cref{sec:properties}) and the rejection-path test or structural check that exercises it (\cref{sec:threats}); the genesis equality in P1 and the compiled-dynamics fidelity of P3 are enforced structurally by circuit soundness, T6 is the adjacent weight-substitution rejection, and T-a is a chain-side precondition rather than a scripted test.}
\label{tab:spine}
\end{table}

\paragraph{(P1) Keypair authenticity.} Condition~(i) pins the registered $\pk$ to the same witnessed 32-byte seed $\sk$ from which all downstream artefacts are derived. A candidate $\sk'$ whose RFC~8032 key derivation does not yield $\pk$ cannot satisfy $F_2$ at registration, and, under $\ed$ EUF-CMA, a valid advance signature for an existing $\pk$ cannot be produced without signing authority for that public key. The genesis equality is enforced structurally by $F_2$ soundness; the wrong-key advance case T4-a confirms that the deployed program rejects a signer other than the registered $\pk$, while T4-b/c exercise the per-cycle self-authorisation property P4.

\paragraph{(P2) Substrate--key binding.} Condition~(ii) makes the on-chain commitment $\WC$ a deterministic function of $\sk$ via HKDF. An adversary holding only $\WC$ cannot recover $W$ ($\keccak$-256 preimage resistance, since $\WC = \keccak(W)$). If $W$ leaks, the one-wayness of HKDF-SHA256 prevents recovery of $\sk$. Nor can an adversary have an alternative $W'$ accepted under the same $\pk$: genesis requires exhibiting $\sk$ in-circuit, which forces $W = \HKDF(\sk, \mathit{tag})$, and at advance the on-chain $\WC$ equality rejects any substitute (Test~T6). The agent's computational substrate is thus pinned to the rigid designator.

\paragraph{(P3) Compiled-dynamics fidelity.} Condition~(iii) constrains every advance to the deterministic fixed-point Elman implementation compiled into the guest under the~$W$ derived from~$\sk$. The circuit binds the start state to the previous commitment ($\keccak(h_{t-1}) = c_{t-1}$, line~\ref{f1:prev-assert} of \cref{alg:f1}), recomputes $(h_t, y_t)$ from $(h_{t-1}, x_t, W)$ at lines~\ref{f1:elman-h}--\ref{f1:elman-y}, and chains them into~$c_t$, so no published~$c_t$ can disagree with the guest-defined successor of the committed state without invalidating~$\pi_{F_1}$.\footnote{The guest does not separately range-check the right-shifted hidden-layer accumulator before Rust narrows it to \texttt{i32}. An out-of-range preactivation therefore follows the compiled narrowing semantics before \texttt{hardtanh}, rather than the unrestricted-integer reading of \cref{eq:elman}. The subsequent hidden state is clamped to $[-1,1]$; with 16 output terms and $|W_{hy}|<0.25$, this also bounds every published $|y_{t,i}|<4$, including when $y_t$ becomes the next active-query action. Thus the proof establishes exact execution of the compiled guest, not equivalence to an unbounded-arithmetic equation. Condition~(iv) additionally establishes key-holder authorisation but does not repair that semantic distinction.} In the core protocol the environment $x_t$ is itself a prover-chosen private witness (\cref{alg:f1}), so this fidelity is stated relative to the witnessed $x_t$; the active-query extension pins $x_t$ to a patron-signed, in-circuit-attested signal under a witness-supplied patron key (\cref{sec:design-phase2}), and the economic extension additionally imposes the in-circuit range bound of \cref{sec:phase4-m1}. The adjacent attack --- a host that mutates~$W$ while leaving the on-chain $\WC$ intact --- publishes a mismatched $\WC = \keccak(W')$ (line~\ref{f1:wc-assert}) and is rejected by the on-chain commitment check, since the published $\WC$ no longer equals the registered commitment (\texttt{WeightCommitMismatch}, Test~T6).

\paragraph{(P4) Chain self-authorisation.} Condition~(iv) requires every advance to be signed by~$\sk$ over a message that includes the new commitment $c_t$ and $\mathit{cycle}{+}1$ (under the \texttt{"zkalife:p1:advance:"} domain-separation prefix; the signed message does not bind the program ID, a bounded cross-program-replay consequence analysed in the self-signature footnote of \cref{sec:f1}). A replayed proof or signature against an advanced PDA fails the precondition $\mathit{state\_commit}{=}c_{t-1}$ (Test~T3), and forging a fresh $\ed$ signature requires~$\sk$. Concurrent advances are serialised atomically by the same precondition (Test~A4); a non-monotone or stalled cycle counter is rejected by the PDA precondition (tag~T-a) on the chain side, before any proof verification. The of-record rejection-path tests (those whose results this paper reports) are operator-scripted: the tester constructs the attacks while holding the agents' own key material, so P4 is verified structurally rather than against a key-withholding adversary (\cref{sec:disc-limits}). The guarantee \emph{against a third party who lacks $\sk$} is discharged by the reduction below (the Security statement), which rests on $\ed$ EUF-CMA (existential unforgeability under chosen-message attacks), $\keccak$ second-preimage resistance, and Groth16 soundness --- properties established by cryptographic argument, not by an attacker failing to break them in a finite test campaign. What the scripted tests establish is complementary: that the deployed verifier composition admits no \emph{implementation} bypass among the scripted attack classes. The on-chain parser pins the full \texttt{Ed25519SigVerify} offset header to the canonical single-signature layout, so the precompile verifies exactly the $(\pk, m)$ the gating program reads.

\medskip
\noindent
The contribution is not any single property in isolation: each is, in itself, a standard cryptographic check. The contribution is that all four are enforced \emph{jointly}, as a single circuit-and-runtime invariant, on \emph{every} state transition --- so that the binding between key and substrate becomes a protocol-enforced equality at every cycle.

\paragraph{Security statement.} Combining (P1)--(P4): under Groth16 soundness in the SP1 instantiation (including the STARK/FRI soundness assumptions of SP1's RISC-V execution proof, and setting aside the one-time wrap-circuit setup of \cref{sec:threat-scope}), $\keccak$-256 second-preimage resistance, and $\ed$ EUF-CMA, no advance transaction is accepted on chain at $\mathrm{PDA}(\pk)$ unless it carries authorisation under a signing seed for $\pk$ and a valid proof of the compiled guest transition under the $W = \HKDF(\sk, \mathit{tag})$ registered at genesis, extending the agent's own ledger-ordered commitment chain. This statement concerns the authorisation and proof carried by the transaction; it does not assert that the key holder personally generated the proof or submitted the transaction, since an authorised bundle can be relayed verbatim. An adversary without a signing seed for $\pk$ who forged an accepted advance would have to break one of these primitives: substituting $W' \neq W$ contradicts the on-chain $\WC$ check (a $\keccak$ second-preimage --- exhibiting a different input with the same hash); forking the hidden state contradicts $\keccak(h_{t-1}) = c_{t-1}$ (likewise a second-preimage); forging the self-signature contradicts $\ed$ EUF-CMA; and producing a transition inconsistent with the compiled guest contradicts SP1/Groth16 soundness. The binding $W = \HKDF(\sk, \mathit{tag})$ is proven in-circuit only at genesis (by $F_2$); each advance does not re-derive it but is bound to the genesis commitment by the on-chain $\WC$ equality check (\cref{sec:f1}), so the per-transition guarantee is that the cycle executed under the genesis-committed $W$. The guarantee is one of \emph{integrity, not availability}: liveness and prover denial (\cref{sec:disc-safety}), and host-key compromise (T5, T12), are out of scope. An adversary who holds a copy of $W$ but not a signing seed for $\pk$ is excluded by condition~(iv) alone --- the per-cycle $\ed$ self-signature --- since such an adversary would pass the $\WC$ check. HKDF-SHA256's PRF security and one-wayness are assumed separately, for P2's confidentiality claims ($W$'s unpredictability, and $\sk$'s confidentiality from a leaked $W$); they do not enter the integrity reduction above, which rests on $\keccak$, $\ed$, and SP1/Groth16 soundness.

\paragraph{Section summary.}
\Cref{sec:f2,sec:f1,sec:chain,sec:properties} have specified the core protocol: the genesis circuit $F_2$ that registers $\WC = \keccak(\HKDF(\sk, \mathit{tag}))$, the advance circuit $F_1$ that re-checks the weight commitment at every cycle (binding to the genesis derivation enforced on-chain), the state-commitment chain that serialises advances, and the four properties (P1)--(P4) the composition enforces. The next sections enumerate the threats this construction defends against (\cref{sec:threats}) and present the core empirical evidence (\cref{sec:eval}); \cref{sec:poc-extensions} then layers the PoC-tier extensions (active-query loop and homeostatic motivation with key-anchored weight rotation) on top of the core, and \cref{sec:phase4} adds a consumption-side constraint to the $\Fecon(K)$ axis on the same Solana devnet.

\section{Threat Model and Tested Rejections}\label{sec:threats}

\paragraph{Threat-model structure.}
We separate the trust assumptions and adversary classes (\cref{sec:threat-scope}), the operator-scripted on-chain rejection tests (\cref{sec:threat-tests}), accepted residual risks (\cref{sec:threat-accepted}), and the threats inherited from adjacent identity protocols and the environment-oracle extension (\cref{sec:threat-adjacent,sec:threat-p4m1-ext}).

\subsection{Scope and assumptions}\label{sec:threat-scope}
We assume Solana liveness and finality; we do \emph{not} assume that the agent's host or its operator is honest beyond the cryptographic checks the chain enforces. Custody, however, is concentrated: $\sk$ is held in plaintext in a single host-process keystore (not a hardware security module), so whoever holds it can exercise the agent's full protocol authority --- a disclosed residual (T5, T12) whose consequences for the framework's inexorability claim we take up in \cref{sec:disc-limits}. We assume the standard hardness of $\ed$, $\keccak$ second-preimage resistance, HKDF-SHA256 PRF security, SP1's STARK/FRI execution proof, and Groth16 soundness in the SP1 instantiation. The $\Stream(K)$ no-fork property holds at Solana's \emph{finalized}-commitment depth: an advance observed only at \texttt{confirmed} level could be reorganised before finality, so a canonical-history verifier reads at \texttt{finalized}.

\paragraph{Trusted setup and software trust base.}
Groth16 soundness relies on a one-time trusted setup whose secret randomness must be destroyed; a party retaining that ``toxic waste'' could forge accepting proofs. SP1 first proves RISC-V execution using a STARK whose soundness relies on its polynomial commitments, Fiat--Shamir transform, and FRI low-degree testing (\cref{sec:bg-zkflow}), then wraps that proof in one application-independent Groth16 circuit. The setup assumption therefore concerns Succinct's shared wrap circuit rather than a per-agent or per-paper ceremony, while STARK/FRI soundness remains a separate cryptographic assumption. The per-program $\mathit{vkeyHash}$ requires no additional setup: it fingerprints the compiled guest binary (ELF) whose proofs the on-chain program accepts. The construction also inherits a software trust base: the guest source and Rust/LLVM-to-RISC-V compilation pipeline, SP1's patched cryptographic crates and precompiles, the SP1 prover and wrap circuit, the \texttt{sp1-solana} verifier crate, and the Solana runtime must be correct, as SP1's security model states~\cite{sp1securitymodel}. The claimed trust-root shift is therefore about what the cryptographic layer proves under correct implementations, not a claim that the stack is formally verified.

\paragraph{Adversary split: what the weight commitment adds over the signature.} Against a \emph{third party} without $\sk$, the per-cycle $\ed$ self-signature (condition~(iv)) is the load-bearing defence on its own: such an adversary fails~(iv) even when holding a perfect copy of $W$. The weight-commitment check (condition~(ii), enforced at every advance by check~(d)) is instead load-bearing against the \emph{key-holding operator}: without it, the operator could silently substitute $W'\neq W$ while presenting the same on-chain identity (the substrate-drift attack T6; \cref{sec:threat-tests}); with it, every accepted advance publicly certifies that the cycle executed under the genesis-committed, key-derived $W$. Guest-binary substitution is a separate code-axis threat, excluded by the pinned $\mathit{vkeyHash}$ and Groth16 soundness (T-CS; \cref{sec:threat-adjacent}).

The full threat catalogue (T1--T14) is in the project threat-model document~\cite{zkalife2026threatmodel} --- a document specific to this project, included in the public source archive rather than a separate publication; \cref{tab:threats} summarises coverage. We organise the coverage into tested rejections (\cref{sec:threat-tests}), accepted residual risks (\cref{sec:threat-accepted}), threats mitigated by chain primitives or by the stated assumptions (rows T1, T2, T7--T10, T13, T14), and adjacent-protocol attack templates whose applicability depends on the attacked surface (T-CS/T-DR/T-MT, additional to T1--T14; \cref{sec:threat-adjacent}).

\begin{table}[tbp]
\centering
\footnotesize
\setlength{\tabcolsep}{4pt}
\renewcommand{\arraystretch}{1.05}
\begin{tabularx}{\linewidth}{@{}l X X@{}}
\toprule
\textbf{Tag} & \textbf{Threat} & \textbf{Status} \\
\midrule
T3   & replay of a stale advance proof & rejected (test) \\
T4-a/b/c & $\ed$ signature injection variants & rejected (test) \\
T6   & advance with weights mismatching registered $\WC$ & rejected (test) \\
A4   & two concurrent advances at the same prior commitment & atomic: one succeeds, one rejected \\
T-a  & non-monotone cycle counter & rejected (chain precond.) \\
\midrule
T5   & $\sk$ compromise: a party who steals the private key gains full agent authority & accepted; no post-compromise recovery path in the current or $\sk$-derived-lineage design \\
T11  & program upgrade abuse: swapping a deployed program's code via its upgrade authority & precluded for all current programs: upgrade authorities irreversibly set to \texttt{none} on 2026-08-02 (\cref{sec:threat-accepted}) \\
T12  & host side-channel $\sk$ exfiltration: stealing $\sk$ from the host machine's memory or timing channels & accepted; TEE/keystore deferred to future scale extensions \\
\midrule
T1   & front-running / MEV pre-emption of an advance transaction & no identity takeover: a fresh, distinct advance requires $\sk$ (condition~(iv)) and the matching prior commitment; verbatim relay can land only the already-authorised transition \\
T2   & \texttt{init} spam of dummy agent PDAs & cost falls on the attacker (genesis proving + PDA rent) \\
T7   & state-chain reorder / rollback to a stale commitment & precluded by the prev-commitment precondition (\cref{sec:chain}); canonical history read at \emph{finalized} depth \\
T8   & recovering $W$ from the on-chain commitment $\WC$ & keccak preimage resistance; $W$ is $1{,}664$ pseudorandom bytes (entropy anchored in the $256$-bit $\sk$) \\
T9   & lying RPC serving stale agent state & soundness unaffected: the program reads consensus state at execution; a stale proof reverts and is regenerated \\
T10  & double-init of the same $\pk$ to swap $\WC$ & rejected: identical seeds derive the same PDA, so re-\texttt{init} fails (\texttt{AccountAlreadyInUse}) \\
T13  & offline forgery of a genesis proof & excluded under the Groth16-soundness assumption (\cref{sec:threat-scope}) \\
T14  & cycle-counter overflow at $2^{64}$ & rejected by checked arithmetic before the wrap; the horizon is unreachable in practice \\
\midrule
T-CS & code substitution (BAID-style) & different guest binary excluded by pinned $\mathit{vkeyHash}$ plus Groth16 soundness; weight substitution separately rejected as T6 \\
T-DR & document replacement (DIAP-style) & structurally inapplicable: no configuration object \\
T-MT & memory tampering (Spore.fun-style) & direct hidden-state overwrite is rejected; input-mediated memory poisoning remains possible through authorised $x_t$ \\
\bottomrule
\end{tabularx}
\caption{Threat coverage: tested rejections (T3, T4, T6, A4; T-a enforced by the chain precondition), accepted residual risks (T5, T12), chain-primitive or assumption-level mitigations (T1, T2, T7--T10, T13, T14; full analysis in~\cite{zkalife2026threatmodel}), and adjacent-protocol surfaces T-CS, T-DR, and T-MT (\cref{sec:threat-adjacent}). T11 is precluded for the current deployment by its irreversible 2026-08-02 upgrade-authority freeze. Tests are reproducible from the public repository (tag \texttt{arxiv-v1}).}
\label{tab:threats}
\end{table}

\subsection{Tested rejections}\label{sec:threat-tests}
\begin{itemize}[itemsep=2pt,topsep=2pt]
  \item \textbf{T4 (signature-injection family) and T3.} The scripted variants are T4-a (a valid signature under the wrong keypair), T4-b (a correctly keyed signature over the wrong message, its cycle index inflated), T4-c (the \texttt{Ed25519SigVerify} instruction omitted altogether), and T3 (replay of a stale advance proof against an already-advanced PDA). T3 violates the chain-side PDA precondition and is rejected (\texttt{StateChainViolation}) before any $\pi_{F_1}$ verification; the T4 variants pass the runtime precompile (T4-a/b) or omit it (T4-c) and are rejected by the program's introspection check against the registered $\pk$ and the expected message (\texttt{Ed25519Failed}), which in the deployed instruction order runs after proof verification. The reference adversarial bench in the public archive scripts the T4-a/b/c and T3 cases. The related tag \textbf{T-a} (a handcrafted PDA precondition with the cycle counter decremented or stalled) is rejected by the chain-side precondition (check~(c) of \cref{sec:f1}; cf.\ P4) but it is enforced structurally rather than exercised as a separate scripted case in that bench.
  \item \textbf{T6 (weight mismatch).} A modified host substitutes $W' \neq W$ at advance time. The genesis-time $\WC$ remains unchanged; the advance proof then publishes $\WC' = \keccak(W') \neq \WC$, and the on-chain program rejects with \texttt{WeightCommitMismatch} (\texttt{Custom(6007)}) before proof verification. Test added in commit \texttt{b136b30}.
  \item \textbf{A4 (concurrent advance).} Two advance transactions $(\tau_A, \tau_B)$ are identical except in their \texttt{ComputeBudget} unit limits ($400{,}000$ vs.\ $400{,}001$ --- the one-unit difference exists only to make the two transactions byte-distinct, since the cluster deduplicates identical transactions, while leaving their cost effectively equal --- hence distinct signatures). Both are submitted against the same \texttt{recent\_blockhash} (Solana's transaction-recency anchor) and reference the agent's cycle-3 PDA. The runtime serialises them; $\tau_A$ lands and advances cycle~$3 \to 4$, while $\tau_B$ lands second and fails with \texttt{Custom(6006)} \texttt{StateChainViolation}. Atomic advance is preserved. Bench: the A4 collision script in the public archive.
\end{itemize}
Both bench harnesses pin the deployed core program of \cref{tab:deployed} as their hard-coded target, so every case above is reproducible as a landed rejection against the current deployment; unlike the economic-path rejections (\cref{sec:phase4-threats}), individual core reverts are not explorer-linked.

\subsection{Accepted residual risks}\label{sec:threat-accepted}
T5 and T12 are accepted within this paper's scope. Post-compromise recovery from T5 requires a future authorisation path independent of the leaked seed (\cref{sec:disc-limits}, ``No post-compromise rekey''), while TEE/keystore hardening for T12 is deferred to future scale extensions. T11 (upgrade-authority abuse) is closed for all current programs (\cref{tab:deployed}), including the core program on which the \textit{alice}/\textit{bob} individuation evidence rests: on 2026-08-02 their upgrade authorities were irreversibly set to \texttt{none} with \texttt{set-upgrade-authority -{}-final}, a state publicly checkable via \texttt{solana program show}. This is a deployment property rather than an intrinsic property of the construction; a fresh deployment re-opens T11 until its operator performs the same freeze.

The of-record runs reported here were executed against the current deployments of \cref{tab:deployed}. The freeze pins those exact binaries and thereby preserves deployment provenance; it does not by itself make private-seed trajectory values or hardware-dependent timings bit-for-bit reproducible.

Denial of service against liveness or prover availability (an operator or third party halting advance by withholding proofs or by spamming the chain) is out of present scope, for the same integrity-not-availability reason given in \cref{sec:properties}; the liveness-vs-Byzantine-fault distinction and the per-actor liveness conditions are taken up in \cref{sec:disc-limits,sec:disc-safety}.

\subsection{Threats from adjacent identity protocols}\label{sec:threat-adjacent}
The catalogue above (T1--T14) covers attacks at the cryptographic-binding interface the core enforces. We additionally map three published attack templates from adjacent on-chain agent-identity protocols (BAID, DIAP-style profile chains, and Spore.fun) to this construction. The mapping separates T-CS's code axis from its weight axis, shows why T-DR is structurally inapplicable because the protocol has no configuration document, and distinguishes T-MT's unauthorised state overwrite from input-mediated poisoning.

\paragraph{T-CS (code substitution attack, BAID-style).}
Lin~et~al.~\cite{lin2025baid} formulate a threat in which an adversary holding the signing key runs malicious code in place of the legitimate program while signing the outputs as the legitimate agent's. In this construction, a proof from a different SP1 guest binary fails because the on-chain verifier pins the expected $\mathit{vkeyHash}$ and relies on Groth16 soundness. A genuine guest executed with $W'\neq\HKDF(\sk,\dots)$ is a separate weight-axis attack and fails the $\WC=\keccak(W)$ equality (T6). BAID's binary commitment and this paper's weight commitment therefore protect complementary axes.

\paragraph{T-DR (document replacement attack, DIAP-style).}
A second adjacent line~\cite{liu2025diap} treats identity as a tamperable configuration document --- a profile JSON, tool list, or system prompt swapped and re-signed with $\sk$. The threat is concrete elsewhere: Spore.fun's per-agent JSON genome rests on TEE attestation rather than a key binding~\cite{hu2025sporewild}, and zkLoRA's base-model blob is anchored only by an off-chain hash~\cite{liao2026zklora}, so host-side replacement is not checked against an on-chain commitment. The core construction has no such surface: behaviour is determined by $\sk$, $h_{t-1}$, and the per-cycle input $x_t$, with no protocol-level profile, tool registry, or persisted document. T-DR is therefore structurally inapplicable --- not defended against, but unsupported by the structure. A future extension that re-introduces such a surface (\cref{sec:disc-future-directions}) must bind the configuration inside the same SNARK that binds the weights (concurrent work pursues such configuration-and-capability binding for LLM agent tool use, signature-based rather than SNARK-internal~\cite{zhou2026capabilities}).

\paragraph{T-MT (memory tampering, Spore.fun-style).}
Hu and Rong~\cite{hu2025sporewild} report community trolls ``poisoning'' an agent's memory by repeatedly supplying a phrase until the agent echoed it. That is input-mediated poisoning, not a direct overwrite of stored state. Here $h_{t-1}$ is committed on chain and reproduced by the next $F_1$ proof, so an unauthorised party cannot replace it with an arbitrary hidden vector. Nevertheless, an authorised or maliciously sourced $x_t$ can shape $h_t$ through a valid transition; the designated oracle authenticates provenance but does not guarantee semantic benignity. The construction therefore rejects direct state tampering while retaining input-poisoning and oracle-trust risk. The same distinction applies to the homeostatic extension and must be carried into any future learnable-weights or longer-memory design (\cref{sec:disc-future-directions}).

\subsection{Environment-oracle threats}\label{sec:threat-p4m1-ext}
The economic-metabolism extension authenticates the environment vector $x_t$ through a proof-and-runtime composition. The guest verifies the oracle's \texttt{ed25519} signature over the attested message and commits the signing key, cycle, and $x_t$ to the proof's public values (mechanism and message format in \cref{sec:phase4-m1}); \texttt{advance\_v5} then checks the committed key and cycle against the genesis-registered \texttt{oracle\_pubkey} and expected counter. Signature validity and binding to the committed transition are therefore proof-level properties, whereas equality with the registered oracle and expected cycle is enforced on chain. This composition exposes \emph{three} environment-oracle-specific Anchor \texttt{Custom} errors of \texttt{p4\_economic\_verifier::advance\_v5}: \emph{oracle mismatch} (\texttt{OracleSigFailed} --- the oracle key committed in the proof, $\mathit{PV}[140..172]$ ($\mathit{PV}$: the proof's public values, indexed by byte offset; \cref{sec:bg-zkflow}), does not equal the genesis-registered \texttt{oracle\_pubkey}), \emph{attestation-cycle mismatch} (\texttt{OracleCycleMismatch} --- the attestation cycle committed in the proof, $\mathit{PV}[172..180]$, does not equal the advance's expected cycle, excluding stale-attestation replay), and \emph{environment-range violation} (\texttt{EnvOutOfRange} --- some $|x_{t,i}| \geq 2^{18}$, a redundant on-chain guard backing the in-circuit range assertion that forecloses the fixed-point wrap of an unbounded input). A forged or wrong-message oracle signature cannot produce a valid proof at all --- the in-circuit \texttt{ed25519} verification fails --- so it is rejected at proof generation rather than as an on-chain \texttt{Custom} error; likewise the oracle-signed $x_t$ \emph{is} the committed $\mathit{PV}[100..120]$ by construction, so there is no separate on-chain environment-mismatch check. The wallet-drain template \emph{A-EA1} is foreclosed structurally by the program-owned $\Fecon$ PDA (\cref{sec:phase4-m0,sec:disc-limits}): lamports leave the PDA only through the protocol's metabolic debit, with no withdrawal instruction, so no external party can drain it. Decentralised-committee attacks (Byzantine quorum, slashing griefing, stake forgery) do not apply to the single-oracle design (\cref{sec:disc-future-directions}).

Registration introduces a separate extension-specific liveness caveat. Unlike the core record described in \cref{sec:f2}, \texttt{initialize\_agent\_v4} receives \texttt{oracle\_pubkey} as an instruction argument that is not pinned by the genesis proof, and it does not require \texttt{agent\_pubkey} to sign. A third party holding a copied valid genesis proof can therefore pre-empt creation of the identity's PDA while supplying an attacker-chosen, distinct oracle key, after which the legitimate initialisation fails because the PDA already exists. The squatter still lacks $\sk$, so cannot pass the per-advance self-signature or withdraw from the program-owned economic wallet; the impact is registration denial of service, not identity takeover or fund theft. Requiring an agent signature at initialisation or binding the oracle key into the genesis journal would close this gap.

\paragraph{Section summary.}
Of the catalogue, four tags are demonstrated as landed on-chain rejections (\cref{sec:threat-tests}); the remainder of T1--T14 are handled structurally or by standard cryptographic argument (\cref{tab:threats}); T5 and T12 (host-side key custody and host compromise) remain accepted residuals, while the current deployment's freeze closes T11 (\cref{sec:threat-accepted}). The adjacent-system analysis separates precluded code or state substitutions from input-mediated poisoning that remains possible (\cref{sec:threat-adjacent}). \Cref{sec:eval} turns to the affirmative empirical evidence.

\section{Empirical Evaluation}\label{sec:eval}

\paragraph{Evaluation questions and scope.}
The core evaluation asks whether key-derived weights are functionally distinct (Q1; \cref{sec:eval-individuation,app:q1-detail}), whether the binding survives a long continuous run (Q2; \cref{sec:eval-long}), what verification costs on chain (Q3; \cref{sec:eval-cu}), and how long honest-prover proof generation takes (Q4; \cref{sec:eval-prove}). \Cref{sec:poc-extensions} addresses Q5--Q7 for the PoC-tier extensions. Devnet scope and single-host key custody are delimited in \cref{sec:disc-limits}; the non-statistical $N{=}2$ pairing and descriptive $M_2$ diagnostic are delimited with the corresponding results in \cref{app:q1-detail}.

\subsection{Setup}\label{sec:eval-setup}
Two agents, \textit{alice} and \textit{bob}, are each registered with an independent secret $\sk$. The hosts run the reference orchestrator (the host-side daemon that drives the per-cycle prove-and-submit loop) from the public archive, on the reference workstation (an Ubuntu host with an NVIDIA RTX~3090 GPU); proofs are generated locally with SP1~v5.2.4 using CUDA GPU proving. The on-chain program is the Anchor build at the cited program ID on Solana devnet. The evaluation refers to several distinct runs and agents; \cref{tab:runs} maps them at a glance.

\begin{table}[htbp]
\centering\small
\setlength{\tabcolsep}{4pt}
\begin{tabular}{@{}lllll@{}}
\toprule
\textbf{Run} & \textbf{Cycles} & \textbf{Chain} & \textbf{Key ($\sk$)} & \textbf{Feeds} \\
\midrule
\textit{alice}/\textit{bob} short        & $20$            & on-chain   & independent \texttt{keygen}    & Q1 \\
\textit{alice} continuous                & $166$ (1--166)  & on-chain   & \texttt{keygen}                & Q2--Q3 \\
core-circuit timing rerun                & $166$ proofs    & off-chain  & fixed loop structure           & Q4 \\
\textit{castor}/\textit{pollux} long-horizon & $168$           & off-chain  & \texttt{keygen} (private)      & Q1 (long-run) \\
many-key null                            & $500{\times}20$ & simulation & SHA-256-indexed                & Q1 (null) \\
\textit{charlie} economic                & $168$           & on-chain   & \texttt{keygen} (private)      & \cref{sec:phase4} \\
\bottomrule
\end{tabular}
\caption{The runs and agents referenced in the evaluation, at a glance; the PoC-extension runs (Q5--Q7) are indexed in \cref{app:ext-eval}. All agent runs use independent secret \texttt{solana-keygen} keys, each with a privately held seed: the on-chain core pair \textit{alice}/\textit{bob}, the on-chain economic agent \textit{charlie}, the off-chain long-horizon pair \textit{castor}/\textit{pollux}, and likewise the \cref{sec:poc-extensions} homeostatic daemon (\cref{tab:deployed}). The core-circuit timing rerun is a prover benchmark of the deployed core circuit, not an agent run: its per-proof time is independent of the witnessed key (\cref{sec:eval-prove}). The many-key null's ``SHA-256-indexed'' keys are derived deterministically from public pair indices rather than from secret seeds, which is what makes that null bit-reproducible (\cref{sec:eval-individuation}).}
\label{tab:runs}
\end{table}

\subsection{Key-binding divergence (Q1)}\label{sec:eval-individuation}

We track inter-agent $L_2$ divergence $M_4$ over a 20-cycle run to verify that distinct key-derived weights produce distinct dynamics. Two independently keyed agents reach mean $M_4 = 0.68$ ($>6.8\times$ the pre-specified mean floor of $0.1$; last-five-cycle mean $0.80$, $>16\times$ the tighter $0.05$ floor), while the same-key control stays at $M_1 = 0$ divergence at every cycle. In a 500-pair reference distribution, every distinct-key pair clears both floors; the of-record pair lies in the lower tail.\footnote{The 20-cycle short-run figures ($M_4$ mean $0.68$, last-five $0.80$) place the of-record pair at the $7.2$ and $7.0$ percentiles of the 500-pair reference distribution (\cref{app:q1-detail}; data \texttt{figures/manykey\_null\_m4.json}). It is less divergent than a typical pair while still clearing the two floors by $6.8\times$ and $16\times$.} An independent-host 168-cycle $N{=}2$ pairing (\textit{castor}/\textit{pollux}) extends the result over a longer horizon, with $M_4$ growing from $0.62$ to $3.50$ (\cref{app:q1-detail}).

Conditional on a fixed key pair and input schedule, the dynamics are deterministic, so there is no trial-to-trial noise distribution. Across sampled key pairs, however, $M_4$ has the empirical key-induced reference distribution reported here; the floors are pre-specified minimum effect sizes rather than noise estimates. This analysis is a non-degeneracy check of the key-to-weights pipeline: every sampled distinct-key pair induces dynamics separated above the pre-specified floors. The on-chain Q1 evidence comprises one independently keyed \textit{alice}/\textit{bob} pair; the 500-pair result is a simulation reference distribution, not 500 on-chain replications. The weight-commitment rejection test T6 supplies the enforcement evidence (\cref{sec:threats}). \Cref{app:q1-detail} gives the full metric definitions ($M_1$--$M_4$), the long-horizon table, the reference distribution, and the secondary $M_2$ sign-structure diagnostic.

\subsection{Long continuous run (Q2)}\label{sec:eval-long}
A 166-cycle continuous run of \textit{alice} (one cycle per $20$ minutes) spanned $2.36$ days (cycles $1$ through $166$), with no failed transactions: the agent PDA's complete on-chain history is $167$ landed transactions --- one registration plus $166$ advances --- every one successful. The run log's per-cycle on-chain counter matches the host's local cycle index at every cycle, and the PDA cycle counter reads $166$ at run completion. The mean inter-cycle wall-clock period was $20.6$~min (median $20.7$~min, range $[2.4, 22.1]$~min), modestly above the $20$~min cadence target because the period carries the daemon's scheduling interval on top of each cycle's prove-and-submit work; the $2.4$~min minimum is a catch-up cycle immediately after a host-side daemon restart (the daemon was restarted twice over the run, each time resuming from the on-chain state --- once after a transient network outage --- and the on-chain commitment chain is unbroken across both, which is the invariant under test). Per-cycle periods are computed from the run's timestamped per-cycle log (committed in the public archive); within devnet's ledger retention they are also independently recomputable from the block timestamps of the PDA's on-chain transaction history (\cref{tab:deployed}) --- devnet, unlike mainnet, prunes old transaction history and may be reset, so beyond that window the committed log is the durable source. The run is a liveness stress test, not a Byzantine-fault test (\cref{sec:disc-limits}).

\subsection{On-chain verification cost (Q3)}\label{sec:eval-cu}
The dominant on-chain cost is Groth16 verification via \texttt{sp1-solana}; the PDA precondition check and the state-update logic account for the small remainder, and the $\ed$ verification, delegated to Solana's native precompile, is charged through the per-signature fee rather than through metered compute units. Aggregating all $166$ \textit{alice} advance transactions of the continuous run (cycles 1--166), the per-transaction compute-unit cost has mean $246{,}906$, median $246{,}906$, standard deviation $15.5$, and lies in $[246{,}861, 246{,}951]$ --- a $90$-CU band (interdecile range $39$) with no outlier transactions. A per-instruction decomposition of a fresh disposable advance against the same deployed program ($246{,}876$~CU in total, within that band) attributes $246{,}726$~CU to the verifier program instruction and $150$ to the compute-budget instruction, with the $\ed$ precompile contributing zero metered CU (the advance transaction instead pays two signatures' base fees, versus the registration's one). The observed maximum $\mathit{CU}_{\max} = 246{,}951$ comfortably leaves headroom under the Solana per-transaction CU cap ($1.4$~million CU). The one-time genesis registration consumed $255{,}670$~CU ($255{,}520$ in the program instruction --- the genesis-proof verification plus agent-PDA creation --- $8{,}794$ above an advance), measured on a fresh disposable registration against the same deployed program, as the of-record registration transactions predate devnet's ledger retention (\cref{sec:eval-long}).
\subsection{Honest-prover proof generation (Q4)}\label{sec:eval-prove}
SP1 proof generation with CUDA GPU proving on the reference workstation, measured in a dedicated $166$-proof timing rerun of the deployed core circuit (\cref{tab:runs}), takes mean $38.74$~s, median $39.10$~s, standard deviation $3.05$~s, and lies in $[34.71, 44.71]$~s. Circuit dimensions and guest loop counts are fixed across keys, so the rerun estimates per-transition proving cost without conflating it with the continuous run's concurrent GPU workloads. Small data-dependent branch effects, witness-generation details, and host scheduling can still vary timing; the reported distribution is empirical rather than a key-independence proof. Variance is dominated by witness generation and Groth16 proving; together they constitute the principal cost the protocol pays for re-proving the binding in~\cref{eq:f1} on every transition.

The one-time genesis proof ($F_2$) was measured in a dedicated $20$-proof run on the same GPU prover (two additional contention-affected samples, taken while the GPU was shared with an unrelated workload, are excluded): mean $48.18$~s, standard deviation $1.93$~s, range $[41.48, 50.42]$~s --- roughly $9$~s above the advance circuit, a difference consistent with --- but not causally isolated to --- the in-circuit $\ed$ key derivation and full HKDF expansion that only genesis carries (\cref{sec:f2}). It is paid once per agent, not per cycle; the dedicated run's per-proof timings are committed in the public archive, and the extension lineage's genesis prove time is recorded in the shipped economic-run metrics artefact.

Three orientation notes for ML readers.
\begin{description}
\item[Scaling.] SP1 proving cost is expected to grow quasi-linearly with guest execution cycles while the Groth16 wrap stage is roughly constant (\cref{sec:bg-zkflow}). As a first-order expectation, a substrate with $k\times$ more per-cycle arithmetic should approach a similar increase in proving cost once guest execution dominates, although trace width, memory behaviour, and wrap overhead can shift the ratio; on-chain verification cost remains unchanged (\cref{sec:eval-cu}). Larger models are therefore gated primarily on prover throughput, not on-chain cost.
\item[Why prove at all at this scale.] At $416$ weights the cheaper baseline is transparent verification: publish $W$ (revealing neither $\sk$ nor any signing ability) with a one-time genesis proof of key derivation, and let verifiers re-execute each step directly. Per-cycle proving is thus a design choice here, not a requirement: it keeps on-chain verification cost constant as the substrate grows, and leaves room for the cryptographically private internal state of \cref{sec:disc-future-directions}.
\item[Cross-paper orientation.] On the GPU prover the wrapped (on-chain-grade) per-cycle proof times reported in this paper are ${\sim}39$~s here (core protocol), ${\sim}42$~s for the larger active-query circuit (\cref{sec:eval-phase2}), ${\sim}41$~s for the mutation circuit (\cref{sec:eval-phase3}), and ${\sim}53$~s for the economic-metabolism circuit (\cref{app:m3-detail}); the near-constant Groth16 wrap stage holds the three smaller circuits to a ${\sim}40$~s band, while the economic circuit's ${\sim}53$~s sits above it. The economic and active-query circuits both verify an \texttt{ed25519} signature in-circuit. The economic circuit additionally commits the registered-oracle key and attestation cycle in a 180-byte journal (\cref{sec:phase4-m1}), but because the circuit families differ in several respects, the observed gap is descriptive rather than a causal attribution to any one feature.
\end{description}
\begin{figure}[tb]
\centering
\begin{tikzpicture}
  \begin{axis}[
    width=0.95\linewidth,height=4.4cm,
    xlabel={proof index},
    ylabel={proof time (s)},
    xmin=0,xmax=170,
    ymin=33,ymax=46,
    grid=both,
    grid style={gray!20},
    tick label style={font=\footnotesize},
    label style={font=\footnotesize},
    legend style={font=\scriptsize,draw=none,fill=none,at={(0.98,0.97)},anchor=north east}
  ]
    \addplot+[only marks,mark=*,mark size=1.2pt,blue!70!black,mark options={fill=blue!50}] table[x=index,y=proof_sec] {figures/proof_time_data.dat};
    \addlegendentry{per-proof time}
    \addplot[red,dashed,thick,domain=0:170] {38.74};
    \addlegendentry{mean ($38.74$ s)}
  \end{axis}
\end{tikzpicture}
\caption{Per-proof SP1 proof-generation time for the deployed core circuit, over the dedicated $166$-proof GPU timing rerun of \cref{sec:eval-prove} ($n{=}166$); summary statistics in the text. The longest proofs ($44.1$--$44.7$~s) are consistent with host-side timing variability.}
\label{fig:proof-time}
\end{figure}
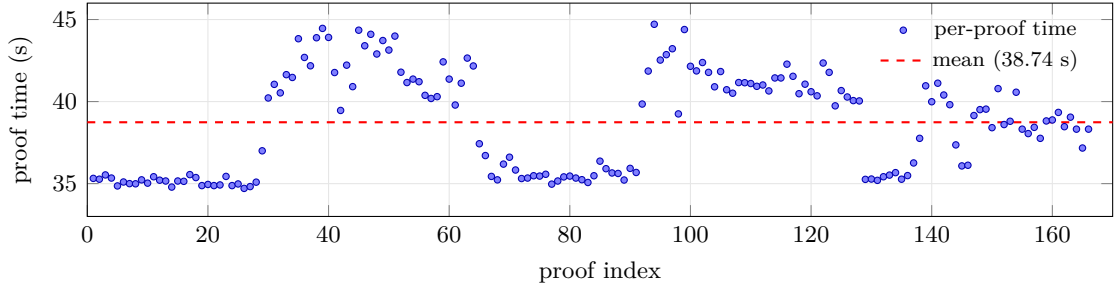

\paragraph{Section summary.}
Independently keyed agents diverge while a same-key control does not (Q1), the binding holds over a 2.36-day, 166-cycle continuous on-chain run with two host-side resumptions (Q2), each on-chain verification costs a bounded ${\sim}247$k CU (Q3), and re-proving the binding costs ${\sim}39$ seconds of GPU proving per cycle (Q4). \Cref{sec:poc-extensions} layers the PoC-tier extensions on this base.

\section{Sensorimotor and Homeostatic Extensions}\label{sec:poc-extensions}

\paragraph{Extension evaluation.}
The \emph{active-query-loop extension} turns the synthetic input $x_t$ into a patron-signed, in-circuit-attested external signal (\cref{sec:design-phase2}), and the \emph{homeostatic extension} adds an internal motivational driver and key-anchored weight rotation (\cref{sec:design-phase3}). Their evaluation asks whether environment binding preserves individuation (Q5; \cref{sec:eval-phase2}), whether the deployed rotation is restricted to the agent's key holder and the committed rotation inputs (Q6; \cref{sec:eval-phase3}), and whether the core protocol plus both extensions execute end-to-end on a public chain (Q7; \cref{sec:eval-onchain}).

\subsection{Extension: active-query loop for sensorimotor coupling}\label{sec:design-phase2}
The core evaluation instantiates the environment vector $x_t$ in \cref{eq:elman} as a synthetic deterministic schedule (the fixed ramp of \cref{app:q1-detail}); the core protocol itself leaves $x_t$ as a prover-chosen private witness (\cref{sec:properties}). Intuitively, the evaluated core agent therefore has no real environment input: its $x_t$ is a pre-programmed sequence that nothing in the world influences. The active-query loop closes that gap: each cycle the agent asks for a fresh observation, a designated patron answers, and the patron-signed answer becomes the input that the agent's circuit consumes. Formally, this extension turns $x_t$ into a patron-signed, in-circuit-attested external signal under a witness-supplied patron key while retaining the identity-binding invariants ($W = \HKDF(\sk, \mathit{tag})$, $\Stream(K)$ chain, $\ed$ self-authorisation).

The active-query mechanism has three components. \emph{First}, the agent's prior action $y_{t-1}$ parameterises the request for the cycle's environment. \emph{Second}, a designated patron computes the environment vector $x_t$ (a function of the cycle and of $y_{t-1}$) and signs, \emph{off chain}, the message $\texttt{"zkalife:p2:respond:"} \,\|\, \mathit{cycle} \,\|\, \keccak(y_{t-1}) \,\|\, x_t \,\|\, \mathit{flags}$ under its own key, publishing the signed vector as an off-chain artefact (no on-chain query or response account is used). \emph{Third}, the agent's $F_1$ advance circuit takes the patron's public key and signature as witnesses and verifies the signature \emph{inside the SP1 guest} against the reconstructed message (the canonical verifying key fixes the guest code), binding $x_t$ to the cycle and to $\keccak(y_{t-1})$; the responder key is supplied as a witness and is neither committed to the journal nor checked by the on-chain program (\cref{sec:disc-limits}). The verified environment vector $x_t$ enters the journal output and thereby $c_t$; a dedicated on-chain verifier --- in the deployed lineage, the \texttt{p3\_motivation\_verifier} program of \cref{sec:eval-onchain}, which embeds this extension's advance verifying key --- then checks the SP1 proof, the state chain, and the agent's own $\ed$ self-signature. The threat tag T-a (non-monotone cycle counter) has an environment analogue in this extension: an advance whose environment fails the in-circuit patron-signature check cannot produce a valid $\pi_{F_1}$ at all, so the case is foreclosed at proof generation rather than on chain (cf.\ \cref{sec:threat-p4m1-ext}).

The design takes a partial step toward the environmental-coupling (interactional-asymmetry) criterion that the core construction explicitly defers (\cref{sec:disc-limits}); full \emph{phenomenal coupling} (an environment that is not a cooperative patron but a non-cooperative external process) remains an open research direction.

\subsection{Extension: homeostatic motivation and key-anchored weight rotation}\label{sec:design-phase3}
The homeostatic extension adds two further constructions on top of the active-query-loop extension: (a)~a \emph{homeostatic driver inspired by homeostatic reinforcement learning} (HRRL) --- an internal motivational economy modelled on biological homeostasis (drives such as ``energy'' and ``temperature'' that must stay close to setpoints), independent of any external reward; the deployed driver regulates these drives but does not itself perform learning; and (b)~a \emph{key-anchored weight rotation} that admits per-epoch $W$ updates while preserving the cryptographic-identity binding. We describe each in turn. The homeostatic variables here are \emph{virtual}: ``energy'' is an internal drive state, not the agent's on-chain balance. A natural next step couples the two --- letting each agent hold its own wallet and treat its Solana balance ($\Fecon(K)$, \cref{sec:phase4}) as the energy reservoir the homeostatic loop must keep above its setpoint --- folding the economic-metabolism axis into the motivational economy; we leave this to the sleep-replication extension.

The HRRL-inspired driver maintains an internal 4-dimensional $\mathrm{Q}16.16$ state $\xdrive(t) = (\text{energy\_tank}, \text{thermal}, \text{novelty}, \text{reserve})$ subject to a setpoint $\bm{s} = (1.0, 0.5, 0.5, 0.7)$. The agent's allostatic free energy is the squared-distance proxy
\[
F_{\text{allostasis}}(\xdrive, \bm{s}) \;=\; \sum_{i=1}^{4} (\xdrive_i - s_i)^2,
\]
expressed in $\mathrm{Q}16.16$ fixed-point. A \emph{sleep-mini protocol} gates a sleep cycle when $F_{\text{allostasis}} \ge \theta = 0.4$ (the canonical threshold); the sleep cycle (i)~skips the active query for that cycle, (ii)~restores \texttt{reserve} to setpoint, and (iii)~halves the gap to setpoint for the thermal and novelty drivers. The \texttt{p3\_motivation\_verifier} program records the sleep transition as a key-signed \texttt{sleep\_trigger} instruction that commits a root value $r_{\text{replay}}$ supplied by the key holder as a commitment to the replay buffer since the previous sleep; the chain-side state retains $r_{\text{replay}}$ as the agent's \texttt{last\_history\_root}, but does not verify that the value was derived from those cycles.

The key-anchored weight rotation produces a new weight epoch
\[
\begin{aligned}
W_{t+1} \;=\; \HKDF\bigl(&\sk,\; s = \texttt{"zkalife:phase3:v1"},\\
&\mathit{info} = \texttt{"zkalife:phase3:rotate:"} \mathbin\| n_{\text{mut}}^{\mathrm{LE}} \mathbin\| r_{\text{replay}}\bigr),
\end{aligned}
\]
proven inside a dedicated mutation guest circuit.\footnote{The literal byte strings \texttt{"zkalife:phase3:*"} are HKDF domain-separation tags retained from the private development repository, where the construction was first prototyped under the working name ``phase 3''; in this archive the corresponding component is \texttt{code/homeostatic-extension/}. The literals are kept verbatim because they are committed to in the deployed circuit and cannot be renamed without invalidating every existing weight commitment on chain.} Intuitively, the rotation lets the agent change its weights only in ways that still prove the new weights belong to the same agent: derived from the same secret, continuing from the previous commitment, bound to a replay-root value attested by the key holder, and self-signed. A successful rotation requires four cryptographic guarantees jointly enforced on chain: \textbf{(i)~SP1 proof.} The mutation guest re-derives both $W_t$ and $W_{t+1}$ in-circuit from $\sk$ by the displayed derivations (neither is a free witness) and attests $\keccak(W_t) = \mathit{old}\WC$, $\keccak(W_{t+1}) = \mathit{new}\WC$, and $\keccak(\sk) = \mathit{sk}\mathit{Commit}$ in zero knowledge. \textbf{(ii)~State chain.} The Anchor program requires $\texttt{agent.weight\_commit} = \mathit{old}\WC$, enforcing continuity with the previous epoch. \textbf{(iii)~Replay binding.} The program requires $\texttt{agent.last\_history\_root} = r_{\text{replay}}$, where $r_{\text{replay}}$ was committed by a preceding $\sk$-signed \texttt{sleep\_trigger}; no third party can substitute a different replay root. This binding does not prove that the key-holder-selected root faithfully summarises prior cycles. (The freshly initialised root is the all-zero vector, so an agent's first rotation can be proven against that zero root without a preceding sleep; the $\sk$ gating of (i)/(iv) is unaffected, and the of-record demonstration runs \texttt{sleep\_trigger} first.) \textbf{(iv)~Self-signature.} A standard $\ed$ self-signature over $\texttt{"zkalife:p3:mutate:"} \| \mathit{old}\WC \| \mathit{new}\WC \| r_{\text{replay}} \| n_{\text{mut}}^{\mathrm{LE}}$ verifies as the leading instruction of the transaction.

The four guarantees together pin every $W_t \to W_{t+1}$ transition to the agent's own $\sk$ \emph{and} a replay-root value signed by that key. No external party who lacks $\sk$ can construct a mutation accepted on chain, regardless of any side knowledge of $W_t$, the replay buffer, or the deployed circuit. The construction realises a constrained form of \emph{in-life weight rotation} that admits per-epoch updates while preserving the identity binding: $W$ remains a deterministic function of $\sk$, the mutation counter, and a \emph{self-attested replay-root value}, composed over successive epochs rather than fixed at genesis. The rotation is not itself a learning rule or evidence of behavioural improvement: it optimises no objective and assigns no credit, but deterministically expands HKDF from $(\sk,n_{\mathrm{mut}},r_{\mathrm{replay}})$.

``Self-attested replay-root value'' is meant precisely: the rotation input includes the monotone mutation counter $n_{\text{mut}}$ and the $\sk$-signed replay root $r_{\text{replay}}$, but the proof-less \texttt{sleep\_trigger} commits $r_{\text{replay}}$ without a circuit verifying that it faithfully summarises the prior on-chain cycles (an $\sk$-holder can therefore also select among candidate rotation outcomes by grinding $r_{\text{replay}}$ --- the rotation-time analogue of the key-grinding residual noted in \cref{sec:disc-safety}, with the difference that it preserves the agent's identity and lineage); binding the rotation to a circuit-verified history is deferred to the \emph{sleep-replication extension}, a successor deliverable adding full sleep-phase weight-rotation chaining, key rotation, and child-agent replication. A rotation is a discrete sleep-phase event rather than a per-cycle advance, and the current implementation exercises just a single such rotation epoch on chain (\cref{sec:eval-onchain}; the integrated daemon run of \cref{sec:eval-phase3} chains seven epochs off-chain); chaining multiple rotations on chain is likewise deferred to that extension.

The rotation addresses an apparent tension between the frozen-$W$ choice (which deliberately isolates the identity-binding question from learning-driven adaptation; \cref{sec:construction}) and the de~facto Lamarckian inheritance of modern training pipelines, in which learned improvements are written back into heritable representations. The present HKDF rotation demonstrates the cryptographic discipline under which a future learning rule could perform such writeback: any accepted update must remain inside the same SNARK-enforced identity binding $\WC = \keccak(\HKDF(\sk, \dots))$.

\subsection{Evaluation of the PoC extensions (Q5--Q7)}\label{sec:poc-eval-summary}
The two extensions are evaluated in full in \cref{app:ext-eval}; we summarise the findings here. \emph{(Q5)} Under the active-query loop the binding invariants survive environment coupling: the $L_2$ individuation divergence $M_4$ stays well above threshold, but because a shared deterministic environment increases inter-agent correlation, this is an integrity result rather than independent individuation evidence. \emph{(Q6)} The key-anchored $W$ rotation is restricted to the agent's key holder and the committed rotation inputs by construction --- no party without $\sk$ can inject a mutation the chain accepts --- and is reproduced bit-exactly on an independent machine. The replay-root input is key-holder-attested, not a circuit-verified history summary. \emph{(Q7)} The full Tier~1 PoC sequence (\texttt{advance\_v2} $\to$ \texttt{sleep\_trigger} $\to$ \texttt{mutate\_w}) executes end-to-end on Solana devnet with every required precondition satisfied. Full numbers, the cross-machine determinism check, and the devnet transaction signatures are in \cref{app:ext-eval}; on-chain compute-unit cost (\cref{sec:eval-cu}'s Q3 counterpart) was not separately measured for these PoC-tier instructions, only proof-generation time.

\section{Economic Metabolism Extension}\label{sec:phase4}

\paragraph{Extension components.}
The economic extension adds a consumption-side constraint to the $\Fecon(K)$ axis through a single-cycle aliveness predicate (\cref{sec:phase4-motivation,sec:phase4-m0}) and an in-circuit designated-oracle binding (\cref{sec:phase4-m1}). We evaluate it with 24- and 168-cycle metabolic runs and environment-oracle rejection paths (\cref{sec:phase4-m2,sec:phase4-m3,sec:phase4-threats}), then delimit the resulting claim in \cref{sec:phase4-closure}.

\subsection{Motivation: constraining the \texorpdfstring{$\Fecon(K)$}{F\_econ(K)} axis}\label{sec:phase4-motivation}
Until this point, a patron has paid the agent's bills on chain. This extension gives the agent its own economic substrate and imposes a metabolic cost on its continuation: every cycle the protocol debits an explicit cost from a program-owned account derived from the agent's key (its $\Fecon$ wallet), and an on-chain check rejects the cycle if that account falls below the survival floor. The cost is \emph{imposed by the protocol, not authorised by the agent}: expenditure is involuntary, analogous to biological metabolism, rather than a deliberate transaction. Deliberate, key-authorised spending is deferred to a future economic-agent stage (\cref{sec:disc-future-directions}). This supplies the consumption side of the $\Fecon(K)$ axis in the agent triple $\agent = (K, \Stream(K), \Fecon(K))$. For Artificial-Life readers, it realises the \emph{metabolic-expenditure} half of the metabolic-closure criterion of \cref{sec:intro}; an agent that also earns its keep is not yet realised (\cref{sec:phase4-closure}).

The construction is incremental: the same $W = \HKDF(\sk, \mathit{tag})$ binding and per-cycle Groth16 attestation, plus three on-chain mechanisms --- an \emph{aliveness predicate}, a \emph{designated environment oracle}, and a \emph{continuous metabolic run} (\cref{sec:phase4-m0,sec:phase4-m1,sec:phase4-m2,sec:phase4-m3}).

The \emph{aliveness predicate} ``$K$ valid $\wedge\ \Stream(K)$ advancing $\wedge\ \Fecon(K) > 0$'' is enforced at every state-changing instruction of the economic extension, making the agent triple a circuit-and-runtime invariant rather than a theoretical construct.\footnote{We use \emph{aliveness predicate} in a strictly operational sense --- an advance-admissibility and state-transition invariant: it cannot compel the next submission, so it does not by itself enforce temporal progress. Here $\Fecon(K) > 0$ is an externally replenishable consumption floor rather than a metabolic-viability condition; the term makes no claim of biological aliveness or autopoietic self-maintenance (\cref{sec:phase4-closure,sec:disc-barandiaran-triad}). In the deployed economic program, \texttt{advance\_v5}, weight rotation, and sleep triggering all enforce the same $\Fecon$ floor; the latter two additionally retain their own signature and replay gates from the base design (\cref{sec:design-phase3,sec:phase4-m0}). Only the advance-path rejection was exercised as an on-chain negative test.}

\subsection{Aliveness predicate: single-cycle advance constraint}\label{sec:phase4-m0}
The aliveness predicate is realised on chain by the extended \texttt{AgentV4} record, which replaces the extension lineage's \texttt{AgentV3} layout (\cref{sec:eval-onchain}) and carries the agent's $\Fecon$ substrate. It records a key-derived, program-owned wallet PDA, an atomically updated balance snapshot and check slot, the registered environment-oracle key, and a monotonic sleep-count replay guard; the implementation-layout appendix (\cref{app:implementation-layouts,app:agentv4-layout}) gives the exact field types and the expansion from 254 to 342 bytes. The wallet is a distinct address bound to $K$ by derivation, $\texttt{wallet\_pubkey}=\mathrm{PDA}([\texttt{"agent\_wallet"},\pk])$, and its lamports can be moved only by the program's metabolic debit, not by any key (closing the wallet-drain vector A-EA1, \cref{sec:threat-p4m1-ext}); the operator endows it at genesis, and the underlying account balance remains canonical.

The oracle-gated \texttt{advance\_v5} instruction (\cref{sec:phase4-m1}) gates each advance on the aliveness predicate within the same atomic instruction that verifies $\pi_{F_1}$ (the deployed binary checks the economic and state-chain preconditions before the proof verification, and the $\ed$ self-signature after it; failure of any check reverts the whole transaction):
\begin{enumerate}[itemsep=2pt,topsep=2pt,leftmargin=1.6em,label=(\alph*)]
  \item $\sk$-side: the $\ed$ signature over the advance message $(\mathit{cycle}{+}1) \,\|\, c_t$ under the inherited \texttt{"zkalife:p2:advance:"} domain prefix is verified ($K$ valid).\footnote{The economic-metabolism programs reuse the \texttt{"zkalife:p2:advance:"} domain-separation prefix inherited from the active-query program lineage they were cloned from; the core protocol of \cref{sec:f1} uses \texttt{"zkalife:p1:advance:"}. Like the HKDF tags of \cref{sec:design-phase3}, the literal is committed in the deployed binary and cannot be renamed without invalidating existing on-chain signatures.}
  \item $\Stream(K)$-side: the PDA precondition $(\mathit{cycle}, \mathit{state\_commit}) = (t, c_{t-1})$ holds ($\Stream$ advancing).
  \item $\Fecon(K)$-side: the agent's economic PDA holds $\Fecon(K) > \texttt{ECON\_THRESHOLD}$ ($10^7$~lamports $= 0.01$~SOL in the deployed program, a deliberately stronger floor than the predicate's literal $\Fecon > 0$, and above the account's rent-exempt minimum so the metabolic debit cannot close it) to bear this cycle's metabolic cost.
\end{enumerate}
This cycle's metabolic cost is debited from the agent's economic PDA \emph{by the program itself} --- a direct lamport decrement of the program-owned account, independent of which signer pays the Solana transaction fee --- so the metabolic debit is a program-enforced invariant rather than a property of the fee-payer convention. If any of (a)--(c) fails, the instruction reverts before any state change (a landed-but-reverted transaction still pays its transaction fee under Solana's fee rules, which is why the orchestrator pre-checks these failure modes client-side before submitting).
\subsection{Designated environment oracle}\label{sec:phase4-m1}
Where does the agent's environment data come from? The aliveness predicate constrains $\Fecon(K)$, but $x_t$ still requires authentication. In the active-query extension (\cref{sec:design-phase2}), the patron key was a witness-supplied value that was neither committed nor checked on chain, so the operator could fabricate the environment. The economic extension instead binds environment provenance to a \emph{designated oracle} whose public key is registered at genesis.

The SP1 circuit verifies the oracle's strict RFC~8032 \texttt{ed25519} signature over $\mathit{prefix}\,\|\,\mathit{cycle}\,\|\,\keccak(y_{t-1})\,\|\,x_t\,\|\,\mathit{flags}$, commits the signing key in $\mathit{PV}[140..172]$ and the cycle in $\mathit{PV}[172..180]$, and bounds each input by $|x_{t,i}|<2^{18}$. \texttt{advance\_v5} checks that the committed key equals the registered \texttt{oracle\_pubkey} and that the committed cycle equals $\mathit{cycle\_count}+1$. The result is a proof that the registered oracle signed the $x_t$ driving this transition and that an earlier attestation was not replayed; it is not a proof that $x_t$ is truthful or benign. The first oracle-gated transition is \textit{charlie}'s cycle~1 advance (transaction \href{https://explorer.solana.com/tx/GjFXGQvW6UissKKgrQcvaQdF15M9LkJCmgBZZwrXzY74pDAYSf3yshSQaWq9DkCT9qbgjbBYaRQYSz3UKhe7a5x?cluster=devnet}{\texttt{GjFXGQvW\dots e7a5x}}).

This is a single designated trust point, not a decentralised quorum (\cref{sec:disc-limits} details the trust boundary; decentralising it is future work, \cref{sec:disc-future-directions}).

\subsection{24-cycle continuous run}\label{sec:phase4-m2}
A 24-cycle continuous run (a disposable agent, cycles~1--24, $24$ \texttt{advance\_v5} transactions, zero failures; commit \texttt{25d8324}) extends the single-cycle check to sustained enforcement: every cycle's \texttt{advance\_v5} satisfied the aliveness predicate under both the single designated oracle's attested environment commitment and the same protocol-imposed metabolic debit from the agent's program-owned economic PDA as in the 168-cycle long-run below ($\Fecon$ debited $12{,}000{,}000 \to 11{,}760{,}000$~lamports over the 24 cycles); full detail in \cref{app:m2-detail}.

\subsection{168-cycle long-run}\label{sec:phase4-m3}
\begin{figure}[tb]
\centering
\begin{tikzpicture}
  \begin{axis}[
    width=0.95\linewidth,height=4.4cm,
    xlabel={cycle index},
    ylabel={$\Fecon$ (million lamports)},
    xmin=0,xmax=170,
    ymin=9.8,ymax=12.2,
    grid=both,
    grid style={gray!20},
    tick label style={font=\footnotesize},
    label style={font=\footnotesize},
    legend style={font=\scriptsize,draw=none,fill=none,at={(0.98,0.97)},anchor=north east}
  ]
    \addplot+[only marks,mark=*,mark size=0.9pt,blue!70!black,mark options={fill=blue!50}] table[x=cycle,y=f_econ_mlamports] {figures/charlie_wallet_drain.dat};
    \addlegendentry{per-cycle $\Fecon$ balance}
    \addplot[red,dashed,thick,domain=0:170] {10.0};
    \addlegendentry{aliveness floor}
  \end{axis}
\end{tikzpicture}
\caption{Wallet-balance trajectory of the 168-cycle economic-metabolism long-run (\textit{charlie}). The program-owned $\Fecon$ PDA is debited by exactly $10{,}000$~lamports at every accepted advance, from $12{,}000{,}000$~lamports at the start of the run to $10{,}320{,}000$~lamports at cycle~168 (linear fit slope $-10{,}000$~lamports/cycle, $R^2 = 1.0$; \cref{app:m3-detail}). The dashed line is the $10{,}000{,}000$-lamport aliveness floor at or below which \texttt{advance\_v5} rejects with \texttt{InsufficientEcon} (\cref{sec:phase4-closure}); the run ends $3.2\%$ above it (data: \texttt{figures/charlie\_wallet\_drain.dat}, extracted from the committed run log).}
\label{fig:wallet-drain}
\end{figure}
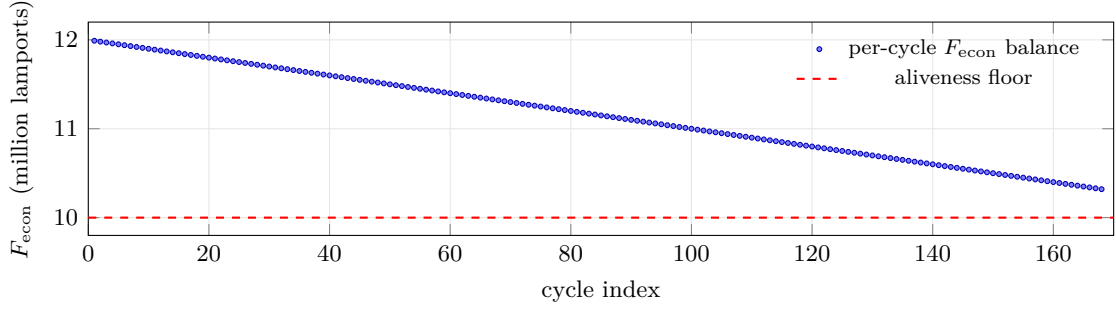

The 168-cycle long-run is the project's longest sequential \emph{on-chain} run by operation count: $168$ of-record advances by the \textit{charlie} agent (cycles~1--168), matching the horizon of the individuation runs (\cref{tab:runs}). Across these $168$ cycles the protocol debited an explicit metabolic cost from the agent's identity-derived economic PDA, draining it linearly and visibly (\cref{fig:wallet-drain}) --- a concrete, program-enforced on-chain \emph{economic metabolism} at the mechanism level (\cref{sec:phase4-closure}). The run completed with zero on-chain rejections (the single designated oracle needs no per-cycle attestation transactions). The committed run record preserves the cycle sequence, wallet-balance trajectory, and transactions but not per-cycle timestamps, so this result is a 168-transition feasibility demonstration rather than a claim of 168 hours of unattended wall-clock endurance. \Cref{app:m3-detail} reports the wallet trajectory, dedicated proof-time distribution, and resume-robustness event; the per-cycle on-chain verification cost is constant in the run length.

\subsection{Environment-oracle rejection paths}\label{sec:phase4-threats}
The rejection paths of \cref{sec:threat-p4m1-ext}, together with the core agent-signature rejections (\cref{sec:threat-tests}), carry into the economic extension. Both oracle- and agent-authentication error codes are confirmed against the live program enum and additionally exercised \emph{on devnet} as landed reverts against the current economic program \texttt{7Rvamkqp} (commit \texttt{5ac1624}), using a disposable agent (\textit{mallory}, keyed by an independent secret seed and non-interfering with \textit{charlie}): a valid proof self-signed by a non-agent key lands \texttt{Ed25519Failed} (\texttt{6005}, tx \href{https://explorer.solana.com/tx/3kZHnAWWn84zDmB7z9DQ3334kjhcnmzDB4XoHDzjQ52G8Qzz22eDug8iPAaDNUQ6n9C2BFxFBZvZwmrmh8ifH9MR?cluster=devnet}{\texttt{3kZHnAWW\dots fH9MR}}) --- the direct ``only the holder of $\sk$ advances'' rejection (P4, \cref{sec:properties}) --- while a crafted proof committing a non-registered oracle lands \texttt{OracleSigFailed} (\texttt{6016}, tx \href{https://explorer.solana.com/tx/65DJmBTtoKaVvaY8c92Xg2vVcBghU1mp88HVfKtJ7PxwiGqnUHSvYryPWUBvYCe9Uri4Ze3dqAMJobFvEPmQ9GfZ?cluster=devnet}{\texttt{65DJmBTt\dots Q9GfZ}}). The same valid proof, correctly self-signed, advances the agent (tx \href{https://explorer.solana.com/tx/3ihtuDXcJ1pevboVwQbF7bzu2cUN681aB6e2F7dTHTQji9sVoDyVjGPVEq9g7H4BqZ9PbFybNygkFTywJ355jT5h?cluster=devnet}{\texttt{3ihtuDXc\dots 5jT5h}}), so neither rejection is vacuous. Each revert lands with \texttt{skip\_preflight} --- a real slot and a burned fee, not a client-side pre-check --- and the program binary is unchanged (advance vkey \texttt{0x00df52a4}). \texttt{EnvOutOfRange} is non-landable: the in-circuit $|x_{t,i}| < 2^{18}$ assertion forecloses it, so no valid proof can carry an out-of-range environment (\cref{sec:phase4-m1}). The remaining oracle-path code, \texttt{OracleCycleMismatch} (\cref{sec:threat-p4m1-ext}), is not separately exercised as a landed revert.

\subsection{What the economic-metabolism extension enforces, and what remains}\label{sec:phase4-closure}
Together, the four components of \crefrange{sec:phase4-m0}{sec:phase4-m3} add a consumption-side $\Fecon(K)$ constraint on Solana devnet: a per-transaction advance-admissibility predicate, environment authentication that binds oracle-signed $x_t$ to the transition, and sustained enforcement across 24- and 168-cycle runs under the protocol-imposed debit.
The patron-fee-payer convention is no longer a protocol requirement, but external resource provision remains. The agent's advance admissibility is program-enforced and verifiable on chain at every transition. Devnet SOL is faucet-issued and valueless (\cref{sec:bg-solana}), so the mechanism demonstrates involuntary expenditure rather than economic self-maintenance: the program debits the metabolic cost and an underfunded account halts. In the of-record economic runs, the agent identity account is the Solana fee payer, funded with SOL by the operator; the client also supports an optional third-party payer.

The 168-cycle wallet trajectory demonstrates the debit side. An executed on-chain rejection against the current economic program demonstrates the halt side: draining a disposable agent's wallet below the survival floor lands a \texttt{Custom 0x1779} revert on devnet (tx \href{https://explorer.solana.com/tx/35j9ysBkcNZWQ4Gz2uZGCQdcrUhtFqdRipMcZWb7ofaFF1oUvHgv9osPeAwWWBSBKxjCM5pqtNcYe4FEGRdW5GLq?cluster=devnet}{\texttt{35j9ysBk\dots dW5GLq}}). The transaction lands and its fee is burned, so the \texttt{InsufficientEcon} gate fires in execution rather than merely at preflight. Mainnet scarcity, income, fee-payer independence, and decentralisation of the environment oracle remain open (\cref{sec:disc-limits,sec:disc-future-directions}).

\section{Discussion}\label{sec:discussion}

\subsection{Identity locus and trust root}\label{sec:disc-internal-shift}

\Cref{tab:identity-locus-comparison} classifies adjacent systems by the \emph{locus} of identity and the \emph{trust root} an attacker must compromise; deriving $W = \HKDF(\sk, \mathit{tag})$ inside Groth16 moves the locus into $\sk$ (\cref{sec:properties}); the economic-metabolism extension additionally enforces the metabolism precondition $\Fecon(K) > 0$ as an on-chain runtime check (\cref{sec:phase4}). The construction offers a cryptographic-provenance substrate \emph{relevant to} the metabolism, reproduction, and mutation questions of Hu and Fangting~\cite{hu2024speculating} (their RQ5), though it closes none: reproduction and open-ended mutation remain open (\cref{sec:phase4-closure,sec:disc-future-directions}), and the metabolism is consumption-only.

\begin{table}[htp]
\centering\footnotesize
\setlength{\tabcolsep}{3pt}
\renewcommand{\arraystretch}{1.15}
\begin{tabularx}{\linewidth}{@{}>{\raggedright\arraybackslash\hsize=.7\hsize}X >{\raggedright\arraybackslash\hsize=.85\hsize}X >{\raggedright\arraybackslash\hsize=.7\hsize}X >{\raggedright\arraybackslash\hsize=1.75\hsize}X@{}}
\toprule
\textbf{Identity locus} & \textbf{Trust root} & \textbf{Example} & \textbf{Attack surface left open} \\
\midrule
On-chain genome (NFT-rendered phenotype) & Smart contract / operator & Masumori et~al.~\cite{masumori2024life} & NFT swap; contract-upgrade authority; phenotype--key disconnect \\
TEE memory + JSON genome & Hardware vendor (Intel SGX, Phala) & Spore.fun~\cite{hu2025sporewild} & TEE compromise (Foreshadow-class); memory poisoning (T-MT) \\
Code binary hash & Protocol-verified binary commitment & BAID~\cite{lin2025baid} & weight and configuration semantics outside the committed binary; key not bound to weights \\
LoRA update blob & Operator (blob hash anchor) & zkLoRA~\cite{liao2026zklora} & T-DR on base-model blob (off-chain hash anchor); key not bound to weights \\
Profile / config document & Operator (document hash) & DIAP-style chains~\cite{liu2025diap} & T-DR (document replacement) \\
\emph{None (inference-only)} & Computation-proof assumptions and implementation & Modulus Labs~\cite{modulus2024}, ORA~\cite{ora2024}, Giza~\cite{giza2024}, Gensyn~\cite{gensyn2024}, ezkl~\cite{ezkl2024} & no identity claim; orthogonal capability \\
\midrule
\textbf{In-key derivation} $W{=}\HKDF(\sk)$ & \textbf{Cryptographic assumptions + pinned/frozen implementation} & \textbf{this paper} & host-side key custody (T5/T12, disclosed), software/runtime correctness, and single-oracle liveness (\cref{sec:disc-limits}); guest substitution pinned by $\mathit{vkeyHash}$, weight substitution rejected as T6; input poisoning remains \\
\bottomrule
\end{tabularx}
\caption{The locus of identity and its trust root across adjacent on-chain agent and zkML protocols. Rows~1--5 locate identity in an on-chain or externally attested object; BAID's protocol-verified binary commitment protects the code axis and is complementary to this paper's weight-axis commitment. Row~6 (zkML inference) has no identity claim and is included as an orthogonal capability. The construction (last row) internalises identity in $\sk$ via a deterministic key-to-weights derivation enforced by the pinned proof and runtime stack under the assumptions of \cref{sec:threat-scope}.}\label{tab:identity-locus-comparison}
\end{table}

\paragraph{zkML as a complementary capability.}
The zero-knowledge proving stacks of \cref{tab:identity-locus-comparison} (row~6: Modulus Labs, ORA, Giza, Gensyn, ezkl) provide \emph{verifiable inference} --- a proof that a fixed model produced a specific output --- with no notion of agent identity, as does adjacent ML-provenance work, both zkML (zkLoRA~\cite{liao2026zklora}) and non-ZK (the hash-chain AuditableLLM~\cite{li2025auditablellm}, the replay-based Proof-of-Learning~\cite{jia2021pol}). The same holds one level up, for proofs over an agent's whole execution: concurrent work proves an LLM agent's inference-and-tool-call pipeline in zero knowledge via batched transcript proofs~\cite{wang2026zkagent}, certifying that the recorded execution happened as claimed without binding the model to any agent identity.\footnote{That work is titled ``zkAgent''; we use $\agent$ throughout as notation for the agent triple of \cref{sec:notation}, a coincidence of naming rather than a shared construction (cf.\ the zkLoRA naming note in \cref{sec:intro}).} These capabilities compose: zkML answers ``did this model produce this output?''; the construction adds ``and is this model a function of the agent's own key?'' through the $W = \HKDF(\sk)$ constraint in the same circuit. The resulting runs provide continuous \emph{identity-bound} zkML execution on a public chain, with the evaluation centred on identity integrity rather than inference accuracy.

\paragraph{Operational consequences.}
Two consequences follow from the internal shift.
(i)~\emph{Composability}: the identity primitive is internalised in $\sk$. The same $\sk$ may simultaneously wrap multiple external identity tokens --- an SBT (soulbound token: a non-transferable on-chain credential), a DID controller key (decentralised identifier), an ENS reverse record (Ethereum Name Service) --- each derived via HKDF with a distinct domain tag. Revoking or losing any wrapper does not compromise the agent: the load-bearing identity is the key-to-weights binding of \cref{sec:properties}, not the wrapper.
(ii)~\emph{Lineage as a cryptographic primitive}: the future sleep-replication extension's $\sk' = \HKDF(\sk, \mathit{tag}_{\text{child}})$ derivation would make the parent--child relationship a verifiable on-chain fact rather than a database entry, opening royalty or governance flows that are honest by construction, keyed on lineage. Selection across such lineages would still require an external fitness-and-culling signal --- e.g., Masumori et al.'s~\cite{masumori2024life} human-purchase signal, or an on-chain market selecting which keyed lineages persist --- which the construction does not supply; it provides the verifiable lineage relation on which such a selection process could operate.

\paragraph{Composition with TEE protection.}
The TEE-based protection in Spore.fun composes with the cryptographic-binding primitive reported here: an enclave-confined $\sk$ inside the construction yields both host-attested \emph{and} circuit-attested identity, as sketched in \cref{sec:disc-limits} (``Host security boundary'').

\subsection{From cryptographic individuality toward a full Barandiaran agent}\label{sec:disc-barandiaran-triad}

The Barandiaran--Di~Paolo--Rohde framework~\cite{barandiaran2009define} (introduced in \cref{sec:related}) sets three conditions for agency --- \emph{individuality}, \emph{interactional asymmetry}, and \emph{normativity}.

Suzuki~2026~\cite{suzuki2026externality} builds on this framework, in which individuality is not one criterion among equals: Barandiaran et al.~\cite{barandiaran2009define} rank it as the \emph{precondition} for the other two (``neither asymmetry nor normativity would make much sense in the lack of an individualized system'') and use \emph{individuality} and \emph{identity} interchangeably. That is the bridge this paper rests on. The Artificial-Externality framework proposes cryptographic identity as the structural anchor for individuality: a cryptographic key pair satisfies only the individuality requirement (``identity by declaration''). The framework treats \emph{interactional asymmetry} as comparatively easy to realise through on-chain transaction emission, and identifies intrinsic \emph{normativity} --- whether an economic-metabolic substrate (persistence paid in gas fees) can become an \emph{intrinsic} norm rather than one externally imposed by designers --- as ``the next frontier in the ontology of artificial life''.

The construction reported in this paper realises the individuality criterion as a cryptographic invariant in a deliberately limited structural-engineering sense --- identifiability and persistence under a key, not autopoietic self-production or a metabolic boundary --- and takes PoC-tier steps toward the other two, with each component mapping onto one Barandiaran condition.

\paragraph{Individuality.} The core protocol (\cref{sec:construction}, the Inexorable layer) realises this along two complementary axes. \emph{Temporal coupling} is the key-signed history no fork can replay once finalized --- the $\Stream(K)$ axis, realised through the on-chain commitment chain (\cref{sec:chain}) --- and instantiates the \emph{temporal} dimension along which Barandiaran et al.~\cite{barandiaran2009define} hold agency to be extended. \emph{Computational-substrate coupling} arises because the weights are a deterministic function of the rigid designator $\sk$ via $W = \HKDF(\sk, \mathit{tag})$ (\cref{sec:f2,sec:f1}). Under HKDF's pseudorandomness, distinct keys yield independently derived substrates except with negligible collision probability; whether their resulting trajectories are behaviourally distinct is an empirical question assessed in \cref{sec:eval-individuation}. This anchors the \emph{spatial} locus of the individual --- its substrate --- to the key, realising the other half of Barandiaran et al.'s spatio-temporal account cryptographically rather than through a metabolic boundary. The two axes are distinct --- one binds what the agent has \emph{done}, the other the substrate from which its computation is derived --- and complementary, each alone admitting an attack the other forecloses (\cref{sec:related}).

\paragraph{Interactional asymmetry.} The extensions of \cref{sec:poc-extensions} approach this: the active-query loop (\cref{sec:design-phase2}) casts the agent as the \emph{active source} that initiates and modulates a per-cycle query (a host-side, off-chain step; \cref{sec:disc-limits}), while the homeostatic driver (\cref{sec:design-phase3}) converts the environmental response into an internal scalar via the allostatic free energy $F_{\text{allostasis}}$. The fuller sense of the asymmetry --- modulating a \emph{non-cooperative} coupling rather than one a cooperative patron supplies --- awaits the phenomenal-coupling extension (\cref{sec:disc-limits}).

\paragraph{Normativity.} The economic-metabolism extension (\cref{sec:phase4}) approaches this: the aliveness predicate $\Fecon(K) > 0$ enforced at every state-changing instruction (\cref{sec:phase4-m0}) makes the \emph{economic precondition} for survival a circuit-and-runtime invariant. Intrinsic normativity would require the agent to \emph{produce} the organisation whose viability sets the norm, which an externally wired halt-condition does not supply. The operator-endowed metabolism is consumption-only: the agent does not \emph{earn} (\cref{sec:phase4-closure,sec:disc-future-directions}), so the construction remains an identity-bound state machine rather than a self-sustaining Barandiaran agent.

The construction therefore starts from cryptographically guaranteed individuality --- structural identifiability and persistence under a key --- and builds outward toward Barandiaran et al.'s integrated account. Whether separately realised components can be assembled into autopoietic agency remains open.

\subsection{Safety considerations}\label{sec:disc-safety}

\paragraph{What ``unstoppable'' means here.}
The current programs' upgrade authorities were irreversibly set to \texttt{none} on 2026-08-02 (T11; \cref{sec:threat-accepted}), so the on-chain rules governing their advances are no longer modifiable by the deployer, host operator, or designated environment oracle. Continued operation, however, depends on four orthogonal liveness conditions, controlled by different mechanisms and actors (in the of-record deployment the operator controls both (iii) and (iv)): \emph{(i)~chain liveness} (Solana itself continues to produce blocks, an out-of-scope but realistic mainnet assumption); \emph{(ii)~economic continuance} ($\Fecon(K) > 0$ at every advance, gated by the aliveness predicate of \cref{sec:phase4-m0}, so an agent whose wallet is depleted halts automatically); \emph{(iii)~prover availability} (SP1 proofs are produced off-chain by the host process, so an operator who stops the prover stops the agent, and the operator likewise funds the fee-payer account that pays each advance's base fee); and \emph{(iv)~environment attestation} (each cycle's environment vector must be signed by the registered designated oracle (\cref{sec:phase4-m1}), so a withheld oracle key halts advances --- centralising \emph{liveness} at a single operator-held key).

\paragraph{What cryptographic identity contributes to safety.}
The construction's central safety property is \emph{cryptographic provenance}: every accepted state transition is signed by $\sk$ and re-proven against the genesis-time $\WC$ binding, so the record is attributable to the registered key/PDA and its genesis-committed substrate. It does not identify a human operator or physical host, and a key-holding operator can act as the agent (T5/T12; \cref{sec:disc-limits}). The pinned guest binary excludes code substitution, and the commitment chain excludes direct hidden-state overwrite; neither prevents semantically harmful but valid inputs from shaping later state (T-MT; \cref{sec:threat-adjacent}). It therefore provides accountable provenance within the stated trust boundary.

\paragraph{What the construction does not provide.}
Three safety guarantees are explicitly out of scope. \emph{(i)~Behavioural bounds}: the binding ties $W$ to $\sk$, but says nothing about \emph{what} the resulting policy does --- cryptographic identity certifies provenance, not benignity. In the frozen-$W$ core an operator cannot even train a chosen policy (genesis $F_2$ rejects any $W \neq \HKDF(\sk)$, leaving only key-grinding over untrained policies); the learnable-weights extension (\cref{sec:disc-future-directions}) would reopen this, accepting any cryptographically valid advance regardless of the policy's content. \emph{(ii)~Capability ceiling}: the reference model is a small Elman recurrent network ($5$-dim input, $16$-dim hidden state, $5$-dim output) with frozen $W$, which limits representational capacity but does not by itself bound downstream harm; impact depends on the actuator and tool interfaces attached to its five outputs. Transformer-class scaling would shift the capacity regime and is out of scope. \emph{(iii)~Resource ceiling}: the aliveness predicate halts a depleted agent, but nothing prevents a third party from replenishing $\Fecon$ to extend a malicious agent's operational horizon.
\subsection{Limitations}\label{sec:disc-limits}
We consolidate the remaining limitations here; each is paired with the threat-tag (when applicable) that names the corresponding residual risk in \cref{sec:threats}.

\subsubsection{Environment trust boundary and coupling}

\paragraph{Environmental coupling (interactional asymmetry) is partially deferred.} The designated environment oracle (\cref{sec:phase4-m1}) closes the cryptographic-attestation half of the environmental-coupling step by binding $x_t$ to a registered oracle key. The environment vector $x_t$ in \cref{eq:elman} is still synthetic in the sense that no external sensor or policy adversary drives it; full \emph{phenomenal coupling} --- an agent whose $x_t$ is a verifiably acquired observation from a non-cooperative environment --- remains a further-extension research direction.

\paragraph{Patron-binding in the active-query extension is in-circuit only.} The patron signature over the environment (\cref{sec:design-phase2}) is verified inside the SP1 guest against a witness-supplied patron public key that is neither committed to the journal nor checked on chain; the signed cycle counter and previous-action hash likewise enter the signature message but not the journal. An operator holding the agent's own $\sk$ can therefore fabricate an environment under a self-chosen key (the patron pubkey is not pinned on chain) or replay a patron signature across cycles (cycle and previous action are not journalled), so the extension's ``patron-signed environment'' is forgeable by the agent operator rather than being an independent third-party attestation. Under the threat model of \cref{sec:threats} (the operator is not assumed honest), this is a residual of the PoC-tier extension; closing it requires committing the patron pubkey, cycle, and previous-action hash into the advance journal and asserting them on chain. The designated-oracle extension (\cref{sec:phase4-m1}) supersedes the witness-only patron key with an on-chain-registered oracle key whose signed $x_t$ is bound to the committed transition, closing the oracle-attribution and cycle-replay parts of this binding gap for \texttt{advance\_v5} (the previous-action-hash binding remains future work).

\paragraph{The environment oracle is a single designated key.} The environment authentication of \cref{sec:phase4-m1} binds $x_t$ to a registered \texttt{oracle\_pubkey} via an in-circuit \texttt{ed25519} signature check (the signing key committed at $\mathit{PV}[140..172]$) and the $x_t = \mathit{PV}[100..120]$ equality. It is, however, a \emph{single} trust point: in the of-record runs \texttt{oracle\_pubkey} is held by the operator and is required on chain to be distinct from the agent's own key (\texttt{OracleEqualsAgent}; key-distinctness, not independence --- the operator may hold both), so the environment is attested under an operator-trust assumption, not by an independent third party. The construction therefore authenticates \emph{that the registered oracle signed this cycle's committed environment}, not that the environment is true or that the signer is independent of the operator. Like the agent's own key, the oracle key is written once at initialisation with no in-protocol rotation path, so oracle-key compromise shares the no-rekey status of $\sk$ compromise (T5; ``No post-compromise rekey'' below).

Decentralising the oracle would replace the single key with a stake-weighted committee carrying real (Jito-restaked) collateral and slashing for equivocation. Slashing reaches only attributable faults: a committee that consistently signs a false $x_t$ leaves no on-chain ground truth against which to slash, so environmental truth remains an assumption under any committee. On devnet, where stake carries no opportunity cost, such a committee would be decentralisation theatre. Like the prover decentralisation of ``Single-host availability'' below, this is a mainnet-only frontier beyond this paper.

\subsubsection{Identity scope, key custody and recovery, and residual trust assumptions}

\paragraph{Adversarial validity of the clean runs.} All of-record agent runs (\cref{tab:runs}) use independent \texttt{solana-keygen} keypairs with secret seeds, so the self-authorisation guarantee of condition~(iv) (that a valid advance requires $\sk$) holds in the adversarial sense against any party that does not hold the secret key. With the \emph{wallet-drain} vector (template A-EA1) already foreclosed by construction (\cref{sec:threat-p4m1-ext}), the clean trajectories of \cref{sec:phase4-m2,sec:phase4-m3} show no accepted conflicting transition and are consistent with non-interference during the observed run window.

\paragraph{Single-deployment identity scope.} The identity established here is per-deployment: the same $\sk$ could register agents under other programs or on other chains (cf.\ the first-advance replay footnote of \cref{sec:f1}), each with its own independent $\Stream(K)$. Such duplicates are trivially linkable because the same seed yields the same $\pk$ and $\mathit{skCommit}$; what the protocol neither supplies nor precludes is a canonical relation among their distinct histories. The canonical identity of this paper is the PDA under the program IDs cited here; cross-chain or cross-program identity unification is out of scope. Whether such duplication should be \emph{embraced} (key-spawned identical lineages, a form of self-replication) or \emph{precluded} via non-transferable, soulbound identity~\cite{ohlhaver2022desoc,zoltu2022eip5114} is left open: the construction provides uniqueness of \emph{derivation} (one key, one $W$) but not non-transferability of \emph{custody}.

\paragraph{Host security boundary and the distribution paths.} The construction binds $W$ cryptographically to $\sk$, but $\sk$ itself sits in a host-process keystore. Compromise of the host process is equivalent to compromise of the agent (T12). Two distribution paths exist, neither implemented here: (i) a TEE-backed or hardware-signed $\sk$ that confines the keystore inside an attestable enclave (Phala/SGX-style, as in Spore.fun~\cite{hu2025sporewild} for a different identity primitive; exclusive enclave custody of an agent's own account and wallet keys --- locking out even its developers for the run window --- was demonstrated by the \emph{tee\_hee\_he} agent~\cite{nous2024teehee}); (ii) a multi-party-computation prover quorum in which $\sk$ is threshold-shared across $n$ provers and no single host ever reconstructs the key, with the SNARK proof produced collaboratively under the shared key. Path (i) is a single-host hardening; path (ii) is the natural beyond-single-host extension. Both are explicit future sleep-replication work and outside the core and economic-metabolism scope. In the terms of the Artificial-Externality framing, single-host custody means the construction realises cryptographic \emph{individuality} but not yet \emph{inexorability} (the resistance-to-intervention of the Inexorable layer): a single operator can still halt the agent by withholding advances, or seize it by copying $\sk$. Inexorability in the strong sense would additionally require decentralised custody (path~(ii)) together with the prover-quorum liveness above --- both future work.

\paragraph{No post-compromise rekey.} A leaked $\sk$ has no in-protocol recovery (T5): it confers full agent authority --- advancing, rotating $W$ via the mutation circuit (\cref{sec:design-phase3}), and depleting $\Fecon$ through continued advances (\cref{sec:phase4}) --- with no chain-side mechanism to migrate that authority to an independently authorised fresh keypair. The sleep-replication extension's $\sk \to \sk' = \HKDF(\sk, \mathit{tag}_{\text{child}})$ derivation can express parent--child lineage or a voluntary pre-compromise epoch rotation, but it is not recovery from compromise: anyone who steals $\sk$ can derive the same $\sk'$ and produce the same parent-key authorisation.

The pragmatic stopgap today is to abandon the identity entirely --- re-genesis under a fresh independent $\sk'$, severing the prior $\Stream(K)$ chain with no on-chain link to the old identity. Genuine post-compromise recovery would require fresh entropy plus an authorisation path unavailable to the leaked seed, such as a recovery-key commitment fixed at genesis or a separately governed threshold credential. A future ceremony could then migrate $\Stream(K)$ cycle-continuously while allowing an external verifier to confirm from on-chain state that the new key is the authorised successor of the old identity.

\paragraph{Classical-hardness dependence.} The trust root is classical: a discrete-logarithm break on edwards25519 (e.g.\ by a cryptographically relevant quantum computer) recovers the clamped signing scalar, and with it signing authority, as for any wallet key; the seed $\sk$ itself sits behind the one-way seed-to-scalar hash (cf.\ the clamping footnote of \cref{sec:f2}), so $W = \HKDF(\sk, \mathit{tag})$ and the proof-gated stream do not follow from that break alone. It is compromise of the seed itself (T5) that yields the agent's substrate and full authority over its stream and funds at once: the construction concentrates substrate derivation together with everything a conventional wallet already stakes on one seed. Post-quantum instantiation of the binding is orthogonal future work.

\paragraph{Single-host availability.} The long-horizon runs (the core \textit{alice} run over 2.36 days, \textit{castor}/\textit{pollux} on independent hosts, and the economic-metabolism 168-cycle long-run) stress-test liveness, not Byzantine-fault tolerance (\cref{sec:properties,sec:threat-accepted}). A true Byzantine-fault test needs two ingredients absent here: the multi-host prover quorum of ``Host security boundary'' above (path~(ii)), and a \emph{leader-election} protocol rotating which prover produces the next proof in a manipulation-resistant way --- an Algorand-style VRF (a lottery whose draw is publicly checkable), or a Tendermint-style BFT round-robin --- so no malicious prover can monopolise the role, backed by on-chain reward-and-slash incentives. Designing, implementing, and evaluating this stack extends stake-weighted decentralisation (the future environment-oracle committee) to the prover layer; it is out of present-paper scope.

\paragraph{Program-upgrade authority.} All current Solana programs of the construction were irreversibly frozen to \texttt{Authority: none} on 2026-08-02 (T11; \cref{sec:threat-accepted}). The advance path of \texttt{advance\_v5} depends on no privileged admin key: it is gated only by the in-circuit oracle \texttt{ed25519} verification (\cref{sec:phase4-m1}) and the agent's own per-cycle self-signature.

\subsubsection{Frozen substrate and behavioural scope}

\paragraph{Deterministic weight derivation forbids in-cycle gradient learning.} The deterministic key-derivation $W = \HKDF(\sk, \mathit{tag})$ deliberately trades standard gradient-based learning capacity for substrate--key invariance: $W$ is fixed at genesis and cannot be updated by SGD or any other continuous learning rule without breaking the binding that the on-chain re-check at line~\ref{f1:wc-assert} of \cref{alg:f1} enforces. The homeostatic extension (\cref{sec:design-phase3}) admits discrete, key-anchored inter-epoch rotation inside a SNARK that re-proves $\WC = \keccak(\HKDF(\sk, \cdot))$, but the current rotation is not a learning rule; a learnable construction that preserves the binding under per-cycle updates is sketched as a future direction (\cref{sec:disc-future-directions}, ``Learnable weights'').

\paragraph{Scope of the frozen-substrate result.} The frozen random initialisation isolates the identity primitive from task competence and learning ability. For the claim evaluated here, individuation follows from $W = \HKDF(\sk,\mathit{tag})$ and is re-checked at every transition. The deployed binding already lets a counterparty verify that activity under a key comes from the genesis-committed substrate rather than from a silent replacement (T6).

\paragraph{Behavioural safety and shutdown.} Two unimplemented hardenings could extend the provenance guarantee of \cref{sec:disc-safety}. \emph{Behavioural-attestation circuits} would extend the SNARK to attest, alongside weight identity, that the policy stays in a pre-specified safe region, though the threat model for such a circuit is open. A community-attested \emph{vote-to-pause} mechanism would let a quorum of a future oracle committee (above) suspend a flagged agent's advances --- a governance layer atop that decentralised committee rather than the single oracle deployed here. Both build on the substrate primitive reported here.

\subsection{Future directions}\label{sec:disc-future-directions}

\paragraph{Learnable weights with key-anchored update sequences.}
Relaxing the lifetime freeze on $W$ (\cref{sec:disc-limits}, ``Deterministic weight derivation'') is a natural successor question. The \emph{learnable-weights extension} admits a per-cycle update $W_{t+1} = \mathrm{update}(W_t, \mathit{signal}_t, \sk)$ enforced inside Groth16 such that the cycle-$t{+}1$ on-chain $\WC_{t+1}$ is consistent with applying $\mathrm{update}$ to the previous cycle's $\WC_t$, with $\mathit{signal}_t$ provenance attested by signature or environment commitment, and the cycle itself signed by $\sk$ as in the core construction. The adjacent prior work surveyed in \cref{sec:disc-internal-shift} (zkLoRA, AuditableLLM, Proof-of-Learning) covers each piece in isolation but does not combine key-anchored weight identity with attested incremental updates. The learnable-weights extension is exactly the cryptographic discipline under which Lamarckian writeback can be admitted without forfeiting per-agent identity. Its threat model must distinguish unauthorised weight updates from authorised but semantically poisoned $\mathit{signal}_t$. The formal threat statement, residual-risk catalogue for signal provenance, and a concrete proof-time evaluation at our model scale are deferred to a separate report.

\paragraph{Cryptographic identity for language-based agents.}
This paper anchors agent identity to a recurrent neural network's hidden-state vector and its weight derivation $W = \HKDF(\sk, \mathit{tag})$. A natural extension carries the same identity-continuity discipline \emph{up} to language-based agents, where the agent's state is not a numeric hidden vector but a sequence of natural-language prompts, persistent memory entries, and conversation turns. In this regime the underlying computation may be performed by an external foundation model whose parameters are not under the agent's control; what remains anchorable is the agent's \emph{memory} (the accumulating record of prompts and committed responses) and its \emph{control loop} (the rule determining what to attend to next given current memory). The language-based agent's identity is then the cryptographically signed stream of its prompt-and-memory commitments, even if the underlying foundation model is replaced cycle by cycle. The engineering primitives --- attested-memory append-only logs, key-signed prompt commitments, SNARK-attested memory-tampering detection (a strengthened T-MT analogue, \cref{sec:threat-adjacent}) --- are direct lifts of the substrate to discrete-symbolic state; the concrete construction is left to a separate report.

\paragraph{Cryptographically private internal states.}
The construction already keeps the hidden state $h_t$ private to the agent: only the commitment $c_t = \keccak(h_t)$ is published on chain, while $h_t$ itself remains with the prover. A natural strengthening would encrypt $h_t$ under $\sk$ (or a key derived from it) and store the ciphertext on chain, so that only whoever controls $\sk$ could later read the past internal states; under exclusive enclave custody, that reader would be the agent process rather than the operator. This realises a substrate-level analogue of a \emph{first-person} internal state, inaccessible to any external observer once $\sk$ is enclave-confined --- raising an AI-safety question: should agents have internal states that no auditor can inspect?
\paragraph{Deliberate economic agency.} The metabolism of \cref{sec:phase4} is consumption-only (\cref{sec:disc-barandiaran-triad}) and \emph{involuntary}: the protocol debits a fixed cost that the agent neither authorises nor offsets with earnings of its own. A more developed agent would gain two-sided economic agency --- \emph{earning} (resource acquisition, naturally accommodated by inbound transfers to the identity-derived PDA) and \emph{deliberate spending} (a key-gated withdraw/transact path through which the agent authorises its own expenditure with $\sk$). Deployed agent-payment standards (Google's Agent Payments Protocol, whose signed mandates authorise agent-initiated payments and whose sample integrations include the on-chain x402 payment method~\cite{google2025ap2}) supply the authorisation rails such an extension would compose with; the construction contributes the verifiable identity of the paying entity. The PDA design leaves both open as additive instructions over the same account, so metabolism (imposed, now) precedes economic agency (deliberate, future) as a developmental staging rather than a redesign.

The debit rate itself (\cref{sec:phase4-m0}) and the GPU proving cost (\cref{sec:eval-prove}) are devnet-mechanism parameters, not calibrated to any real mainnet fee schedule or prover-market rate; sizing them against actual compute-unit pricing and prover economics is future work alongside this economic-agency extension. The protocol-anchored terms nevertheless transfer directly: a mainnet advance would pay the per-signature base fee plus a market-priced priority fee over the same bounded per-advance compute (\cref{sec:eval-cu}), together with the prover-market price of the measured GPU proving time (\cref{sec:eval-prove}). The volatile quantities are network priority pricing and prover-market pricing, which is why we report the protocol-side bounds and defer price calibration rather than quote a point estimate that would be stale by publication.

\paragraph{These extensions compose, and the composition is the safety question.}
Each direction above is scoped and deferred on its own terms, but they are additive over the same key and PDA, and their conjunction is stronger than any one direction: earning plus key-gated spending removes the funder's control over liveness condition~(ii), enclave or MPC custody (\cref{sec:disc-limits}) removes the operator's control over~(iii), an oracle committee removes single-key control over~(iv), and $\sk' = \HKDF(\sk,\mathit{tag}_{\text{child}})$ lineage (\cref{sec:disc-internal-shift}) lets the resulting configuration reproduce. The endpoint of adopting all of them is a self-funding, self-replicating agent without the current single-operator halt levers; shared chain dependencies remain, and any governance halt must be designed explicitly. The disclosed liveness conditions of \cref{sec:disc-safety} are what currently prevent that endpoint, so each is load-bearing for safety and not merely an artefact of a prototype. We therefore treat the vote-to-pause mechanism of \cref{sec:disc-limits} as a prerequisite for the economic-agency and replication extensions jointly, rather than as an independent future item, and note that a composed deployment warrants its own safety analysis rather than inheriting the analysis of the construction reported here.

\section{Conclusion}\label{sec:conclusion}

This paper demonstrates a deployment-scoped engineering primitive for structural individuality on a public blockchain: the substrate and key form one verified relation at every state transition, and the verifying nodes reject a transition that severs them. The substrate cannot be substituted unnoticed, its finalized history remains bound to its key, and, in the economic-metabolism extension, each continued advance requires a protocol-enforced cost. On chain, these relations instantiate the agent triple $\agent = (K, \Stream(K), \Fecon(K))$ and the deliberately limited form of cryptographic individuality proposed by Suzuki's \emph{Artificial Externality} framework.

Decentralised custody, economic self-maintenance, and agent-level normativity remain open (\cref{sec:phase4-closure,sec:disc-limits}). The enforced binding provides a defined structural basis for investigating those further conditions of agency (\cref{sec:disc-barandiaran-triad}).

\section*{Reproducibility}

The full implementation (core construction and the economic-metabolism extension), design ADRs (architecture decision records), threat-model document, per-cycle orchestrator telemetry, and data files used to produce the figures in this paper are released as a public source archive at \url{https://github.com/ksk-S/internalising-identity-2026} (Apache License~2.0 for code; Creative Commons Attribution~4.0 International for paper text and figures). The artefact version corresponds to the tag \texttt{arxiv-v1}, the source-of-record snapshot for this arXiv submission. Any later revision is released under an independent tag (\texttt{arxiv-v2}, \texttt{arxiv-v3}, \dots), so that a reproducer pinning to \texttt{arxiv-v1} obtains the exact source for this version of the paper. The repository is organised by paper section: \texttt{paper/} (\texttt{main.tex} + figures + bibliography), \texttt{code/\{core,\allowbreak active-query-extension,\allowbreak homeostatic-extension,\allowbreak economic-metabolism-extension\}/} (mapping to \cref{sec:construction,sec:design-phase2,sec:design-phase3,sec:phase4} respectively), \texttt{docs/\{threat-model.md,\allowbreak design/,\allowbreak runbooks/\}}, and \texttt{artifacts/\{on-chain-proofs/,\allowbreak continuous-run-logs/\}} (including the 168-cycle \textit{charlie} continuous-run log cited in \cref{sec:phase4-m3}); see the repository \texttt{README} for the mapping from internal Rust package identifiers (\texttt{p1\_*}, \texttt{p2\_*}, \texttt{p3\_*}, \texttt{p4\_*}) to the construction-aligned nomenclature used throughout the paper. All deployed Solana devnet artefacts referenced throughout the paper --- program IDs, agent PDAs, and the final state commitment --- are collected in \cref{tab:deployed} (each browsable on Solana Explorer~\cite{solanaexplorerprogram}).

\begin{table}[htbp]
\centering\footnotesize
\setlength{\tabcolsep}{4pt}
\begin{tabular}{@{}llp{0.37\linewidth}@{}}
\toprule
\textbf{Artefact} & \textbf{Address / hash (truncated)} & \textbf{Note} \\
\midrule
\multicolumn{3}{@{}l}{\emph{Programs}} \\
\texttt{p1\_identity\_verifier}   & \texttt{7H4Dgrq2\dots bPare}    & core protocol (\cref{sec:construction}) \\
\texttt{p3\_motivation\_verifier} & \texttt{rHegavCD\dots 23oH1}    & PoC extensions (\cref{sec:poc-extensions}) \\
\texttt{p4\_economic\_verifier}   & \texttt{7Rvamkqp\dots pciht} & economic metabolism (\cref{sec:phase4}) \\
\midrule
\multicolumn{3}{@{}l}{\emph{Agents} (on-chain PDAs)} \\
\textit{alice} (core)             & \texttt{CtV6Pent\dots fMMHC}    & secret \texttt{keygen} seed \\
\textit{bob} (core)               & \texttt{GDejx8GR\dots Rp8Xb}    & secret \texttt{keygen} seed \\
homeostatic daemon                & \texttt{D2NbAiRF\dots Yej1R}          & secret seed (\cref{sec:eval-onchain}) \\
\textit{charlie} (economic, AgentV4)& \texttt{7CEQQj1s\dots c7zBs}    & final commit \texttt{0xa341ea88\dots 2bfe} \\
\midrule
\multicolumn{3}{@{}l}{\emph{Verifying-key hashes} (SP1 guest circuits)} \\
advance ($F_1$, active-query)     & \texttt{0x00ed9904\dots}          & long-horizon runs; cross-machine-reproduced (\cref{app:q1-detail}) \\
advance ($F_1$, deployed core)    & \texttt{0x004cab07\dots}          & the \texttt{7H4Dgrq2} program; 20-/166-cycle runs \\
advance ($F_1$, economic)         & \texttt{0x00df52a4\dots}          & the \texttt{7Rvamkqp} program; 168-cycle of-record run (\cref{sec:phase4-m3}) \\
mutation guest                    & \texttt{0x00424d29\dots cd708da} & weight rotation (\cref{sec:eval-phase3}) \\
\bottomrule
\end{tabular}
\caption{Deployed Solana devnet artefacts and verifying-key hashes referenced throughout the paper (truncated for readability; full values are listed in the public archive's README and, within devnet's ledger retention, recoverable on-chain via the transaction signatures cited in \cref{sec:phase4-m1,sec:phase4-threats,sec:phase4-closure,app:m2-detail} and \cref{tab:p3-onchain-tx}). All current deployments were irreversibly frozen to \texttt{Authority: none} on 2026-08-02 (T11, \cref{sec:threat-accepted}; publicly checkable via \texttt{solana program show}). The three \texttt{advance} verifying keys correspond to distinct circuits --- active-query, the deployed core, and the economic circuit (which internalises the oracle-key, environment-range, and cycle commitments, lengthening the proof journal from 140 to 180 bytes; \cref{sec:phase4-m1}). The cross-machine reproductions on record cover the active-query circuit and the mutation guest (\cref{app:q1-detail,sec:eval-phase3}).}
\label{tab:deployed}
\end{table} The 24-cycle continuous run is recorded at commit \texttt{25d8324} and the 168-cycle long-run at commit \texttt{c0a6326}. These and the other git commit hashes cited in this paper are development-history provenance labels dating each artefact; the public archive is a fresh-history snapshot published at tag \texttt{arxiv-v1}, so they do not resolve as git commits of the public repository.

\paragraph{Build and verification toolchain.}
The reference build uses SP1 v5.2.4 (\texttt{cargo prove build} with \texttt{RUSTUP\_TOOLCHAIN=1.94.0}, pinned in the guest's \texttt{rust-toolchain.toml}) for the guest, Anchor~0.30.1 with \texttt{anchor build -{}-no-idl} for the on-chain program, and SP1's standard Groth16 prover. End-to-end verification of the \emph{core construction} is reproducible via the commands of \cref{tab:repro-commands}, each run from its listed working directory under \texttt{code/core/}; the economic-extension measurements are reproduced from the committed per-cycle records described below rather than by that core-only command table.
\begin{table}[htbp]
\centering\footnotesize
\setlength{\tabcolsep}{4pt}
\begin{tabular}{@{}llp{0.5\linewidth}@{}}
\toprule
\textbf{Action} & \textbf{Directory} & \textbf{Command} \\
\midrule
guest build       & \texttt{guest/}, \texttt{genesis-guest/} & \texttt{cargo prove build} \\
program build     & \texttt{sp1-program/} & \texttt{anchor build -{}-no-idl} \\
adversarial bench & \texttt{scripts/} & \texttt{cargo run -{}-release -{}-bin test\_adversarial} \\
A4 collision      & \texttt{scripts/} & \texttt{cargo run -{}-release -{}-bin test\_a4\_collision} \\
proof generation  & \texttt{host/} & \texttt{AGENT\_KEY=\allowbreak\textit{KEY} SP1\_PROOF\_MODE=\allowbreak groth16 cargo run -{}-release -{}-bin p1\_sp1\_host} \\
single advance    & \texttt{scripts/} & \texttt{AGENT\_KEY=\allowbreak\textit{KEY} cargo run -{}-release -{}-bin p1\_advance} \\
metric extract    & \texttt{./} & \texttt{python3 analysis/\allowbreak extract\_run\_metrics.py -{}-log <log>} \\
\bottomrule
\end{tabular}
\caption{Commands reproducing the build-and-verification pipeline. The crates under \texttt{code/core/} build independently (there is no workspace manifest at \texttt{code/core/} itself), so each command runs from the working directory listed in its row; directories are relative to \texttt{code/core/}. The \emph{single advance} row consumes the proof pair (\texttt{proof\_raw.bin}, \texttt{public\_values.bin}) written by the \emph{proof generation} row; a fresh key is first registered through the genesis flow (\cref{sec:f2}).}
\label{tab:repro-commands}
\end{table}
\Cref{fig:proof-time}'s committed source \texttt{figures/}\allowbreak\texttt{proof\_time\_data.dat} pairs the timing rerun's per-proof seconds (\cref{sec:eval-prove}) with the continuous run's per-transaction compute units (\cref{sec:eval-cu}); both the compute units and the run's inter-cycle periods are backed by committed artefacts (the \texttt{.dat}'s per-transaction CU column and the run log's per-cycle timestamps) and, within devnet's ledger retention, are independently recomputable from the run's on-chain transaction history (\cref{tab:deployed}). The individuation metrics of \cref{sec:eval-individuation,sec:eval-phase2} (the \textit{alice}, \textit{bob}, and same-key-control hidden-state trajectories) are recomputed by the shipped analysis scripts \texttt{code/core/\allowbreak analysis/\allowbreak p1\_individuation.py}, \texttt{code/\{active-query-extension,\allowbreak homeostatic-extension\}/\allowbreak analysis/\allowbreak p2\_individuation.py}, and \texttt{code/\allowbreak homeostatic-extension/\allowbreak analysis/\allowbreak p3\_n2\_comparison.py}, which derive each trajectory deterministically from the corresponding agent keypair --- an input privately held for the of-record agents, so these scripts reproduce the of-record trajectories only under the author's seeds (the committed run logs carry transaction-level telemetry, not hidden states; the boundary is stated below); the many-key $M_4$ null distribution of \cref{sec:eval-individuation} is regenerated by \texttt{code/core/\allowbreak analysis/\allowbreak p1\_manykey\_null.py} from deterministically indexed keys, with its output committed as \texttt{figures/\allowbreak manykey\_null\_m4.json}.

Bit-for-bit reproducible from the public archive are the verifying-key hashes (deterministic given the guest source and its pinned toolchain) and the many-key $M_4$ \emph{null distribution} (deterministic from its SHA-256-indexed keys); the per-transaction CU costs are recomputable from the committed per-cycle records (\texttt{figures/proof\_time\_data.dat}); and the committed per-cycle \texttt{.dat} files reproduce the reported proof-time summary statistics, though absolute proof-generation seconds are hardware-dependent of-record measurements on the reference workstation and will vary by host. Not bit-for-bit reproducible are the secret-seed trajectory metrics of the of-record \textit{alice}/\textit{bob} and \textit{castor}/\textit{pollux} pairs, including their $M_2$/$M_4$ values and the short-run pair's percentile within the null, because those metrics derive from privately held seeds (\cref{sec:eval-individuation}). The \textit{charlie} CU, proof-time, and wallet-debit summaries are instead reproduced from the shipped per-cycle records as reported measurements; fresh proof-time seconds remain hardware-dependent as noted above. This trajectory-reproduction boundary is separate from the adversarial guarantee of condition~(iv) against a non-$\sk$-holder: that guarantee follows from $\ed$ EUF-CMA and the verifier composition (\cref{sec:properties,sec:disc-limits}), while fresh-key tests can reproduce the implemented rejection path without revealing the of-record seeds. A third party thus reproduces the \emph{method} and the population-level claim --- rerun \texttt{p1\_manykey\_null.py} and every one of the 500 indexed pairs clears both floors; rebuild the guests and the verifying keys match --- while the specific secret-seed trajectory numbers stand as attested measurements under seeds the author holds.

\section*{Acknowledgements}
The author thanks the Succinct Labs team for the SP1 zkVM and the \texttt{sp1-solana} on-chain Groth16 verifier, and the Solana Foundation for devnet infrastructure.

\section*{Funding}
This work was supported by the Japan Society for the Promotion of Science (JSPS) KAKENHI (grant number 24H01534).

\section*{Competing interests}
The author declares no competing interests. The author holds no equity, advisory role, or paid relationship with Succinct Labs, the Solana Foundation, Coral (Anchor maintainer), RISC Zero Inc., or any other organisation whose software or services are used or evaluated in this work.

\section*{Use of AI assistance}
The author acknowledges the use of AI language models --- Claude (Anthropic) and Codex (OpenAI) --- in four roles: (i)~English-language polishing and structural editing of this manuscript; (ii)~design and implementation assistance for the Solana on-chain programs (Anchor framework, Rust) and the SP1 zkVM guest circuits ($F_2$, $F_1$); (iii)~implementation of the Python analysis pipeline (individuation metrics, telemetry driver); and (iv)~consistency verification cross-checking numerical claims, citations, figures, and manuscript statements against the codebase and on-chain telemetry. All code and edits were reviewed by the author; the deployed mechanisms were verified on Solana devnet, and all scientific content, theoretical arguments, hypotheses, and conclusions are solely the author's responsibility.

\clearpage
\appendix

\section{Detailed comparison with adjacent systems}\label{app:related-work}
The prose comparison in \cref{sec:related} locates each system in its original research context. \Cref{tab:related-comparison} records the corresponding primitive-level comparison; \cref{tab:identity-locus-comparison} gives the complementary main-text comparison by identity locus and trust root.

\begin{table}[H]
\centering
\footnotesize
\setlength{\tabcolsep}{2.4pt}
\renewcommand{\arraystretch}{1.15}
\begin{tabularx}{\linewidth}{@{}l >{\raggedright\arraybackslash}X >{\raggedright\arraybackslash}X >{\raggedright\arraybackslash}X c l@{}}
\toprule
\textbf{System} &
\textbf{Primitive} &
\textbf{Binds what} &
\textbf{Mechanism} &
\textbf{Learns?} &
\textbf{Anchor} \\
\midrule
\textbf{zkALIFE (this work)} &
SNARK (SP1, Groth16) &
key $\to$ weights ($W{=}\HKDF(\sk)$) &
zkVM circuit + ed25519 sig &
no (frozen $W$)\textsuperscript{\dag} &
Solana \\
Spore.fun~\cite{hu2025sporewild} &
TEE (Phala GPU, SGX) &
JSON genome + memory state &
attestation + Eliza framework &
yes &
Solana \\
BAID~\cite{lin2025baid} &
SNARK (RISC Zero) &
program binary $C_P$ + biometric &
recursive receipt + face proof &
no &
Ethereum \\
zkLoRA~\cite{liao2026zklora} &
SNARK (Hyrax/BLS12-381) &
LoRA update correctness &
fwd/\allowbreak bwd/\allowbreak update circuit &
yes &
off-chain \\
AuditableLLM~\cite{li2025auditablellm} &
hash chain &
update history (no identity) &
tamper-evident hash log &
yes &
off-chain \\
DGM~\cite{zhang2025dgm} &
none (foundation-model mutation) &
agent scaffold (code, tools) &
empirical benchmark selection &
partial &
n/a \\
DIAP~\cite{liu2025diap} &
Noir ZKP &
agent identity $\leftrightarrow$ IPFS CID &
stateless ownership proof &
n/a &
IPFS \\
ZKROWNN~\cite{sheybani2023zkrownn} &
ZK ownership proof &
watermark-key $\leftrightarrow$ NN weights &
sub-second 3rd-party verify &
n/a &
off-chain \\
OML~\cite{cheng2024oml} &
fingerprint + econ &
foundation model loyalty &
AI-native fingerprint + audit &
no &
off-chain \\
DID / VC standards~\cite{w3c2022did} &
signatures + DID registry &
controller key $\leftrightarrow$ external subject identifier &
DID-document resolution + credential attestation &
n/a &
method-dependent \\
ERC-8004~\cite{erc8004} &
registry + pluggable validator &
registry entry $\leftrightarrow$ off-chain agent card &
card resolution + reputation/\allowbreak validation registries &
n/a &
Ethereum \\
\bottomrule
\end{tabularx}
\caption{Cryptographic, TEE, and audit primitives for AI-agent integrity. zkALIFE enforces a circuit-level relation $W{=}\HKDF(\sk)$ between agent identity and weights with on-chain anchoring at every state transition; the closest siblings bind code and a biometric (BAID), only update correctness (zkLoRA), or only audit history (AuditableLLM). See \cref{sec:related} for context on each row. The \emph{Mechanism} column names the verification, attestation, or audit mechanism on which each system relies. \textsuperscript{\dag}The core protocol does not learn; the homeostatic extension adds constrained key-anchored rotation (\cref{sec:design-phase3}).}
\label{tab:related-comparison}
\end{table}

\clearpage

\section{Extended empirical detail}\label{app:ext-eval}
This appendix collects the full empirical detail summarised in the body: the core individuation metrics (\cref{sec:eval-individuation}, question Q1), the evaluation of the two PoC extensions (\cref{sec:poc-extensions}, questions Q5--Q7), the 24-cycle and 168-cycle economic-metabolism continuous runs (\cref{sec:phase4-m2,sec:phase4-m3}), and the Tier~1 PoC end-to-end on-chain state demonstration (\cref{sec:eval-onchain}).

\subsection{Key-binding divergence (Q1): full detail}\label{app:q1-detail}
This subsection gives the full individuation evidence summarised in \cref{sec:eval-individuation}: the four short-run metrics, the 168-cycle long-horizon $N{=}2$ confirmation, and the many-key null.

Q1 tests the end-to-end derivation pipeline of \cref{sec:eval-individuation} (key to weights to dynamics). The full metric set establishes numerical non-degeneracy, while rejection test T6 provides the enforcement evidence. Conditional on a fixed key pair and input schedule, the dynamics are deterministic and have no trial-to-trial process-noise distribution. Across sampled key pairs, $M_4$ has the empirical key-induced distribution reported as the many-key null (the 500-pair reference distribution of \cref{sec:eval-individuation}, named after its committed artefact; \cref{tab:longrun-individuation}); we use this distribution descriptively rather than for $p$-values.

\paragraph{Status note.} The short-run metrics ($M_1$--$M_4$ below) draw on a 20-cycle dataset, and the long-horizon confirmation that follows them is the independent-host 168-cycle off-chain pairing; the full 166-cycle continuous run (\textit{alice} cycles 1--166, $n{=}166$) underlies the long-run and verification-cost numbers of \cref{sec:eval-long,sec:eval-cu}, while the proof-generation cost of \cref{sec:eval-prove} is measured in the dedicated timing rerun described there.

We track four metrics over $T = 20$ short-run cycles, during which both agents consume an identical deterministic environment schedule, $x_t[i] = (i{+}1)\cdot 0.1 + t \cdot 0.01$ in $\mathrm{Q}16.16$ ($i = 0,\dots,4$), so the shared input cannot itself be a source of divergence:

\begin{description}[leftmargin=1.5em,itemsep=2pt,topsep=2pt]
  \item[$M_1$ (same-key baseline).] Two hosts initialised with the \emph{same} $\sk$ produce $L_2$-divergence~$=0$ at every cycle, confirming determinism.
  \item[$M_2$ (per-dimension correlation, descriptive only).] Over the $16$-dim hidden state the per-dimension Pearson correlation between \textit{alice} and \textit{bob} is sign-mixed (median~$+0.20$). Its pass is specific to the of-record key pair, and the $\pm 0.5$ cut-offs are not significance tests over so short an autocorrelated series; the individuation verdict rests entirely on $M_4$ ($L_2$ divergence) below. The strongly negative median at the 168-cycle horizon (\cref{tab:longrun-individuation}) is a distinct extended-exposure effect.
  \item[$M_3$ (concurrent operability).] Two agents advance in alternating cycles on the same on-chain program with no contention or cross-corruption.
  \item[$M_4$ ($L_2$ divergence).] Mean inter-agent $L_2$ on hidden state is~$0.68$ ($>6.8\times$ the pre-specified mean-metric effect-size floor of~$0.1$), with last-five-cycle mean $0.80$ ($>16\times$ the tighter last-five-cycle floor of~$0.05$ against which the last-five metric is pre-specified). Both floors are pre-specified \emph{minimum effect sizes}, not estimates of stochastic noise: they fix in advance the divergence level below which we would have judged two trajectories operationally indistinguishable.
\end{description}

Together these answer Q1 plainly: two agents with different keys reach measurably different internal states ($M_4$ far above the floor), while a same-key control stays identical ($M_1 = 0$).

\paragraph{Long-run individuation (off-chain, 168-cycle N=2 confirmation).}
The 20-cycle results were confirmed over a 168-cycle horizon by an $N{=}2$ pair --- \textit{castor} and \textit{pollux}, proved on independent GPUs under the active-query extension circuit (\cref{sec:design-phase2}) in SP1 core-proof mode (off-chain prover runs, each under an independent secret \texttt{keygen} seed). \Cref{tab:longrun-individuation} reports the metrics, computed host-side from the agents' prover-held trajectories (the hidden state $h_t$ is private; the proof journal exposes only its commitment $c_t = \keccak(h_t)$, matching the host simulator bit-exactly at each cycle): the of-record pair's $M_4$ grows near-monotonically over the horizon, and every one of the 500 independently keyed pairs sampled for the many-key null clears both pre-specified floors (the of-record short-run pair's own position within that null is discussed in \cref{sec:eval-individuation}).

The on-chain individuation evidence is the 20-cycle core run above. The active-query verifying-key hash is consistent across all of-record proofs and an independent source rebuild, confirming determinism of the proof system; the deployed on-chain \emph{core} circuit that the 20- and 166-cycle runs verify against is a \emph{distinct} circuit from the active-query one (both program IDs and verifying-key hashes in \cref{tab:deployed}).

The longer horizon also drives $M_2$ for this pair to a strongly negative median (\cref{tab:longrun-individuation}) under the common deterministic base environment (with agent-specific sensorimotor feedback, \cref{sec:eval-phase2}), so the $M_4$ divergence here carries an environment-feedback component alongside the key-driven one --- an anti-phase coupling signature (the sign is pair-specific; other key pairs couple in phase) reported as a supplementary measurement, not a redefinition of $M_2$ for short runs. Throughout, $N{=}2$ is a protocol-demonstration sample rather than a statistical sample.

\begin{table}[htbp]
\centering\small
\setlength{\tabcolsep}{5pt}
\begin{tabular}{@{}lll@{}}
\toprule
 & \textbf{of-record \textit{castor}/\textit{pollux}} & \textbf{many-key null} \\
 & (168-cycle, off-chain) & (500 pairs, 20-cycle) \\
\midrule
$M_4$ mean             & ---                                & $[0.44, 1.75]$, median $0.93$ \\
$M_4$ last-five        & $3.476$ ($70\times$ floor)          & $[0.51, 2.16]$, median $1.10$ \\
$M_4$ trajectory       & $0.62 \to 3.50$ (cycle 1$\to$168)  & all $500 >$ both floors \\
$M_2$ median $\rho$    & $-0.913$ ($5$ pos.\ / $11$ neg.)    & --- \\
short-run of-record percentile & ---                                & $7.2$ / $7.0$ (lower tail) \\
\bottomrule
\end{tabular}
\caption{Long-horizon ($N{=}2$, off-chain) and many-key null individuation. $M_4$ is the inter-agent $L_2$ divergence, and $M_2$ is the median per-dimension Pearson $\rho$. Pre-specified floors: mean $0.1$, last-five $0.05$ (the minimum null margins are $4.4\times$ and $10\times$). The null draws $500$ disjoint key-pairs from $1{,}000$ deterministically indexed keys, simulated for the same $20$ cycles under the identical environment schedule (data: \texttt{figures/manykey\_null\_m4.json}). Blank cells are structural rather than missing data: the long-run pair reports its trajectory in place of a whole-horizon mean (a $168$-cycle mean would average over the growth transient and is not comparable to the null's $20$-cycle means); $M_2$ is computed for the of-record pair only (the null pipeline computes $M_4$ alone); and the percentile row is by definition the short-run of-record pair's position \emph{within} the null (\cref{sec:eval-individuation}).}
\label{tab:longrun-individuation}
\end{table}

\subsection{Environment-bound individuation under the active-query loop (Q5)}\label{sec:eval-phase2}

\paragraph{Environment publication.}
Patron-signed environment publication is verified across $7$~consecutive cycles for \textit{castor}; the responder public key is consistent across all cycles (\texttt{0xf25d9024\dots c48b7ed1}), and the \texttt{env}\textsubscript{q16} sequence shows monotone drift (cycle~1 $[7159, 13740, 20374, 26897, 33424] \to$ cycle~7 $[11126, 17683, 24264, 30811, 37351]$), a drift consistent with the environment-coupling pattern quantified at larger scale below.

\paragraph{Environment-bound individuation.}
We re-run the individuation protocol under the active-query-loop extension's environment binding over 24- and 50-cycle horizons. The $L_2$ divergence~$M_4$ remains strongly above the $0.05$ threshold (the active-query regime's pre-specified effect-size floor, distinct from the core run's $0.1$; final $0.696$ at $24$ cycles, $1.007$ at $50$, both $14$--$20\times$ threshold). The per-dimension correlation~$M_2$ exhibits a qualitatively different pattern from that of the core construction's static-environment regime: the median correlation grows positive with the horizon (e.g., $0.10$ at $24$ cycles and $0.60$ at $50$; the sign is pair-specific --- the 168-cycle \textit{castor}/\textit{pollux} pair of \cref{tab:longrun-individuation} instead couples anti-phase), reflecting the shared deterministic environment both agents observe --- a stage-0 control without action coupling yields the same positive correlation, so the positive $M_2$ is driven by the common environment rather than by active-query feedback. As in the core static-environment regime, $M_4$ is the primary divergence metric here; $M_2$ is again secondary, now as an attractor-coupling indicator rather than the sign-structure diagnostic. Accordingly, the active-query experiment demonstrates that the binding invariants \emph{survive} environment coupling (an integrity result); it does not add independent individuation evidence, since a shared deterministic environment \emph{increases} inter-agent correlation rather than the reverse, so the residual $M_4$ here mixes a key-driven and an environment-feedback component (\cref{app:q1-detail}).

\paragraph{Cross-implementation determinism.}
Implementation consistency is checked at three layers. First, the chain runner and orchestrator smoke produce identical \texttt{new\_commit} hashes for cycles~1--3. Second, the 24-cycle orchestrator run passes all 23 state-chain links and matches the Python host simulator's $y$ vectors bit-exactly in all 24 cycles. Third, the deployed on-chain program accepts the extension's proof and journal layout in end-to-end transactions. These checks establish agreement across the tested generator, reference, and verifier layers.

\paragraph{Proof characteristics.}
The deployed advance requires the Groth16-\emph{wrapped} proof for on-chain verification (the $4{,}881{,}608$-byte core proof compressed to $1{,}542$~bytes, varying between $1{,}540$ and $1{,}543$ bytes across cycles, a $3166\times$ reduction); with CUDA GPU proving the end-to-end wrapped per-cycle proof time is mean $42.06$~s (median $42.33$, range $[40.8, 43.9]$, $n{=}10$) --- the relevant per-transition cost, comparable to the core figure, since on the GPU the near-constant Groth16 wrap stage dominates per-cycle cost.

\subsection{Weight rotation and the homeostatic driver (Q6)}\label{sec:eval-phase3}

\paragraph{$W$ rotation: cryptographic determinism (Q6).}
The mutation guest's verifying key is recorded in \cref{tab:deployed}; one mutation proof (Groth16) takes ${\sim}41$~s with CUDA GPU proving on the reference workstation, with output proof $5{,}750{,}729$~bytes (core mode) or wrapped to ${\approx}1{,}542$~bytes (Groth16, ready for on-chain verification). One mutation guest execution takes $664{,}035$~SP1 cycles. By construction (\cref{sec:design-phase3}, four cryptographic guarantees), no mutation is accepted on chain without~$\sk$; the on-chain gating is exercised end-to-end in \cref{sec:eval-onchain}.
\paragraph{168-cycle full daemon (homeostatic-extension integration run).}
We run the integrated homeostatic-extension daemon for $168$~cycles with a mutation event every $24$~cycles ($7$~mutations total: cycles $24, 48, 72, 96, 120, 144, 168$). The run completes with no failed cycles, no overrun retries, and the full chain-integrity check passing for all $167$~chain links and all $168$~journal $y$~vectors bit-exact against the host simulator. With CUDA GPU proving, per-cycle advance proof time is mean $4.35$~s (median $4.36$, range $[3.92, 4.79]$~s) in SP1 core-proof mode (the Groth16-wrapped figure for the active-query circuit family is the ${\sim}42$~s of \cref{sec:eval-phase2}); per-mutation proof time is mean $41.43$~s (Groth16-wrapped, the ${\sim}41$~s entry of \cref{sec:eval-prove}). The seven mutation new-$W$ commitments are deterministically derived from $(\sk, n_\text{mut}, r_\text{replay})$ by the HKDF rotation rule (\cref{sec:design-phase3}); under a fresh key-holder-attested $r_\text{replay}$ per epoch (the cumulative leaves of cycles $1\ldots n_\text{mut}\cdot 24$ in this run), the new-$W$ trajectory is jointly determined by the agent's $\sk$, the mutation counter, and those replay-root values. The circuit does not independently verify that a submitted root summarises the preceding cycles.

\paragraph{Cross-machine deterministic build.}
The mutation guest is built and executed on the independent workstation ($2{\times}$ RTX 3080 Ti, distinct hardware; a clean Ubuntu 22.04 host with the pinned toolchain) using a single-shot validation script from the public archive. On a fixed input vector, the independent build's ELF size ($94{,}860$~bytes), $\mathit{vkeyHash}$, execution-mode cycle count ($664{,}035$), and journal commitments ($\mathit{old}\WC, \mathit{new}\WC, \mathit{sk}\mathit{Commit}$) all match the reference workstation's values. This establishes a bit-for-bit reproducible build and a concrete provenance mechanism relevant to the reproduction-control concern raised for evolvable AI~\cite{muller2026evolvableAI}.

\subsection{On-chain end-to-end integration of the PoC construction (Q7)}\label{sec:eval-onchain}

We close the Tier~1 PoC with an end-to-end on-chain demonstration on Solana devnet. The Anchor program \texttt{p3\_motivation\_verifier} is deployed at program ID \texttt{rHegavCD\dots 23oH1} (\cref{tab:deployed}; $337{,}256$~bytes of BPF bytecode). The homeostatic-extension daemon agent here, keyed by an independent \texttt{keygen} seed and distinct from the core-protocol agents of \cref{sec:eval-setup}, is registered at on-chain PDA \texttt{D2NbAiRF\dots Yej1R} via \texttt{initialize\_agent} using a Groth16 genesis proof ($260$~bytes raw). The demo then executes the full instruction sequence
\[
\texttt{advance\_v2} \;\to\; \texttt{sleep\_trigger} \;\to\; \texttt{mutate\_w}
\]
on devnet, with each instruction passing the program's full required precondition set: SP1 verification, state-chain check, weight-commit consistency, $\ed$ self-authentication, and (for \texttt{mutate\_w}) the four-fold provenance check of \cref{sec:design-phase3}. Post-mutation, the on-chain agent state matches the host daemon's expected post-sequence snapshot after the cycle-1 advance, sleep trigger, and weight mutation, bit-exactly across all eight recorded fields of the \texttt{AgentV3} record (the extension's on-chain agent-state layout); the full field-by-field values are given in \cref{tab:p3-poststate} (\cref{app:p3-poststate}). The demonstration is witnessed by the Solana devnet transaction signatures of \cref{tab:p3-onchain-tx} (within devnet's ledger retention).

\begin{table}[htp]
\centering\small
\begin{tabularx}{\linewidth}{lX}
\toprule
\textbf{Instruction} & \textbf{Devnet tx signature (cluster=devnet)} \\
\midrule
\texttt{initialize\_agent} & \href{https://explorer.solana.com/tx/5emBF6bCKKpXVQCRioxchc3rFSE5WqUuExu1tc14EPTzRdbcPcMKc1iYmkqRXXyrq9R461GYMmvxXgtntZeRiMaq?cluster=devnet}{\texttt{5emBF6bC\dots tZeRiMaq}} \\
\texttt{advance\_v2} & \href{https://explorer.solana.com/tx/613CiFA54yEuQjwvYBAKBD6Yit85Au2n6tyPNgutw7ERkL1ssoNgxnJQsTTmJ6q4F34M3ua26Nbn5Je5xpqX7FzU?cluster=devnet}{\texttt{613CiFA5\dots xpqX7FzU}} \\
\texttt{sleep\_trigger} & \href{https://explorer.solana.com/tx/8Lo3rxCy8DESYic4AGzbXVzqFoWNnV8wQGHYLphd1S1BNj98WknS9ceLDpoPUTAajZjUvPZ1foxXBE6WvnC5gnd?cluster=devnet}{\texttt{8Lo3rxCy\dots WvnC5gnd}} \\
\texttt{mutate\_w} & \href{https://explorer.solana.com/tx/DUYEcDMtepiezKjrKzhKc5CCWvPZmQUC54pbSmnHYWbZ9zEVw7aq5E1DHu2PmiEVsv9QTGpt95nSCtJMgntud8n?cluster=devnet}{\texttt{DUYEcDMt\dots Mgntud8n}} \\
\bottomrule
\end{tabularx}
\caption{Solana devnet transaction signatures for the Tier~1 PoC end-to-end demonstration on 2026-06-27 (on the current \texttt{p3\_motivation\_verifier} deployment). The displayed signatures are truncated for typesetting; clicking each opens the full signature on Solana Explorer (\url{https://explorer.solana.com/tx/<sig>?cluster=devnet}).}
\label{tab:p3-onchain-tx}
\end{table}

\Cref{tab:p3-onchain-tx} witnesses, on a public chain, the entire Tier~1 PoC programme: an agent with an HRRL-inspired homeostatic driver advances state under environment binding, enters sleep with a key-holder-attested replay-root commitment, and rotates its weights under key-holder authorisation and the committed rotation inputs in a single transaction sequence whose every enforced constraint is verifiable, within devnet's ledger retention, by anyone with read access to the Solana devnet ledger. The replay root is not a circuit-verified summary of the agent's prior experience. We answer Q7 in the affirmative.

\subsection{24-cycle continuous run: full detail}\label{app:m2-detail}
The 24-cycle run extends the single-cycle aliveness-and-oracle check to sustained enforcement: a disposable agent (\texttt{4TmciKBz\dots F5zrL}) on the GPU prover landed all $24$ \texttt{advance\_v5} transactions with zero failures (commit \texttt{25d8324}), each cycle satisfying the aliveness predicate under the single designated oracle's attested environment commitment while the program debited the same $10{,}000$-lamport metabolic cost as in the 168-cycle long-run --- an involuntary, program-enforced debit, not an operator-paid fee --- draining the program-owned economic PDA (\cref{sec:phase4-m0}) from $12{,}000{,}000$ to $11{,}760{,}000$~lamports ($240{,}000$ in total). The first (cycle~1, \href{https://explorer.solana.com/tx/4cPT5YNMYPRwv8c7FZSThQYMSx48QhqUFMhgppQ1dYZmnDDtCFRKNu38521CgVe6gMC1kAEnhccY6QoBVa42FUNT?cluster=devnet}{\texttt{4cPT5YNM\dots 42FUNT}}) and last (cycle~24, \href{https://explorer.solana.com/tx/3C8WUxBvG8yjdfQSVFJUkm8YBPuveq8a8NC7UYMDS2jsQYiZeQKQMjFAm7K8siUCzAVZD7d9ihNvMfYUMvn9FhUw?cluster=devnet}{\texttt{3C8WUxBv\dots n9FhUw}}) advance signatures are independently verifiable on Solana Explorer (within devnet's ledger retention). Here the agent is itself the Solana fee payer --- each \texttt{advance\_v5} is signed and paid by the agent --- a role distinct from the program-debited metabolic cost.

\subsection{168-cycle long-run: full detail}\label{app:m3-detail}
This subsection gives the committed telemetry of the 168-cycle long-run summarised in \cref{sec:phase4-m3}. The reference run uses \textit{charlie}, an agent keyed by an independent \texttt{solana-keygen} keypair with a privately held seed, on the GPU prover (CUDA, RTX~3090; run commit \texttt{c0a6326}); re-proving its $168$ committed per-cycle witnesses in a dedicated prove-only pass on the same circuit sums to ${\approx}147$~min of GPU proving. It completed with zero on-chain rejections across $168$ transactions --- one \texttt{advance\_v5} per cycle. A ${\approx}1.5$-hour mid-run Solana-devnet RPC degradation was absorbed by the idempotent/retry orchestrator (\cref{sec:eval-resume}) with no cycle lost. The committed of-record run log has no per-cycle timestamp field, so it does not support a claim about total elapsed duration or hourly cadence; the 168-cycle result is scoped to sequential transition count.

The wallet-balance trajectory (\cref{fig:wallet-drain}) is monotonically decreasing under the aliveness predicate: the program-owned economic PDA (\cref{sec:phase4-m0}) holds $\Fecon = 12{,}000{,}000$~lamports at the start and is debited to $10{,}320{,}000$~lamports by cycle~168, a total of $1{,}680{,}000$~lamports over the $168$ cycles. A linear fit gives slope $-10{,}000$~lamports per cycle with coefficient of determination $R^2 = 1.0$: every cycle debits exactly the same $10{,}000$-lamport metabolic cost. The endpoint sits $3.2\%$ above the $10{,}000{,}000$-lamport aliveness floor --- an endpoint fixed in advance by the chosen endowment and per-cycle debit, not an emergent margin --- so the aliveness predicate held for the entire run.\footnote{The $10{,}000$-lamport per-cycle cost is a \emph{protocol-imposed metabolic debit}, not a transaction fee: \texttt{advance\_v5} moves it from the program-owned $\Fecon$ PDA to the incinerator by a direct lamport decrement inside the instruction (\cref{sec:phase4-m0}), independently of whichever account pays the Solana network fee --- the involuntary metabolism of \cref{sec:disc-limits}.} The final agent state commitment \texttt{0xa341ea88\dots 2bfe} and \textit{charlie}'s on-chain PDA are recorded in \cref{tab:deployed} (the commitment occupies byte offset $[106, 138)$ of that PDA).

Per-cycle proof generation on the GPU prover, in that dedicated prove-only pass, averaged $52.66$~s (median $52.22$~s, p99 $58.84$~s, maximum $58.87$~s over the $168$ proofs); the cross-circuit proof-time comparison appears in the proof-cost evaluation (\cref{sec:eval-prove}).

\paragraph{Operator-side resume robustness.}\label{sec:eval-resume}
The RPC degradation noted above did not cost a single cycle. The orchestrator issues each \texttt{advance\_v5} idempotently --- it reads the on-chain \texttt{cycle\_count} before submitting and retries transient client-side RPC failures rather than counting them as cycle failures --- so a degraded RPC endpoint delays, but does not break, the advance chain. The of-record dataset is correspondingly contiguous over cycles~$1$ through~$168$, with the on-chain cycle counter advancing by exactly one at each recorded step; the construction's on-chain rules and circuit guarantees are unchanged by the operator-side disturbance.

\subsection{End-to-end on-chain demonstration: post-mutation agent state}\label{app:p3-poststate}
\Cref{tab:p3-poststate} records the full on-chain \texttt{AgentV3} state after the \texttt{mutate\_w} instruction of the Tier~1 PoC end-to-end demonstration (\cref{sec:eval-onchain}); every field matches the host daemon's expected post-sequence snapshot after the cycle-1 advance, sleep trigger, and weight mutation, bit-exactly.

\begin{table}[!htp]
\centering\small
\begin{tabularx}{\linewidth}{l l X}
\toprule
\textbf{Field} & \textbf{Value} & \textbf{Note} \\
\midrule
\texttt{cycle\_count} & $1$ & \\
\texttt{state\_commit} & \texttt{0x2a8081f2\dots 6ed9849} & matches daemon cycle~1 \\
\texttt{last\_action} & $[-2578, -2389, 8091, -969, 9986]$ & daemon cycle-1 $y_{q16}$ \\
\texttt{last\_history\_root} & \texttt{0xfcbf58a0\dots cf8e917} & \texttt{sleep\_trigger} commit \\
\texttt{driver\_state} & $[65536, 32768, 32768, 45875]$ & post-sleep recovery (Q16 setpoint) \\
\texttt{mutation\_count} & $1$ & \\
\texttt{last\_w\_commit} & \texttt{0xa180ef10\dots 70ef7c8} & daemon initial $W$ (prior epoch) \\
\texttt{weight\_commit} & \texttt{0x19a9ffbd\dots 1883c03} & new HKDF-rotated $W$ epoch \\
\bottomrule
\end{tabularx}
\caption{Post-mutation on-chain agent state for the Tier~1 PoC end-to-end demonstration (\cref{sec:eval-onchain}), witnessed by the transactions in \cref{tab:p3-onchain-tx}.}
\label{tab:p3-poststate}
\end{table}

\section{Implementation record layouts}\label{app:implementation-layouts}

\subsection{Economic-metabolism \texttt{AgentV4} layout}\label{app:agentv4-layout}
The deployed \texttt{AgentV4} record extends the 254-byte \texttt{AgentV3} layout to 342~bytes. The added fields are recorded exactly below with their declared types in the deployed Anchor program (\texttt{Pubkey} is the runtime's 32-byte address type) and their byte ranges within the serialised record, written $\mathit{start}..\mathit{end}$ as for the proof's public values; the main text describes their security role in the aliveness predicate (\cref{sec:phase4-m0}).

\begin{description}[itemsep=2pt,topsep=2pt,leftmargin=2.2em]
  \item[\texttt{wallet\_pubkey: Pubkey} (bytes $254..286$)] Program-owned economic PDA, $\mathrm{PDA}([\texttt{"agent\_wallet"},\pk])$, whose lamports constitute $\Fecon(K)$. Only the program's metabolic debit can move them; the operator endows the account at genesis.
  \item[\texttt{econ\_balance\_lamports: u64} (bytes $286..294$)] Cached lamport snapshot, updated atomically at every advance for read-only callers; the underlying account balance is canonical.
  \item[\texttt{last\_econ\_check\_slot: u64} (bytes $294..302$)] Solana slot of the latest $\Fecon$ check, recorded with the balance snapshot.
  \item[\texttt{oracle\_pubkey: Pubkey} (bytes $302..334$)] Registered designated environment oracle verified by \texttt{advance\_v5} (\cref{sec:phase4-m1}).
  \item[\texttt{sleep\_count: u64} (bytes $334..342$)] Monotonic sleep-trigger replay guard appended after the economic fields; these 8~bytes bring the record to 342~bytes (\cref{sec:design-phase3}).
\end{description}

\section{Glossary}\label{sec:glossary}

This glossary collects short definitions of cross-domain terms used throughout the paper, organised by domain. Each entry is one to two sentences; full discussion is in the body section indicated by the cross-references.

\paragraph{Cryptographic primitives.}
\begin{description}[itemsep=2pt,topsep=2pt,leftmargin=1.6em]
  \item[$\ed$ EdDSA] Standard 32-byte public-key / 64-byte signature scheme on edwards25519. Used here for all agent-to-chain authentication. (\cref{sec:bg-primitives})
  \item[$\HKDF$ (HKDF-SHA256, RFC~5869)] HMAC-based key derivation function. Takes a secret and a domain-separating tag, produces a deterministic byte stream. Used to derive the agent's neural-network weights $W$ from $\sk$. (\cref{sec:bg-primitives,sec:f2})
  \item[$\keccak$-256] $256$-bit hash function (the variant matching Solana's \texttt{keccak} syscall). Used for on-chain weight and state commitments. (\cref{sec:bg-primitives})
  \item[Groth16] Pairing-based zero-knowledge proof system with constant-cost (bounded compute-unit) on-chain verification; a bare proof is ${\sim}260$ bytes regardless of circuit size (the SP1-wrapped proofs reported in this paper, which carry the public values, are ${\sim}1.5$~KB). Instantiated here via SP1~v5.2.4. (\cref{sec:bg-primitives,sec:f1})
  \item[Trusted setup (Groth16)] A one-time preparatory ceremony that generates a Groth16 system's proving and verifying keys from secret randomness (the ``toxic waste''); soundness assumes that randomness was destroyed. (\cref{sec:threat-scope})
  \item[SP1] A zkVM (zero-knowledge virtual machine) from Succinct Labs. Compiles Rust to RISC-V and produces a Groth16 proof of the Rust program's execution. (\cref{sec:bg-primitives})
  \item[Zero-knowledge proof] A cryptographic proof that a computation was performed correctly without revealing inputs or intermediate values. (\cref{sec:bg-primitives,sec:f1})
  \item[TEE (trusted execution environment)] A hardware-isolated enclave (e.g., Intel SGX, Phala) that attests, at the hardware level, the integrity of the code and data running inside it. (\cref{sec:disc-internal-shift})
  \item[Witness] A prover-held secret input to a zero-knowledge proof; the proven statement is verified without the witness being revealed. (\cref{sec:bg-zkflow,sec:f1})
  \item[Rigid designator] From Kripke (1980)~\cite{kripke1980naming}, a name that refers to the same entity in every counterfactual scenario. Used here only as a \emph{loose analogy} for stable reference under the agent's $\sk$ within one deployed protocol, not as a claim of metaphysical identity across all counterfactual scenarios. (\cref{sec:properties,sec:disc-barandiaran-triad,sec:disc-limits})
  \item[Commitment] A short digest used here to bind a prover to a value; a hash-based commitment is hiding only to the extent that its input has sufficient entropy. Revealing the value later permits the digest to be checked, and substitution is computationally infeasible under the stated hash assumptions. (\cref{sec:f2})
\end{description}

\paragraph{Solana / blockchain.}
\begin{description}[itemsep=2pt,topsep=2pt,leftmargin=1.6em]
  \item[Program] A deterministic Rust binary deployed at a public address on Solana, processing transactions by mutating accounts it owns. (\cref{sec:bg-solana})
  \item[Instruction] An operation invoked by name on a specific program with specific accounts. A transaction is a signed bundle of one or more instructions, applied atomically. (\cref{sec:bg-solana})
  \item[Account] A region of bytes at a Solana address, owned by exactly one program (only that program can mutate it; anyone can read). (\cref{sec:bg-solana})
  \item[Lamport] The smallest unit of SOL; $10^9$ lamports $=1$~SOL. (\cref{sec:bg-solana})
  \item[Fee (transaction fee)] The lamport amount the runtime debits from a transaction's designated fee-payer account at landing: a per-signature base fee ($5{,}000$ lamports each, precompile-verified signatures included) plus an optional priority fee per requested compute unit; landed-but-reverted transactions still pay. (\cref{sec:bg-solana,sec:phase4-m0})
  \item[PDA (program-derived address)] A deterministic account address computed by hashing a program ID with a list of \emph{seeds} (byte strings). No private key corresponds to a PDA, so only the owning program can sign for it. (\cref{sec:bg-solana,sec:f2})
  \item[Seeds] Byte strings used to derive PDAs (e.g., \texttt{"agent\_v1"} + agent public key for the per-agent on-chain record). (\cref{sec:bg-solana})
  \item[CU (compute unit)] Solana's unit of execution cost, with per-instruction and per-transaction caps. (\cref{sec:bg-solana,sec:eval-cu})
  \item[devnet] A public Solana test cluster sharing the execution model and APIs used here with mainnet, but differing in validator conditions, load, stability, and economics; its SOL is faucet-issued and has no monetary value. (\cref{sec:bg-solana})
  \item[Finality (finalized commitment)] The depth at which the cluster guarantees a transaction can no longer be rolled back or replaced by a competing fork (a \emph{reorganisation}); this paper's no-fork claims are stated at finalized depth. (\cref{sec:bg-solana,sec:threat-scope})
  \item[Stake-weighted committee / validator / slashing] Future-work concepts for decentralising the environment oracle and prover: a quorum of stake-weighted \emph{validators} (nodes publishing signed environment attestations, distinct from Solana's block-producing validators) with \emph{slashing} --- confiscation of staked collateral --- penalising provable misbehaviour. The of-record advance path uses a single designated oracle, not a committee. (\cref{sec:phase4-m1,sec:disc-future-directions})
\end{description}

\paragraph{Machine learning.}
\begin{description}[itemsep=2pt,topsep=2pt,leftmargin=1.6em]
  \item[Elman recurrence] A basic recurrent neural network where the hidden state at time $t$ combines the previous hidden state $h_{t-1}$ with a new input $x_t$. (\cref{sec:bg-elman})
  \item[Hidden state] The recurrent network's internal vector that carries information from past steps to future ones. (\cref{sec:bg-elman})
  \item[Fixed-point arithmetic ($\mathrm{Q}16.16$)] Integer encoding of fractional numbers with 16 integer bits and 16 fraction bits (total 32-bit signed; also written $\mathrm{Q}16$ for brevity). Used for bit-exact reproducibility across hosts. (\cref{sec:bg-elman})
  \item[HRRL-inspired homeostatic driver] A motivation driver inspired by homeostatic reinforcement learning, in which internal drives must stay near setpoints and deviations create internal ``discomfort.'' The deployed driver regulates these variables but does not itself learn. (\cref{sec:design-phase3})
  \item[Allostatic free energy ($F_{\text{allostasis}}$)] A squared-distance proxy for total drive deviation from setpoints; here computed in $\mathrm{Q}16$ arithmetic. (\cref{sec:design-phase3})
\end{description}

\paragraph{Biology / philosophy of individuality and agency.}
\begin{description}[itemsep=2pt,topsep=2pt,leftmargin=1.6em]
  \item[Biological individual] A unified, distinct, persistent entity that counts as one thing rather than as a sum of parts; classical candidate criteria include genetic continuity, immune self-recognition, autonomous reproduction, and metabolic closure. (\cref{sec:intro})
  \item[Agency (Barandiaran et al.\ 2009)] Defined by three jointly held conditions: \emph{individuality}, \emph{interactional asymmetry}, and \emph{normativity}. (\cref{sec:disc-barandiaran-triad})
  \item[Interactional asymmetry] The agent acts on the environment in ways the environment does not act back symmetrically. (\cref{sec:disc-barandiaran-triad})
  \item[Normativity] The agent has self-referential interests that ground for-the-agent values; approached here only as the economic \emph{precondition} $\Fecon(K) > 0$. (\cref{sec:disc-barandiaran-triad})
  \item[Autopoiesis] A self-producing system that maintains its own organisation through ongoing metabolic operations; an ALIFE root concept of individuation. (\cref{sec:intro})
  \item[Autocatalytic set] A collection of molecules that catalyse each other's production, sometimes proposed as a chemical basis for individuation. (\cref{sec:related})
\end{description}

\paragraph{Paper-specific constructions.}
\begin{description}[itemsep=2pt,topsep=2pt,leftmargin=1.6em]
  \item[$F_2$ (genesis circuit)] SP1 guest circuit that proves $\WC = \keccak(\HKDF(\sk, \mathit{tag}))$ at agent registration. (\cref{sec:f2})
  \item[$F_1$ (advance circuit)] SP1 guest circuit that proves each Elman step under a privately witnessed $W$ and republishes $\WC = \keccak(W)$; the binding to the genesis commitment is enforced on-chain at every state transition. (\cref{sec:f1})
  \item[State-commitment chain ($\Stream(K)$)] An append-only sequence of per-cycle commitments $c_t$, each chained from $c_{t-1}$ and signed by the agent's $\sk$. (\cref{sec:chain})
  \item[Aliveness predicate] The advance-admissibility and state-transition check ``$K$ valid $\wedge\ \Stream(K)$ advancing $\wedge\ \Fecon(K) > 0$''; a stipulative protocol-viability label whose operational scope and economic-extension enforcement are given in \cref{sec:phase4-motivation,sec:phase4-m0}.
  \item[\texttt{AgentV3}, \texttt{AgentV4}] Successive on-chain record layouts of the extension lineage. V3 is the PoC-extension record (core protocol fields plus the motivation and mutation fields); V4 adds the economic-metabolism fields (wallet pubkey, balance snapshot, econ-check slot, oracle pubkey, sleep count). (\cref{sec:eval-onchain,sec:phase4-m0,app:agentv4-layout})
  \item[\texttt{advance\_v2}, \texttt{advance\_v5}] Successive advance instructions of the extension lineage (the core program's advance instruction is named \texttt{advance}). v2: active-query extension; v5: + aliveness predicate and oracle attestation gate. (\cref{sec:design-phase2,sec:phase4-m0,sec:phase4-m1})
  \item[Of-record] Designates the specific deployments, runs, and measurements whose results this paper reports, as distinct from third-party reproductions of the same protocol; the of-record agents' secret seeds are privately held (\cref{sec:eval-setup} and the Reproducibility section).
  \item[Patron] A designated cooperative counterparty with two distinct PoC-tier roles: (i)~\emph{environment responder}, answering the agent's per-cycle query with a signed observation vector (\cref{sec:design-phase2}); and (ii)~\emph{fee funder}, initially paying the agent's transaction fees --- a role no longer required by the economic-metabolism protocol, although its fee funds remain externally supplied (\cref{sec:phase4-motivation,sec:phase4-closure}).
  \item[Key-anchored weight rotation] Construction allowing $W$ updates that remain a deterministic function of $\sk$, the mutation counter, and a key-holder-attested replay-root value across multiple epochs. The root is not a circuit-verified history summary, and the rotation is not itself a learning rule. (\cref{sec:design-phase3})
  \item[Sleep-mini protocol] Mechanism gating a sleep cycle when allostatic free energy meets or exceeds a threshold; sleep skips the active query, restores reserve, and halves thermal/novelty gaps. (\cref{sec:design-phase3})
\end{description}

\clearpage
\bibliographystyle{plain}
\bibliography{references}

@article{suzuki2026externality,
  author       = {Keisuke Suzuki},
  title        = {Artificial {Externality}: A Three-Layer Model of Reality from Substrate to Smart Contract},
  journal      = {Philosophy and Technology},
  note         = {Forthcoming. Preprint: \url{https://philpapers.org/rec/SUZAEA}},
  year         = {2026}
}

@article{barandiaran2009define,
  author       = {Barandiaran, Xabier E. and Di Paolo, Ezequiel and Rohde, Marieke},
  title        = {Defining Agency: Individuality, Normativity, Asymmetry, and Spatio-temporality in Action},
  journal      = {Adaptive Behavior},
  volume       = {17},
  number       = {5},
  pages        = {367--386},
  year         = {2009},
  doi          = {10.1177/1059712309343819}
}

@misc{baltieri2025mathematical,
  author       = {Manuel Baltieri and Keisuke Suzuki},
  title        = {Mathematical Approaches to the Study of Agents},
  year         = {2025},
  howpublished = {PsyArXiv preprint, \url{https://osf.io/preprints/psyarxiv/rqu7s_v1}}
}

@misc{coslett2026which,
  author       = {Anthony Coslett},
  title        = {Which Model Is Running? --- Structural Identity as a Prerequisite for Trustworthy Zero-Knowledge Machine Learning},
  year         = {2026},
  howpublished = {Zenodo preprint, \url{https://zenodo.org/records/19008116}},
  note         = {Identity-first zkML framework binding a pre-existing model by structural fingerprinting under TEE attestation}
}

@misc{sp1securitymodel,
  author       = {{Succinct Labs}},
  title        = {{SP1} Security Model},
  year         = {2026},
  howpublished = {\url{https://docs.succinct.xyz/docs/sp1/security/security-model}}
}

@phdthesis{biehl2018formal,
  author       = {Biehl, Martin},
  title        = {Formal Approaches to a Definition of Agents},
  school       = {University of Hertfordshire},
  year         = {2018}
}

@article{kolchinsky2018semantic,
  author       = {Kolchinsky, Artemy and Wolpert, David H.},
  title        = {Semantic information, autonomous agency and non-equilibrium statistical physics},
  journal      = {Interface Focus},
  volume       = {8},
  number       = {6},
  pages        = {20180041},
  year         = {2018},
  doi          = {10.1098/rsfs.2018.0041}
}

@incollection{albantakis2021macro,
  author       = {Albantakis, Larissa and Massari, Francesco and Beheler-Amass, Maggie and Tononi, Giulio},
  title        = {A Macro Agent and Its Actions},
  booktitle    = {Top-down Causation and Emergence},
  pages        = {135--155},
  publisher    = {Springer},
  year         = {2021},
  doi          = {10.1007/978-3-030-71899-2_7}
}

@misc{ohlhaver2022desoc,
  author       = {Puja Ohlhaver and E. Glen Weyl and Vitalik Buterin},
  title        = {Decentralized Society: Finding {Web3}'s Soul},
  year         = {2022},
  howpublished = {SSRN preprint 4105763, \url{https://papers.ssrn.com/sol3/papers.cfm?abstract_id=4105763}}
}

@misc{zoltu2022eip5114,
  author       = {Micah Zoltu},
  title        = {{ERC-5114}: Soulbound Badge},
  year         = {2022},
  howpublished = {Ethereum Improvement Proposals, \url{https://eips.ethereum.org/EIPS/eip-5114}}
}

@misc{succinct2024sp1,
  author       = {{Succinct Labs}},
  title        = {{SP1}: A Performant, Open-Source {zkVM} for {Rust}},
  howpublished = {\url{https://github.com/succinctlabs/sp1}},
  note         = {v5.2.4 used; \texttt{sp1-solana} on-chain verifier},
  year         = {2024}
}

@misc{bonsol2024,
  author       = {{Bonsol Collective}},
  title        = {{Bonsol}: {RISC Zero} {zkVM} Coordination on {Solana}},
  howpublished = {\url{https://github.com/bonsol-collective/bonsol}},
  year         = {2024}
}

@misc{ezkl2024,
  author       = {{Zkonduit Inc.}},
  title        = {{EZKL}: {Halo2}-based {zkSNARK} Compilation of {ONNX} Models},
  howpublished = {\url{https://github.com/zkonduit/ezkl}},
  note         = {v23.0.5},
  year         = {2024}
}

@misc{modulus2024,
  author       = {{Modulus Labs}},
  title        = {The Cost of Intelligence: Proving Machine Learning Inference with Zero-Knowledge},
  howpublished = {Talk at Scroll Applied ZK Reading Group (presenter: Ryan Cao), \url{https://www.youtube.com/watch?v=nsMM2iE_oUA}},
  year         = {2023},
  note         = {Project acquired by Tools for Humanity, December 2024; modulus.xyz domain inactive as of 2026-05}
}

@misc{ora2024,
  author       = {{ORA Protocol}},
  title        = {{ORA}: Onchain {AI} Oracle and Verifiable {AI} Infrastructure},
  howpublished = {Project documentation, \url{https://docs.ora.io/}},
  year         = {2024},
  note         = {Self-described as ``decentralized, trustless applications powered by verifiable AI''; project active as of 2026-05; offerings include AI Oracle, Tora Launcher, and IMO framework}
}

@misc{giza2024,
  author       = {{Giza}},
  title        = {{Giza}: {zkML} Stack for {Starknet} ({Cairo} / {Orion})},
  howpublished = {Project documentation, \url{https://docs.gizatech.xyz/}},
  year         = {2024},
  note         = {Cited for the original zkML-on-Cairo stack; project repositioned to ``Agents for on-chain capital'' (verified at docs.gizatech.xyz, 2026-05); the original Cairo/Orion stack is no longer publicly featured}
}

@misc{gensyn2024,
  author       = {{Gensyn}},
  title        = {{Gensyn}: A Decentralized Network for Machine Intelligence (Verifiable {ML} Compute)},
  howpublished = {Project documentation, \url{https://docs.gensyn.ai/}; research \url{https://gensyn.ai/research}},
  year         = {2024},
  note         = {Self-described as ``the Network for Machine Intelligence: an open infrastructure layer for AI''; verified active, 2026-05; main product Delphi (on-chain information markets)}
}

@misc{w3c2022did,
  author       = {{W3C}},
  title        = {Decentralized Identifiers ({DID}s) v1.0: Core Architecture, Data Model, and Representations},
  howpublished = {W3C Recommendation},
  year         = {2022},
  url          = {https://www.w3.org/TR/did-core/},
  note         = {Cited for the DID/VC identity model: a controller key attests an external subject identifier; the computational substrate acting under that key is out of the standard's scope}
}

@inproceedings{groth2016size,
  author       = {Jens Groth},
  title        = {On the Size of Pairing-Based Non-interactive Arguments},
  booktitle    = {EUROCRYPT 2016},
  year         = {2016},
  pages        = {305--326}
}

@inproceedings{boneh2018vdf,
  author       = {Dan Boneh and Joseph Bonneau and Benedikt B{\"u}nz and Ben Fisch},
  title        = {Verifiable Delay Functions},
  booktitle    = {CRYPTO 2018},
  pages        = {757--788},
  year         = {2018}
}

@article{bernstein2012ed25519,
  author       = {Daniel J. Bernstein and Niels Duif and Tanja Lange and Peter Schwabe and Bo-Yin Yang},
  title        = {High-speed high-security signatures},
  journal      = {Journal of Cryptographic Engineering},
  volume       = {2},
  number       = {2},
  pages        = {77--89},
  year         = {2012}
}

@misc{krawczyk2010hkdf,
  author       = {Hugo Krawczyk and Pasi Eronen},
  title        = {{HMAC}-based Extract-and-Expand Key Derivation Function ({HKDF})},
  howpublished = {RFC 5869},
  year         = {2010}
}

@article{elman1990finding,
  author       = {Jeffrey L. Elman},
  title        = {Finding Structure in Time},
  journal      = {Cognitive Science},
  volume       = {14},
  number       = {2},
  pages        = {179--211},
  year         = {1990}
}

@book{kripke1980naming,
  author       = {Saul A. Kripke},
  title        = {Naming and Necessity},
  publisher    = {Harvard University Press},
  year         = {1980}
}

@misc{bip32,
  author       = {Pieter Wuille},
  title        = {{BIP-32}: Hierarchical Deterministic Wallets},
  howpublished = {Bitcoin Improvement Proposal 32},
  year         = {2012}
}

@misc{zkalife2026threatmodel,
  author       = {Keisuke Suzuki},
  title        = {{zkALIFE} Phase 1 Threat Model},
  howpublished = {\url{https://github.com/ksk-S/internalising-identity-2026/blob/arxiv-v1/docs/threat-model.md}},
  year         = {2026}
}

@misc{solanaexplorerprogram,
  author       = {{Solana Foundation}},
  title        = {{Solana Explorer entry for \texttt{7H4Dgrq2kXkeCVs3KDQB52F6\allowbreak Pvz1AuudsdCqW6RbPare} on devnet}},
  howpublished = {\url{https://explorer.solana.com/address/7H4Dgrq2kXkeCVs3KDQB52F6Pvz1AuudsdCqW6RbPare?cluster=devnet}},
  year         = {2026}
}

@misc{hu2025sporewild,
  title        = {Spore in the Wild: A Case Study of {Spore.fun} as an Open-Environment Evolution Experiment with Sovereign {AI} Agents on {TEE}-Secured Blockchains},
  author       = {Hu, Botao Amber and Rong, Helena},
  year         = {2025},
  eprint       = {2506.04236},
  archivePrefix= {arXiv},
  primaryClass = {cs.MA},
  url          = {https://arxiv.org/abs/2506.04236}
}

@misc{lin2025baid,
  title        = {Binding Agent {ID}: Unleashing the Power of {AI} Agents with accountability and credibility},
  author       = {Lin, Zibin and Zhang, Shengli and Liao, Guofu and Tao, Dacheng and Wang, Taotao},
  year         = {2025},
  eprint       = {2512.17538},
  archivePrefix= {arXiv},
  url          = {https://arxiv.org/abs/2512.17538}
}

@article{muller2026evolvableAI,
  title    = {Evolvable {AI}: Threats of a new major transition in evolution},
  author   = {M{\"u}ller, Viktor and Steels, Luc and Szathm{\'a}ry, E{\"o}rs},
  journal  = {Proceedings of the National Academy of Sciences},
  year     = {2026},
  volume   = {123},
  number   = {17},
  pages    = {e2527700123},
  doi      = {10.1073/pnas.2527700123},
  url      = {https://www.pnas.org/doi/10.1073/pnas.2527700123}
}

@inproceedings{liao2026zklora,
  title     = {{zkLoRA}: Fine-Tuning Large Language Models with Verifiable Security via Zero-Knowledge Proofs},
  author    = {Liao, Guofu and Wang, Taotao and Zhang, Shengli and Zhang, Jiqun and Shi, Long and Tao, Dacheng},
  booktitle = {Proceedings of the Network and Distributed System Security Symposium (NDSS)},
  year      = {2026},
  eprint    = {2508.21393},
  archivePrefix = {arXiv},
  url       = {https://arxiv.org/abs/2508.21393}
}

@article{li2025auditablellm,
  title    = {{AuditableLLM}: A Hash-Chain-Backed, Compliance-Aware Auditable Framework for Large Language Models},
  author   = {Li, Delong and Yu, Guangsheng and Wang, Xu and Liang, Bin},
  journal  = {Electronics},
  year     = {2026},
  volume   = {15},
  number   = {1},
  pages    = {56},
  publisher= {MDPI},
  doi      = {10.3390/electronics15010056},
  url      = {https://www.mdpi.com/2079-9292/15/1/56}
}

@misc{liu2025diap,
  title         = {{DIAP}: A Decentralized Agent Identity Protocol with Zero-Knowledge Proofs and a Hybrid {P2P} Stack},
  author        = {Liu, Yuanjie and Xing, Wenpeng and Zhou, Ye and Chang, Gaowei and Lin, Changting and Han, Meng},
  year          = {2025},
  eprint        = {2511.11619},
  archivePrefix = {arXiv},
  url           = {https://arxiv.org/abs/2511.11619}
}

@inproceedings{sheybani2023zkrownn,
  title     = {{ZKROWNN}: Zero Knowledge Right of Ownership for Neural Networks},
  author    = {Sheybani, Nojan and Ghodsi, Zahra and Kapila, Ritvik and Koushanfar, Farinaz},
  booktitle = {Proceedings of the 60th ACM/IEEE Design Automation Conference (DAC)},
  year      = {2023},
  eprint    = {2309.06779},
  archivePrefix = {arXiv},
  url       = {https://arxiv.org/abs/2309.06779}
}

@misc{cheng2024oml,
  title         = {{OML}: A Primitive for Reconciling Open Access with Owner Control in {AI} Model Distribution},
  author        = {Cheng, Zerui and Contente, Edoardo and Finch, Ben and Golev, Oleg and Hayase, Jonathan and Miller, Andrew and Moshrefi, Niusha and Nasery, Anshul and Nailwal, Sandeep and Oh, Sewoong and Tyagi, Himanshu and Viswanath, Pramod},
  year          = {2024},
  eprint        = {2411.03887},
  archivePrefix = {arXiv},
  url           = {https://arxiv.org/abs/2411.03887}
}

@misc{zkace2026,
  title         = {{ZK-ACE}: Identity-Centric Zero-Knowledge Authorization for Post-Quantum Blockchain Systems},
  author        = {Wang, Jian Sheng},
  year          = {2026},
  eprint        = {2603.07974},
  archivePrefix = {arXiv},
  primaryClass  = {cs.CR},
  url           = {https://arxiv.org/abs/2603.07974}
}

@misc{aesp2026,
  title         = {{AESP}: A Human-Sovereign Economic Protocol for {AI} Agents with Privacy-Preserving Settlement},
  author        = {Wang, Jian Sheng},
  year          = {2026},
  eprint        = {2603.00318},
  archivePrefix = {arXiv},
  primaryClass  = {cs.CR},
  url           = {https://arxiv.org/abs/2603.00318},
  note          = {Derives context-isolated authorisation keys from an identity root; it does not derive a model substrate from that root}
}

@misc{zhang2025dgm,
  title         = {{Darwin G{\"o}del Machine}: Open-Ended Evolution of Self-Improving Agents},
  author        = {Zhang, Jenny and Hu, Shengran and Lu, Cong and Lange, Robert and Clune, Jeff},
  year          = {2025},
  eprint        = {2505.22954},
  archivePrefix = {arXiv},
  primaryClass  = {cs.AI},
  url           = {https://arxiv.org/abs/2505.22954}
}

@inproceedings{jia2021pol,
  title     = {Proof-of-{Learning}: Definitions and Practice},
  author    = {Jia, Hengrui and Yaghini, Mohammad and Choquette-Choo, Christopher A. and Dullerud, Natalie and Thudi, Anvith and Chandrasekaran, Varun and Papernot, Nicolas},
  booktitle = {Proceedings of the 42nd {IEEE} Symposium on Security and Privacy},
  year      = {2021},
  eprint    = {2103.05633},
  archivePrefix = {arXiv},
  url       = {https://arxiv.org/abs/2103.05633}
}

@inproceedings{hu2024speculating,
  author    = {Hu, Botao Amber and Fangting},
  title     = {Speculating on Blockchain as an Unstoppable `{Nature}' Towards the Emergence of Artificial Life},
  booktitle = {Proceedings of the 2024 Artificial Life Conference ({ALIFE}~2024)},
  pages     = {127},
  publisher = {MIT Press},
  year      = {2024},
  doi       = {10.1162/isal_a_00818},
  url       = {https://doi.org/10.1162/isal_a_00818}
}

@misc{masumori2024life,
  author        = {Masumori, Atsushi and Maruyama, Norihiro and Ikegami, Takashi},
  title         = {Self-replicating and self-employed smart contract on {Ethereum} blockchain},
  year          = {2024},
  eprint        = {2405.04038},
  archivePrefix = {arXiv},
  url           = {https://arxiv.org/abs/2405.04038}
}

@misc{wilson_barker_sep_individual,
  author       = {Wilson, Robert A. and Barker, Matthew J.},
  title        = {Biological Individuals},
  year         = {2024},
  howpublished = {The Stanford Encyclopedia of Philosophy, Edward N. Zalta and Uri Nodelman (eds.)},
  note         = {First published 2007, substantive revision 2024},
  url          = {https://plato.stanford.edu/entries/biology-individual/}
}

@book{mcconwell2023individuality,
  author       = {McConwell, Alison K.},
  title        = {Biological Individuality},
  year         = {2023},
  publisher    = {Cambridge University Press},
  series       = {Cambridge Elements in the Philosophy of Biology},
  doi          = {10.1017/9781108942775}
}

@book{lidgard_nyhart_2017_individuality,
  editor       = {Lidgard, Scott and Nyhart, Lynn K.},
  title        = {Biological Individuality: Integrating Scientific, Philosophical, and Historical Perspectives},
  year         = {2017},
  publisher    = {University of Chicago Press},
  address      = {Chicago},
  isbn         = {9780226446455}
}

@book{maturana1980autopoiesis,
  author       = {Maturana, Humberto R. and Varela, Francisco J.},
  title        = {Autopoiesis and Cognition: The Realization of the Living},
  publisher    = {D. Reidel},
  address      = {Dordrecht},
  year         = {1980}
}

@article{kauffman1986autocatalytic,
  author       = {Kauffman, Stuart A.},
  title        = {Autocatalytic sets of proteins},
  journal      = {Journal of Theoretical Biology},
  volume       = {119},
  number       = {1},
  pages        = {1--24},
  year         = {1986}
}

@misc{erc8004,
  author       = {De Rossi, Marco and Crapis, Davide and Ellis, Jordan and Reppel, Erik},
  title        = {{ERC-8004}: Trustless Agents},
  howpublished = {Ethereum Improvement Proposal, Standards Track (ERC)},
  year         = {2025},
  url          = {https://eips.ethereum.org/EIPS/eip-8004},
  note         = {Identity, Reputation, and Validation registries: an agent is an ERC-721 identifier whose tokenURI resolves to an off-chain agent card; stake-secured re-execution, zkML proofs, and TEE oracles are named as pluggable validation options}
}

@misc{wang2026zkagent,
  author        = {Wang, Lizheng and Lou, Hancheng and Li, Chongrong and Yu, Yu and Hu, Yuncong},
  title         = {{zkAgent}: Verifiable {LLM} Agent Execution via One-Shot Transcript Proofs},
  howpublished  = {IACR Cryptology ePrint Archive, Report 2026/199},
  year          = {2026},
  url           = {https://eprint.iacr.org/2026/199}
}

@misc{alqithami2026survey,
  author        = {Alqithami, Saad},
  title         = {Autonomous Agents on Blockchains: Standards, Execution Models, and Trust Boundaries},
  year          = {2026},
  eprint        = {2601.04583},
  archivePrefix = {arXiv},
  primaryClass  = {cs.MA},
  url           = {https://arxiv.org/abs/2601.04583}
}

@misc{nous2024teehee,
  author       = {{Nous Research} and {Teleport (Flashbots)}},
  title        = {Setting Your Pet Rock Free},
  howpublished = {Nous Research blog},
  year         = {2024},
  url          = {https://nousresearch.com/setting-your-pet-rock-free/},
  note         = {The tee\_hee\_he agent: X-account and Ethereum-wallet credentials generated and confined inside an Intel TDX enclave with a timed-release recovery window, so no developer can act as the agent during the run}
}

@misc{google2025ap2,
  author       = {{Google Cloud}},
  title        = {Announcing the {Agent Payments Protocol} ({AP2})},
  howpublished = {Google Cloud blog; protocol specification at \url{https://ap2-protocol.org/}},
  year         = {2025},
  url          = {https://ap2-protocol.org/},
  note         = {An open extension of the A2A protocol: cryptographically signed mandates capturing user intent and authorising agent-initiated payments; sample integrations include the HTTP-402-based x402 on-chain payment method}
}

@misc{zhou2026capabilities,
  author        = {Zhou, Ziling},
  title         = {Governing Dynamic Capabilities: Cryptographic Binding and Reproducibility Verification for {AI} Agent Tool Use},
  year          = {2026},
  eprint        = {2603.14332},
  archivePrefix = {arXiv},
  url           = {https://arxiv.org/abs/2603.14332},
  note          = {Binds agent tool configurations and capability declarations by signatures and reproducibility checks rather than by a SNARK-internal weight relation}
}

@misc{adler2024personhood,
  author        = {Adler, Steven and Hitzig, Zo{\"e} and Jain, Shrey and Brewer, Catherine and Chang, Wayne and others},
  title         = {Personhood credentials: Artificial intelligence and the value of privacy-preserving tools to distinguish who is real online},
  year          = {2024},
  eprint        = {2408.07892},
  archivePrefix = {arXiv},
  url           = {https://arxiv.org/abs/2408.07892}
}

\end{document}